\documentstyle[12pt]{article}

\def\doit#1#2{\ifcase#1\or#2\fi}

\doit0{
\skewchar\fivmi='177 \skewchar\sixmi='177 \skewchar\sevmi='177
\skewchar\egtmi='177 \skewchar\ninmi='177 \skewchar\tenmi='177
\skewchar\elvmi='177 \skewchar\twlmi='177 \skewchar\frtnmi='177
\skewchar\svtnmi='177 \skewchar\twtymi='177
\def\@magscale#1{ scaled \magstep #1}
}

\doit{0}{
\def\framingfonts#1{
\doit{#1}{\font\twfvmi  = ammi10   \@magscale5 
\skewchar\twfvmi='177 \skewchar\fivsy='60 \skewchar\sixsy='60
\skewchar\sevsy='60 \skewchar\egtsy='60 \skewchar\ninsy='60
\skewchar\tensy='60 \skewchar\elvsy='60 \skewchar\twlsy='60
\skewchar\frtnsy='60 \skewchar\svtnsy='60 \skewchar\twtysy='60
\font\twfvsy  = amsy10   \@magscale5 
\skewchar\twfvsy='60
\font\go=font018			
\font\sc=font005			
\def\Go#1{{\hbox{\go #1}}}	
\def\Sc#1{{\hbox{\sc #1}}}	
\def\Sf#1{{\hbox{\sf #1}}}	
\font\oo=circlew10	      
\font\ooo=circle10			
\font\ro=manfnt				
\def\kcl{{\hbox{\ro 6}}}		
\def\kcr{{\hbox{\ro 7}}}		
\def\ktl{{\hbox{\ro \char'134}}}	
\def\ktr{{\hbox{\ro \char'135}}}	
\def\kbl{{\hbox{\ro \char'136}}}	
\def\kbr{{\hbox{\ro \char'137}}}	
}}  
}

\catcode`@=11
\catcode`@=12

\let\du=\d			

\def\a{\alpha} \def\b{\beta}  \def\d{\delta}
\def\e{\epsilon}  \def\g{\gamma}
   \def\k{\kappa}
\def\l{\lambda} \def\m{\mu} \def\n{\nu} \def\o{\omega}
  \def\r{\rho} \def\s{\sigma}
\def\t{\tau}   
  \def\G{\Gamma} 
\def\L{\Lambda}  \def\P{\Pi} 
\def\S{\Sigma}  

\def\bo{{\raise-.46ex\hbox{\large$\Box$}}}		
\def\pr{\prod}						
\def\TH{{\raise.2ex\hbox{$\displaystyle \bigodot$}\mskip-4.7mu \llap H \;}}
\def\face{{\raise.2ex\hbox{$\displaystyle \bigodot$}\mskip-2.2mu \llap {$\ddot
	\smile$}}}					

\def\sp#1{{}^{#1}}				
\def\Tilde#1{{\widetilde{#1}}\hskip 0.015in}	 
\def\Hat#1{\widehat{#1}}			
\def\Bar#1{\overline{#1}}			
\def\leftrightarrowfill{$\mathsurround=0pt \mathord\leftarrow \mkern-6mu
	\cleaders\hbox{$\mkern-2mu \mathord- \mkern-2mu$}\hfill
	\mkern-6mu \mathord\rightarrow$}
\def\dvec#1{\vbox{\ialign{##\crcr
	\leftrightarrowfill\crcr\noalign{\kern-1pt\nointerlineskip}
	$\hfil\displaystyle{#1}\hfil$\crcr}}}		

\def\frac#1#2{{\textstyle{#1\over\vphantom2\smash{\raise.20ex
	\hbox{$\scriptstyle{#2}$}}}}}			
\def\sfrac#1#2{{\vphantom1\smash{\lower.5ex\hbox{\small$#1$}}\over
	\vphantom1\smash{\raise.4ex\hbox{\small$#2$}}}}	
\def\bfrac#1#2{{\vphantom1\smash{\lower.5ex\hbox{$#1$}}\over
	\vphantom1\smash{\raise.3ex\hbox{$#2$}}}}	
\def\afrac#1#2{{\vphantom1\smash{\lower.5ex\hbox{$#1$}}\over#2}}    

\newskip\humongous \humongous=0pt plus 1000pt minus 1000pt
\def\caja{\mathsurround=0pt}
\def\eqalign#1{\,\vcenter{\openup2\jot \caja
	\ialign{\strut \hfil$\displaystyle{##}$&$
	\displaystyle{{}##}$\hfil\crcr#1\crcr}}\,}
\newif\ifdtup
\def\panorama{\global\dtuptrue \openup2\jot \caja
	\everycr{\noalign{\ifdtup \global\dtupfalse
	\vskip-\lineskiplimit \vskip\normallineskiplimit
	\else \penalty\interdisplaylinepenalty \fi}}}
\def\li#1{\panorama \tabskip=\humongous				
	\halign to\displaywidth{\hfil$\displaystyle{##}$
	\tabskip=0pt&$\displaystyle{{}##}$\hfil
	\tabskip=\humongous&\llap{$##$}\tabskip=0pt
	\crcr#1\crcr}}

\doit0{
\def\ref#1{$\sp{#1)}$}
}

\parskip=\medskipamount			
\thicklines			    

\def\border{						
	\setlength{\unitlength}{1mm}
	\newcount\xco
	\newcount\yco
	\xco=-24
	\yco=12
	\begin{picture}(140,0)
	\put(\xco,\yco){$\ktl$}
	\advance\yco by-1
	{\loop
	\put(\xco,\yco){$\kcl$}
	\advance\yco by-2
	\ifnum\yco>-240
	\repeat
	\put(\xco,\yco){$\kbl$}}
	\xco=158
	\yco=12
	\put(\xco,\yco){$\ktr$}
	\advance\yco by-1
	{\loop
	\put(\xco,\yco){$\kcr$}
	\advance\yco by-2
	\ifnum\yco>-240
	\repeat
	\put(\xco,\yco){$\kbr$}}
        \put(-20,11){\tiny University of Maryland Elementary Particle
Physics University of Maryland Elementary Particle Physics University of
Maryland Elementary Particle Physics}
	\put(-20,-241.5){\tiny University of Maryland Elementary
Particle Physics University of Maryland Elementary Particle Physics
University of Maryland Elementary Particle Physics}
	\end{picture}
	\par\vskip-8mm}
\def\bordero{						
	\setlength{\unitlength}{1mm}
	\newcount\xco
	\newcount\yco
	\xco=-24
	\yco=12
	\begin{picture}(140,0)
	\put(\xco,\yco){$\ktl$}
	\advance\yco by-1
	{\loop
	\put(\xco,\yco){$\kcl$}
	\advance\yco by-2
	\ifnum\yco>-240
	\repeat
	\put(\xco,\yco){$\kbl$}}
	\xco=158
	\yco=12
	\put(\xco,\yco){$\ktr$}
	\advance\yco by-1
	{\loop
	\put(\xco,\yco){$\kcr$}
	\advance\yco by-2
	\ifnum\yco>-240
	\repeat
	\put(\xco,\yco){$\kbr$}}
	\put(-20,12){\ooo bacdefghidfghghdhededbihdgdfdfhhdheidhdhebaaahjhhdahbahgdedgehgfdiehhgdigicba}
	\put(-20,-241.5){\ooo ababaighefdbfghgeahgdfgafagihdidihiidhiagfedhadbfdecdcdfagdcbhaddhbgfchbgfdacfediacbabab}
	\end{picture}
	\par\vskip-8mm}
\def\headpic{						
	\indent
	\setlength{\unitlength}{.4mm}
	\thinlines
	\par
	\begin{picture}(29,16)
	\put(165,16){\line(1,0){4}}
	\put(170,16){\line(1,0){4}}
	\put(180,16){\line(1,0){4}}
	\put(175,0){\line(1,0){4}}
	\put(180,0){\line(1,0){4}}
	\put(185,0){\line(1,0){4}}
	\put(169,0){\line(0,1){16}}
	\put(170,0){\line(0,1){16}}
	\put(179,0){\line(0,1){16}}
	\put(180,0){\line(0,1){16}}
	\put(184,0){\line(0,1){16}}
	\put(185,0){\line(0,1){16}}
	\put(169,16){\oval(8,32)[bl]}
	\put(170,16){\oval(8,32)[br]}
	\put(179,0){\oval(8,32)[tl]}
	\put(185,0){\oval(8,32)[tr]}
	\end{picture}
	\par\vskip-6.5mm
	\thicklines}

\def\endtitle{\end{quotation}\newpage}			

\def\sect#1{\bigskip\medskip \goodbreak \noindent{\bf {#1}} \nobreak \medskip}
\def\refs{\sect{References} \footnotesize \frenchspacing \parskip=0pt}
\def\Item{\par\hang\textindent}

\def\[{\lfloor{\hskip 0.35pt}\!\!\!\lceil}
\def\]{\rfloor{\hskip 0.35pt}\!\!\!\rceil}
\def\delsl{{{\partial\!\!\! /}}}

\def\du#1#2{_{#1}{}^{#2}}
\def\ud#1#2{^{#1}{}_{#2}}
\def\dud#1#2#3{_{#1}{}^{#2}{}_{#3}}

\def\calM{{\cal M}}
\def\calR{{\cal R}}

\def\rma{{\rm a}} \def\rmb{{\rm b}} \def\rmc{{\rm c}} \def\rmd{{\rm d}} 
\def\rme{{\rm e}} \def\rmf{{\rm f}} \def\rmg{{\rm g}}

\def\plpl{{{\hskip0.03em}{}{+{\hskip -0.563em}{\raise -0.03em\hbox{$_+$}}
{\hskip 0.05pt}}{}{\hskip0.03em}}} 
\def\mimi{{{\hskip0.03em}{}{-{\hskip -0.563em}{\raise -0.05em\hbox{$_-$}}
{\hskip 0.05pt}}{}{\hskip0.03em}}}

\def\order#1#2{{\cal O}({#1}^{#2})}

\def\pl#1#2#3{Phys.~Lett.~{\bf {#1}B} (19{#2}) #3}
\def\np#1#2#3{Nucl.~Phys.~{\bf B{#1}} (19{#2}) #3}
\def\prl#1#2#3{Phys.~Rev.~Lett.~{\bf #1} (19{#2}) #3}
\def\pr#1#2#3{Phys.~Rev.~{\bf D{#1}} (19{#2}) #3}
\def\cqg#1#2#3{Class.~and Quant.~Gr.~{\bf {#1}} (19{#2}) #3}

\def\ap#1#2#3{Ann.~of Phys.~{\bf {#1}} (19{#2}) #3} 
\def\prep#1#2#3{Phys.~Rep.~{\bf {#1}C} (19{#2}) #3}

\def\ijmp#1#2#3{Int.~Jour.~Mod.~Phys.~{\bf A{#1}} (19{#2}) #3}
\def\nc#1#2#3{Nuovo Cim.~{\bf {#1}} (19{#2}) #3}

\def\zp#1#2#3{Zeit.~f\"ur Phys.~{\bf{#1}C} (19{#2}) {#3}} 
\def\jgtp#1#2#3{Jour.~of Group Theory for Physicists, {\bf{#1}} (19{#2}) {#3}}

\def\hepth#1{hep-th/{#1}}

\def\ul{\underline} 
\def\un{\underline} 
  
\def\Check#1{{\raise0.6pt\hbox{\Large\v{}}{\hskip -10pt}{#1}}}
 
\def\Pisl{{\Pi\!\!\!\! /}}
\def\eqques{{~\,={\hskip -11.5pt}\raise -1.8pt\hbox{\large ?}
{\hskip 4.5pt}\,}}
\def\fracm#1#2{\,\hbox{\large{${\frac{{#1}}{{#2}}}$}}\,}
\def\fracmm#1#2{\,{{#1}\over{#2}}\,}
\def\half{{\fracm12}}

\def\frac#1#2{{\textstyle{#1\over\vphantom2\smash{\raise -.20ex
	\hbox{$\scriptstyle{#2}$}}}}}			

\def\sqrttwo{{\sqrt2}}
\def\scst{\scriptstyle}

\def\.{.$\,$}
\def\-{{\hskip 1.5pt}\hbox{-}}
\def\kd#1#2{\d\du{#1}{#2}}
\def\footnotew#1{\footnote{{\hsize=7.0in {\lineskip=-5pt #1}}}
\baselineskip 16pt\oddsidemargin=0.03in 
\evensidemargin=0.01in\hsize=6.5in\textwidth=6.5in} 
\def\low#1{\hskip0.01in{\raise -3pt\hbox{${\hskip 1.0pt}\!_{#1}$}}}

\def\Dot#1{\buildrel{\hskip2.5pt_{\scriptscriptstyle\bullet}}\over{#1}}

\def\det{{\rm det}\,}

\begin{document}

\font\tenmib=cmmib10
\font\sevenmib=cmmib10 at 7pt 
\font\fivemib=cmmib10 at 5pt  
\font\tenbsy=cmbsy10
\font\sevenbsy=cmbsy10 at 7pt 
\font\fivebsy=cmbsy10 at 5pt  
\def\BMfont{\textfont0\tenbf \scriptfont0\sevenbf
                              \scriptscriptfont0\fivebf
            \textfont1\tenmib \scriptfont1\sevenmib
                               \scriptscriptfont1\fivemib
            \textfont2\tenbsy \scriptfont2\sevenbsy
                               \scriptscriptfont2\fivebsy}
\def\rlx{\relax\leavevmode}                  
\def\BM#1{\rlx\ifmmode\mathchoice
                      {\hbox{$\BMfont#1$}}
                      {\hbox{$\BMfont#1$}}
                      {\hbox{$\scriptstyle\BMfont#1$}}
                      {\hbox{$\scriptscriptstyle\BMfont#1$}}
                 \else{$\BMfont#1$}\fi}

\font\tenmib=cmmib10
\font\sevenmib=cmmib10 at 7pt 
\font\fivemib=cmmib10 at 5pt  
\font\tenbsy=cmbsy10
\font\sevenbsy=cmbsy10 at 7pt 
\font\fivebsy=cmbsy10 at 5pt  
\def\BMfont{\textfont0\tenbf \scriptfont0\sevenbf
                              \scriptscriptfont0\fivebf
            \textfont1\tenmib \scriptfont1\sevenmib
                               \scriptscriptfont1\fivemib
            \textfont2\tenbsy \scriptfont2\sevenbsy
                               \scriptscriptfont2\fivebsy}
\def\BM#1{\rlx\ifmmode\mathchoice
                      {\hbox{$\BMfont#1$}}
                      {\hbox{$\BMfont#1$}}
                      {\hbox{$\scriptstyle\BMfont#1$}}
                      {\hbox{$\scriptscriptstyle\BMfont#1$}}
                 \else{$\BMfont#1$}\fi}

\def\inbar{\vrule height1.5ex width.4pt depth0pt}
\def\sinbar{\vrule height1ex width.35pt depth0pt}
\def\ssinbar{\vrule height.7ex width.3pt depth0pt}
\font\cmss=cmss10
\font\cmsss=cmss10 at 7pt
\def\ZZ{\rlx\leavevmode
             \ifmmode\mathchoice
                    {\hbox{\cmss Z\kern-.4em Z}}
                    {\hbox{\cmss Z\kern-.4em Z}}
                    {\lower.9pt\hbox{\cmsss Z\kern-.36em Z}}
                    {\lower1.2pt\hbox{\cmsss Z\kern-.36em Z}}
               \else{\cmss Z\kern-.4em Z}\fi}
\def\Ik{\rlx{\rm I\kern-.18em k}}  
\def\IC{\rlx\leavevmode
             \ifmmode\mathchoice
                    {\hbox{\kern.33em\inbar\kern-.3em{\rm C}}}
                    {\hbox{\kern.33em\inbar\kern-.3em{\rm C}}}
                    {\hbox{\kern.28em\sinbar\kern-.25em{\rm C}}}
                    {\hbox{\kern.25em\ssinbar\kern-.22em{\rm C}}}
             \else{\hbox{\kern.3em\inbar\kern-.3em{\rm C}}}\fi}
\def\IP{\rlx{\rm I\kern-.18em P}}
\def\IR{\rlx{\rm I\kern-.18em R}}
\def\IN{\rlx{\rm I\kern-.20em N}}
\def\Ione{\rlx{\rm 1\kern-2.7pt l}}

\def\unredoffs{} \def\redoffs{\voffset=-.31truein\hoffset=-.59truein}
\def\speclscape{\special{ps: landscape}}

\newbox\leftpage \newdimen\fullhsize \newdimen\hstitle \newdimen\hsbody
\tolerance=1000\hfuzz=2pt\def\fontflag{cm}
\catcode`\@=11 
\doit0
{
\def\bigans{b }
\message{ big or little (b/l)? }\read-1 to\answ
\ifx\answ\bigans\message{(This will come out unreduced.}
}
\hsbody=\hsize \hstitle=\hsize 
\doit0{
\else\message{(This will be reduced.} \let\l@r=L
\redoffs \hstitle=8truein\hsbody=4.75truein\fullhsize=10truein\hsize=\hsbody
\output={\ifnum\pageno=0 
  \shipout\vbox{\speclscape{\hsize\fullhsize\makeheadline}
    \hbox to \fullhsize{\hfill\pagebody\hfill}}\advancepageno
  \else
  \almostshipout{\leftline{\vbox{\pagebody\makefootline}}}\advancepageno
  \fi}
}
\def\almostshipout#1{\if L\l@r \count1=1 \message{[\the\count0.\the\count1]}
      \global\setbox\leftpage=#1 \global\let\l@r=R
 \else \count1=2
  \shipout\vbox{\speclscape{\hsize\fullhsize\makeheadline}
      \hbox to\fullhsize{\box\leftpage\hfil#1}}  \global\let\l@r=L\fi}
\fi
\def\nolabels{\def\wrlabeL##1{}\def\eqlabeL##1{}\def\reflabeL##1{}}
\def\writelabels{\def\wrlabeL##1{\leavevmode\vadjust{\rlap{\smash%
{\line{{\escapechar=` \hfill\rlap{\sevenrm\hskip.03in\string##1}}}}}}}%
\def\eqlabeL##1{{\escapechar-1\rlap{\sevenrm\hskip.05in\string##1}}}%
\def\reflabeL##1{\noexpand\llap{\noexpand\sevenrm\string\string\string##1}}}
\nolabels
%
\global\newcount\secno \global\secno=0
\global\newcount\meqno \global\meqno=1
\def\newsec#1{\global\advance\secno by1\message{(\the\secno. #1)}
\global\subsecno=0\eqnres@t\noindent{\bf\the\secno. #1}
\writetoca{{\secsym} {#1}}\par\nobreak\medskip\nobreak}
\def\eqnres@t{\xdef\secsym{\the\secno.}\global\meqno=1\bigbreak\bigskip}
\def\sequentialequations{\def\eqnres@t{\bigbreak}}\xdef\secsym{}
\global\newcount\subsecno \global\subsecno=0
\def\subsec#1{\global\advance\subsecno by1\message{(\secsym\the\subsecno. #1)}
\ifnum\lastpenalty>9000\else\bigbreak\fi
\noindent{\it\secsym\the\subsecno. #1}\writetoca{\string\quad
{\secsym\the\subsecno.} {#1}}\par\nobreak\medskip\nobreak}
\def\appendix#1#2{\global\meqno=1\global\subsecno=0\xdef\secsym{\hbox{#1.}}
\bigbreak\bigskip\noindent{\bf Appendix #1. #2}\message{(#1. #2)}
\writetoca{Appendix {#1.} {#2}}\par\nobreak\medskip\nobreak}
%
%
\def\eqnn#1{\xdef #1{(\secsym\the\meqno)}\writedef{#1\leftbracket#1}%
\global\advance\meqno by1\wrlabeL#1}
\def\eqna#1{\xdef #1##1{\hbox{$(\secsym\the\meqno##1)$}}
\writedef{#1\numbersign1\leftbracket#1{\numbersign1}}%
\global\advance\meqno by1\wrlabeL{#1$\{\}$}}
\def\eqn#1#2{\xdef #1{(\secsym\the\meqno)}\writedef{#1\leftbracket#1}%
\global\advance\meqno by1$$#2\eqno#1\eqlabeL#1$$}
%
\newskip\footskip\footskip14pt plus 1pt minus 1pt 
\def\footnotefont{\ninepoint}\def\f@t#1{\footnotefont #1\@foot}
\def\f@@t{\baselineskip\footskip\bgroup\footnotefont\aftergroup\@foot\let\next}
\setbox\strutbox=\hbox{\vrule height9.5pt depth4.5pt width0pt}
\global\newcount\ftno \global\ftno=0
\def\foot{\global\advance\ftno by1\footnote{$^{\the\ftno}$}}
%
\newwrite\ftfile
\def\footend{\def\foot{\global\advance\ftno by1\chardef\wfile=\ftfile
$^{\the\ftno}$\ifnum\ftno=1\immediate\openout\ftfile=foots.tmp\fi%
\immediate\write\ftfile{\noexpand\smallskip%
\noexpand\item{f\the\ftno:\ }\pctsign}\findarg}%
\def\footatend{\vfill\eject\immediate\closeout\ftfile{\parindent=20pt
\centerline{\bf Footnotes}\nobreak\bigskip\input foots.tmp }}}
\def\footatend{}
%
%
\global\newcount\refno \global\refno=1
\newwrite\rfile
%
\def\ref{[\the\refno]\nref}%
\def\nref#1{\xdef#1{[\the\refno]}\writedef{#1\leftbracket#1}%
\ifnum\refno=1\immediate\openout\rfile=refs.tmp\fi%
\global\advance\refno by1\chardef\wfile=\rfile\immediate%
\write\rfile{\noexpand\Item{#1}\reflabeL{#1\hskip.31in}\pctsign}%
\findarg\hskip10.0pt}%
\def\findarg#1#{\begingroup\obeylines\newlinechar=`\^^M\pass@rg}
{\obeylines\gdef\pass@rg#1{\writ@line\relax #1^^M\hbox{}^^M}%
\gdef\writ@line#1^^M{\expandafter\toks0\expandafter{\striprel@x #1}%
\edef\next{\the\toks0}\ifx\next\em@rk\let\next=\endgroup\else\ifx\next\empty%
\else\immediate\write\wfile{\the\toks0}\fi\let\next=\writ@line\fi\next\relax}}
\def\striprel@x#1{} \def\em@rk{\hbox{}}
\def\lref{\begingroup\obeylines\lr@f}
\def\lr@f#1#2{\gdef#1{\ref#1{#2}}\endgroup\unskip}
\def\semi{;\hfil\break}
\def\addref#1{\immediate\write\rfile{\noexpand\item{}#1}} 
\def\listrefs{\footatend\vfill\supereject\immediate\closeout\rfile\writestoppt
\baselineskip=14pt\centerline{{\bf References}}\bigskip{\frenchspacing%
\parindent=20pt\escapechar=` \input refs.tmp\vfill\eject}\nonfrenchspacing}
%
\def\listrefsr{\immediate\closeout\rfile\writestoppt
\baselineskip=14pt\centerline{{\bf References}}\bigskip{\frenchspacing%
\parindent=20pt\escapechar=` \input refs.tmp\vfill\eject}\nonfrenchspacing}
\def\listrefsrsmall{\immediate\closeout\rfile\writestoppt
\baselineskip=11pt\centerline{{\bf References}}
\font\smallreffonts=cmr9 \font\it=cmti9 \font\bf=cmbx9%
\bigskip{ {\smallreffonts%
\parindent=15pt\escapechar=` \input refs.tmp\vfill\eject}}}
\def\startrefs#1{\immediate\openout\rfile=refs.tmp\refno=#1}
\def\xref{\expandafter\xr@f}\def\xr@f[#1]{#1}
\def\refs#1{\count255=1[\r@fs #1{\hbox{}}]}
\def\r@fs#1{\ifx\und@fined#1\message{reflabel \string#1 is undefined.}%
\nref#1{need to supply reference \string#1.}\fi%
\vphantom{\hphantom{#1}}\edef\next{#1}\ifx\next\em@rk\def\next{}%
\else\ifx\next#1\ifodd\count255\relax\xref#1\count255=0\fi%
\else#1\count255=1\fi\let\next=\r@fs\fi\next}
\def\figures{\centerline{{\bf Figure Captions}}\medskip\parindent=40pt%
\def\fig##1##2{\medskip\item{Fig.~##1.  }##2}}
%
\newwrite\ffile\global\newcount\figno \global\figno=1
\def\fig{fig.~\the\figno\nfig}
\def\nfig#1{\xdef#1{fig.~\the\figno}%
\writedef{#1\leftbracket fig.\noexpand~\the\figno}%
\ifnum\figno=1\immediate\openout\ffile=figs.tmp\fi\chardef\wfile=\ffile%
\immediate\write\ffile{\noexpand\medskip\noexpand\item{Fig.\ \the\figno. }
\reflabeL{#1\hskip.55in}\pctsign}\global\advance\figno by1\findarg}
\def\listfigs{\vfill\eject\immediate\closeout\ffile{\parindent40pt
\baselineskip14pt\centerline{{\bf Figure Captions}}\nobreak\medskip
\escapechar=` \input figs.tmp\vfill\eject}}
\def\xfig{\expandafter\xf@g}\def\xf@g fig.\penalty\@M\ {}
\def\figs#1{figs.~\f@gs #1{\hbox{}}}
\def\f@gs#1{\edef\next{#1}\ifx\next\em@rk\def\next{}\else
\ifx\next#1\xfig #1\else#1\fi\let\next=\f@gs\fi\next}
\newwrite\lfile
{\escapechar-1\xdef\pctsign{\string\%}\xdef\leftbracket{\string\{}
\xdef\rightbracket{\string\}}\xdef\numbersign{\string\#}}
\def\writedefs{\immediate\openout\lfile=labeldefs.tmp \def\writedef##1{%
\immediate\write\lfile{\string\def\string##1\rightbracket}}}
\def\writestop{\def\writestoppt{\immediate\write\lfile{\string\pageno%
\the\pageno\string\startrefs\leftbracket\the\refno\rightbracket%
\string\def\string\secsym\leftbracket\secsym\rightbracket%
\string\secno\the\secno\string\meqno\the\meqno}\immediate\closeout\lfile}}
\def\writestoppt{}\def\writedef#1{}
\def\seclab#1{\xdef #1{\the\secno}\writedef{#1\leftbracket#1}\wrlabeL{#1=#1}}
\def\subseclab#1{\xdef #1{\secsym\the\subsecno}%
\writedef{#1\leftbracket#1}\wrlabeL{#1=#1}}
\newwrite\tfile \def\writetoca#1{}
\def\leaderfill{\leaders\hbox to 1em{\hss.\hss}\hfill}
\def\writetoc{\immediate\openout\tfile=toc.tmp
   \def\writetoca##1{{\edef\next{\write\tfile{\noindent ##1
   \string\leaderfill {\noexpand\number\pageno} \par}}\next}}}
\def\listtoc{\centerline{\bf Contents}\nobreak\medskip{\baselineskip=12pt
 \parskip=0pt\catcode`\@=11 \input toc.tex \catcode`\@=12 \bigbreak\bigskip}}
\catcode`\@=12 
%

\def\kd#1#2{\d\du{#1}{#2}}
\def\jgtp#1#2#3{Jour.~of Group Theory for Physicists {c{#1}} (19{#2})
{#3}}

\def\sqrttwo{{\sqrt2}}
\def\hata{\hat a} \def\hatb{\hat b}  \def\hatc{\hat c}  \def\hatd{\hat d}
\def\hate{\hat e} \def\hatf{\hat f}  \def\hatg{\hat g}  \def\hath{\hat h}
\def\Pup{P_\uparrow} \def\Pdown{P_\downarrow} 
\def\Pupdown{P_{\uparrow\downarrow}} \def\Pdownup{P_{\downarrow\uparrow}}  
\def\HatF{\Hat F} \def\HatR{\Hat R}

\def\footnotew#1{\footnote{{\baselineskip 1pt\hsize=7.0in {#1}}}
\baselineskip 16pt\oddsidemargin=0.03in
\evensidemargin=0.01in\hsize=6.5in\textwidth=6.5in} 

\def\na{n_a} \def\nb{n_b}
\def\nc{n_c} \def\nd{n_d} \def\ne{n_e} \def\nf{n_f}
\def\ng{n_g} \def\ma{m_a} \def\mb{m_b} \def\mc{m_c}
\def\md{m_d} \def\me{m_e} \def\mf{m_f} \def\mg{m_g}  

\def\Du#1#2{\low{#1}{}^{#2}} \def\uD#1#2{^{#1}{}\low{#2}} 

\def\phia{\nabla_a\varphi} \def\phib{\nabla_b\varphi} 
\def\phic{\nabla_c\varphi} \def\phid{\nabla_d\varphi}
\def\phie{\nabla_e\varphi} \def\phif{\nabla_f\varphi} 
\def\phig{\nabla_g\varphi}
\def\phim{D_m\varphi} \def\phin{D_n\varphi} 
\def\phir{D_r\varphi} \def\phis{D_s\varphi} 
\def\phimu{\partial_\m\varphi} \def\phinu{\partial_\n\varphi}
\def\phiroh{\partial_\r\varphi} \def\phisigma{\partial_\s\varphi}
\def\phitau{\partial_\t\varphi} \def\philambda{\partial_\l\varphi}
\def\phiomega{\partial_\o\varphi}  \def\phipsi{\partial_\psi\varphi} 

\def\tilphim{D_m\Tilde\varphi} \def\tilphin{D_n\Tilde\varphi} 
\def\tilphir{D_r\Tilde\varphi} \def\tilphis{D_s\Tilde\varphi} 
\def\tilphia{\nabla_a\Tilde\varphi} \def\tilphib{\nabla_b\Tilde\varphi} 
\def\tilphic{\nabla_c\Tilde\varphi} \def\tilphid{\nabla_d\Tilde\varphi}
\def\tilphie{\nabla_e\Tilde\varphi} \def\tilphif{\nabla_f\Tilde\varphi} 
\def\tilphig{\nabla_g\Tilde\varphi}
\def\phimu{\partial_\m\varphi} \def\phinu{\partial_\n\varphi}
\def\tilphiroh{\partial_\r\Tilde\varphi} 
\def\tilphisigma{\partial_\s\Tilde\varphi}
\def\tilphitau{\partial_\t\Tilde\varphi} 

\def\nsl{{n\!\!\!\!\hskip1.2pt/}\,} \def\msl{{m\!\!\!\!/}\hskip1.8pt}

\def\Pisl{\Pi\!\!\!\!/\hskip2.0pt}   
\def\nablasl{\nabla\!\!\!\!/} 
\def\calM{{\cal M}}

\def\parenth#1{\left({#1}\right)} 
\def\brack#1{\left[ \,{#1} \, \right]} 

\def\Check#1{{\raise0.6pt\hbox{\Large\v{}}{\hskip -10pt}{#1}}}
\def\fracm#1#2{\,\hbox{\large{${\frac{{#1}}{{#2}}}$}}\,}
\def\fracmm#1#2{\,{{#1}\over{#2}}\,}
\def\rma{\rm a} \def\rmb{\rm b} \def\rmc{\rm c} \def\rmd{\rm d} 
\def\rme{\rm e} \def\rmf{\rm f} \def\rmg{\rm g} 
\def\rmh{\rm h} \def\rmi{\rm i} \def\rmj{\rm j} \def\rmk{\rm k}
\def\rml{\rm l} \def\rmm{\rm m} 
 
\def\fermionsquare{\order\psi 2}  
\def\TildeM{\Tilde{\cal M}}

\def\framing#1{\doit{#1}
{\framingfonts{#1}
\border\headpic 
}}

\framing{0}
~~~

\baselineskip 12pt 

\doit0
{\bf Preliminary Version (FOR YOUR EYES ONLY!) \hfill \today\\
}
\vskip 0.07in

{\hbox to\hsize{July 1998
\hfill UMDEPP 98--126}}
{\hbox to\hsize{~~~~~ ~~~~~
\doit1{\hfill ~hep-th/9807199}%
}

\vskip -0.035in
\hfill {(Revised Version)}\\

\begin{center}
\vglue 0.2in

\baselineskip 18pt 

{\large\bf Supergravity Theories in $~D\ge 12$} \\
{\large\bf Coupled to Super p-Branes}$\,$\footnote
{This work is supported in part by NSF grant \# PHY-93-41926.} \\  

\baselineskip 9pt 

\vskip 0.5in

Hitoshi~ N{\small ISHINO}

\vskip 0.08in

{\it Department of Physics} \\[.015in]
{\it University of Maryland} \\[.015in]
{\it College Park, MD 20742-4111, USA} \\[.020in]   
{E-Mail: nishino@nscpmail.physics.umd.edu}

\vskip 2.3in

{\bf Abstract} \\[0.18in]  
\end{center}

\begin{quotation}

\vbox{\baselineskip 16pt 
~~~We construct supergravity theories in twelve and thirteen dimensions with the
respective signatures $~(10,2)$~ and $~(11,2)$~ with some technical details.  
Starting with $~N=1$~
supergravity in 10+2 dimensions coupled to Green-Schwarz superstring, we give
$~N=2$~ chiral supergravity in 10+2 dimensions with its couplings to
super ~$(2+2)\-$brane.  We  also build an $~N=1$~ supergravity in 11+2 dimensions,
coupled to supermembrane.  All of these formulations utilize scalar
(super)fields intact  under supersymmetry, replacing
the null-vectors introduced in their original formulations.  This method 
makes all the equations $~SO(10,2)$~ or $~SO(11,2)$~ 
Lorentz covariant, up to modified Lorentz generators.  
We inspect the internal consistency of these formulations, in
particular with the usage of the modified Lorentz generators for the extra
coordinates.       
} 

\endtitle

\oddsidemargin=0.03in
\evensidemargin=0.01in
\hsize=6.5in
\textwidth=6.5in

\vskip 0.1in                                                       
\centerline{\bf 1.~~Introduction}          
\bigskip
                                                                        
There has been strong indication that higher-dimensional theories of extended
objects in higher than eleven dimensions (11D) \ref\cjs{E.~Cremmer, B.~Julia and N.~Scherk, \pl{76}{78}{409}; E.~Cremmer and
B.~Julia, \pl{80}{78}{48}; \np{159}{790}{141}.}, such as F-theory
\ref\ftheory{C.~Vafa, \np{469}{96}{403}.} or S-theory \ref\stheory{I.~Bars,
\pr{55}{97}{2373}.}, all with multiple time  coordinates \ref\twotimes{I.~Bars
and C.~Kounnas, \pl{402}{97}{25}; \pr{56}{97}{3664}.} have many
promising features.  In particular, these theories may well provide the guiding
principle for understanding non-perturbative features  or vacuum structures of
superstring \ref\gsw{M.~Green,
J.H.~Schwarz and E.~Witten, {\it `Superstring Theory'}, Vols.~I and II, 
Cambridge University Press (1987).} or supermembrane \ref\bst{E.~Bergshoeff, 
E.~Sezgin and P.K.~Townsend, \pl{189}{87}{75}; \ap{185}{88}{330}.} theories 
{\it via} M-theory \ref\km{D.~Kutasov and
E.~Martinec, \np{477}{96}{652}.}\ref\bfss{T.~Banks, W.~Fischler, S.H.~Shenker 
and L.~Susskind, \pr{55}{97}{5112}.}\ref\bars{I.~Bars, 
\pr{54}{96}{5203}.}\ref\mtheoryrev{{\it For reviews, 
see e.g.,} P.K.~Townsend, 
{\it `M-Theory from its Superalgebra'}, \hepth{9712004};  
A.~Bilal, {\it `M(atrix) Theory: A Pedagogical Introduction'}, \hepth{9710136};
J.H.~Schwarz, {\it `Beyond Gauge Theories'}, \hepth{9807195}; 
{\it and references in them}.}   
in terms of supersymmetry algebra, {\it e.g.,} in $~D=10+2$~ 
\stheory\footnotew{In the expression $~D=s+t$~ 
the number $~s$~ (or $~t$) denotes that of spacial (or time) coordinates.  
When clear from the context, we also use 
the expression $~12D$~ or $~D=12$.} or ~$D=11+3$~ \ref\barsfourteen{I.~Bars,
\pl{403}{97}{257}.}.   

Motivated by this philosophy, explicit field theoretic formulations of a
supersymmetric Yang-Mills theory in $~D=10+2$~ 
\ref\ns{H.~Nishino and E.~Sezgin, \pl{388}{96}{569}.}, or 
in $~D=11+3~$ \ref\sezgin{E.~Sezgin, \pl{403}{97}{265}.}, and of
an $~N=1$~ supergravity theory \ref\nishone{H.~Nishino, \pl{428}{98}{85}.}, or
of an $~N=2$~ supergravity theory \ref\nishtwo{H.~Nishino, {\it `N=2~
Chiral Supergravity in (10+2)-Dimensions as Consistent Background for Super
(2+2)$\-$Brane'}, hep-th/9706148, to appear in Phys.~Lett.~B.} 
have been recently presented.  Further developed are an invariant lagrangian for 
supersymmetric Yang-Mills theory
in $~D=10+2$, as well as a set of Lorentz covariant field equations for the
first time \ref\lag{H.~Nishino, \pl{426}{98}{64}.}, in all the even dimensions
higher than $~D=12$~  \ref\symall{H.~Nishino, \np{523}{98}{450}.}, or 
more general supersymmetry algebras \ref\rss{I.~Rudychev,
E.~Sezgin and P.~Sundell, Nucl.~Phys.~Proc.~Suppl.~{\bf 68} (1998) 285.}.   

All of these previous
formulations were based on null-vectors that are common in these dimensions with
more than a single time coordinate.  The existence of such supergravity theories
had been also vaguely predicted in various contexts, such as the categorization
of Clifford algebra in arbitrary dimensions \ref\kt{T.~Kugo and P.K.~Townsend,
\np{211}{83}{157}.}, due to the smallness of the freedoms in the Majorana-Weyl
spinors in 12D, when there are two time coordinates \kt\ref\bd{M.~Blencowe and
M.~Duff, \np{310}{88}{387}.}.  However, there is also a certain no-go theorem
\ref\nahm{W.~Nahm, \np{135}{78}{149}.}\ref\nogo{L.~Castellani, P.~Fr\'e,
F.~Giani, K.~Pilch and P.~van Nieuwenhuizen, \pr{26}{82}{1481}.} that prohibits
`conventional' supergravity theories in such higher-dimensions.  A recent new
technique in \lag\ introducing scalar fields intact under supersymmetry, seems
to bypass (but not overcome) this no-go theorem by making the higher-dimensional
supergravity/supersymmetry formulations manifestly $~SO(10,2)$~ Lorentz
covariant, up to modified Lorentz generators.  

In this present paper, we give improved 
formulations of higher-dimensional supergravity in which all the
null-vectors in the previous formulations \ns\nishone\nishtwo\ are replaced by
the gradients of scalar (super)fields which are invariant under
supersymmetry, both in superspace and component.  By this prescription, all of
these higher-dimensional supergravity theories will become
formally Lorentz covariant, leaving the non-covariant nature to the
modified Lorentz generators.  

This paper is organized as follows.  We start with the $~N=1$~ supergravity in
$~D=10+2$~ in superspace \nishone, where the consistency of the system is
guaranteed by the satisfaction of all the Bianchi identities, based on the
improved method using the gradient of scalar superfields making the system
$~SO(10,2)$~ Lorentz covariant as manifest as possible.  We give rather
detailed construction of this theory which was not given enough in our previous
paper \nishone, that will be common features of other supergravity
theories.  As a probe for the validity of this supergravity theory, 
we confirm fermionic invariance of Green-Schwarz superstring put in this
12D supergravity background.  
We next give the component formulation of $~N=2$~ supergravity in
$~D=10+2$~ \nishtwo\ predicted by F-theory \ftheory, which now looks
straightforward, once the $~N=1$~ case is understood.  We further confirm the
fermionic symmetry on the $~(2+2)\-$dimensional world-supervolume of super 
$~(2+2)\-$brane coupled to this $~N=2$~ supergravity theory in 12D.  Based on
the experience in 12D supergravity, we build an $~N=1$~ supergravity in 
$~D=11+2$, which
can be consistently coupled to supermembranes \bst\ with fermionic symmetries. 
Appendix A and B are devoted for useful identities in 12D and 13D, while in
Appendix C, we inspect the consistency of our modified Lorentz generators.  In
Appendix D, we study the consistency of our extra constraints in component with
supersymmetry.

\bigskip\bigskip


\oddsidemargin=0.03in
\evensidemargin=0.01in
\hsize=6.5in
\textwidth=6.5in
\vsize=8.3in
\baselineskip 15.5pt

\centerline{\bf 2.~~$N=1$~ Supergravity in $~D=10+2$}
\bigskip

\leftline{\bf 2.1~~Notations}

We first establish all the notational foundation, in order to deal with 
our $~N=1$~ supergravity in $~D=10+2$.  
We first fix our metric to be
$~\big(\eta_{a b}\big) = \hbox{diag}.~(-,+, \cdots, +,+,-)$, where we use the 
indices $~{\scst a,~b,~\cdots~=~(0),~(1),~\cdots,~(9),~(11),~(12)}$~ 
for local Lorentz
coordinates, while $~{\scst m,~n,~\cdots~=~0,~1,~\cdots,~9,~11,~12}$~ for 
curved coordinates, in this section of superspace.\footnotew{The reason 
we use two different index systems in superspace 
and component formulations in this 
paper is due to their proper advantages.  For example, the indices 
$~{\scst \a,~\b,~\cdots}$~ are more convenient for frequently-used 
spinorial components in superspace, while in component formulation these 
spinorial indices are usually implicit and suppressed.}  
Accordingly, our Clifford 
algebra is ~$\{ \g_a,\g_b\} = +2 \eta_{a b}$.  Relevantly, we have 
$~\e^{012\cdots 9\,11\,12}= +1$, and $~\g\low{(13)}\equiv \g\low{(0)} \g\low{(1)} 
\cdots\g\low{(9)} \g\low{(10)} \g\low{(12)}$.  Compared with the notation in 
ref.~\nishone, the only difference is the usage of $~\g^a$~ instead of
$~\s^a~$ for ~$\g\-$matrices.    
We next setup two null-vectors, 
which have zero norm, and are orthogonal to each other \ns:
$$\li{
&\left( n^a \right)
  = (0, 0, \cdots, 0 , + \frac1{\sqrt2}, - \frac1{\sqrt2}) ~~, ~~~~ 
\left( \na \right) = (0,0,\cdots,0, + \frac1{\sqrt2},+\frac1{\sqrt2})~~, \cr  
&\left( m^a \right)
  = (0, 0, \cdots, 0 , + \frac1{\sqrt2}, + \frac1{\sqrt2}) ~~, ~~~~ 
  \left(\ma \right) = (0, 0, \cdots, 0 , + \frac1{\sqrt2}, -\frac1{\sqrt2})
  ~~.      
&(2.1.1) \cr }$$  
It is also convenient to use $~{\scst\pm}\-$indices \nishone\nishtwo, 
in order to handle the extra dimensions:
$$ V_{\pm} \equiv \frac1{\sqrt2} \big( V_{(11)} \pm V_{(12)} \big) ~~.
\eqno(2.1.2) $$  
Accordingly, we have $~n_+ = m^+ = +1,~n_- = m^- = 0 $, and therefore  
$$  n^a\na = m^a\ma = 0 ~~,
~~~~ m^a\na = m^+ n_+ = m_- n^- = + 1 ~~.   
\eqno(2.1.3) $$ 
Now the necessity of $~\pm1/{\sqrt2}$~ in (2.1.1) is now obvious,  
maintaining the normalization $~m^a n_a=+1$.  

Other important quantities to be defined are the projection operators 
in the space of two extra coordinates, satisfying the usual 
ortho-normality conditions \nishone:
$$ \li{& \Pup \equiv \frac12
\nsl\msl = \frac12\g^+ \g^- ~~, ~~~~ \Pdown \equiv \frac12 \msl\nsl 
     = \frac12 \g^- \g^+ ~~,  
&(2.1.4\rm a) \cr 
& \Pup\Pup = +\Pup~~, ~~~~ \Pdown\Pdown = + \Pdown~~, ~~~~ 
\Pup + \Pdown = + I~~, 
&(2.1.4\rm b) \cr  
&\Pupdown \equiv \Pup - \Pdown = \g^{+ -}~~, 
&(2.1.4\rmc) \cr } $$
where as usual $~\msl\equiv m^a\g_a$~ and $~\nsl\equiv n^a\g_a$.  
The following symmetry properties are also useful for the manipulations 
of $~\g\-$matrices: 
$$\eqalign{&(\nsl)_{\a\Dot\b} = - (\nsl)_{\Dot\b\a} ~~, ~~~~
(\msl)_{\a\Dot\b} = - (\msl)_{\Dot\b\a} ~~,  \cr 
&(\Pup)_{\a\b} = - (\Pdown)_{\b\a} ~~, ~~~~ (\Pupdown)_{\a\b} = +
(\Pupdown)_{\b\a} ~~. \cr }   
\eqno(2.1.5) $$ 
Note that in our signature convention in 12D, the {\it dotted} (or {\it
undotted}) spinors have positive (or negative) chirality under $~\g_{13}$~
\nishone, as opposed to the usual convention.    
We also use the collective spinorial indices 
$~{\scst \un\a~\equiv~(\a,\Dot\a),~\un\b~\equiv~(\b,\Dot\b),
~\cdots}$~ to symbolize both chiralities, for the 
chiral spinorial indices $~{\scst \a,~\b,~\cdots~=~1,~2,~\cdots,~32}$~ and
$~{\scst \Dot\a,~\Dot\b,~\cdots~=~\Dot 1,~\Dot 2,~\cdots,~\Dot{32}}$.

We next study various features
of our modified Lorentz generators introduced in \nishone.  
These modified Lorentz generators are defined by \nishone
$$\li{&\big({\Tilde{\cal M}}_{a b}\big)^{c d} 
\equiv + {\Tilde\d}\du{\[a} c \, {\Tilde\d}\du{b\]} d ~~, 
&(2.1.6\rm a)\cr ~~~~ 
& \big({\Tilde{\cal M}}_{a b}\big) \du\a\b \equiv +\half \big(
\g_{a b}  \Pup \big) \du\a\b~~,  ~~~~ \big( \Tilde{\cal M}_{a b}\big)
\du{\Dot\a}{\Dot\b} \equiv +\half\big(\Pdown\g_{a b}\big)\du{\Dot\a}{\Dot\b}~~. 
&(2.1.6\rm b) \cr} $$
Here $~\Tilde\d$~ is defined by 
$$ \Tilde\d\du a b \equiv \d\du a b -m_a n^b 
     =\cases{ \d\du i j ~~& \hbox{(for ~${\scst a~=~i,~~b~=~j}$)}~~,  \cr 
     \d\du++ = 1 ~~&\hbox{(for ~${\scst a~=~+,~~b~=~+}$)} ~~, \cr
     0~~&\hbox{(otherwise)} ~~. \cr } 
\eqno(2.1.7) $$
Here $~{\scst i,~j,~\cdots}$~ are purely 10D indices, and in particular,
$~\Tilde\d\du-- =0$.  This is to be consistent with the spinorial representation
(2.1.6b) satisfying $~\big( \TildeM_{-i} \big)  \du{\un\a}{\un\b}=0$.  
The vectorial representation (2.1.6a) implies that  $~\big(\TildeM_{+-}\big)^{c
d}=0$, causing no problem with $~\big(\TildeM_{+-}\big)
\du{\un\a}{\un\b} 
\neq0$, because as long as $~\TildeM_{a b}$ ~ is
always accompanied by $~\phi\du A{a b}$, the combination $~\phi\du A{+-} 
\TildeM_{+ -}$~ vanishes due to the extra constraint $~\phi\du A{+-} =0$, to be
systematically given in (2.3.6).  Note also that the only effect of
(2.1.7) is to get rid of the unwanted generators $~\big(\TildeM_{-b}\big)^{c
d}$~ in the vectorial representation which does not vanish in the combination
$~\phi\du A{-b}\TildeM_{-b} $~ even with the extra constraint 
$~\phi\du A {+b}=0$~
on $~\phi$.  As is seen in (C.3) in Appendix C, we emphasize 
that these modified Lorentz generators satisfy the usual Jacobi identities
among $~\TildeM$'s, which is the foundation of the Bianchi identities in
superspace.  In the next subsection, we will confirm these 
Bianchi identities at engineering dimensions $~d=1$~
and $~d=3/2$.\footnotew{The engineering dimensions are defined in the usual way
in superspace, {\it i.e.,} we put dimension 1/2 (or 1) for a spinorial
derivative $~\nabla_\a$~ (or  bosonic derivative ~$\nabla_a$), which determine
all the dimensions of  torsion/curvatures, {\it e.g.,} the dimension of
$~T\du{\a\b}\g$~ is 1/2.  In this paper, we do not use differential forms in
order to avoid confusing expressions especially for index-contractions.}  

In order to see the internal consistency of our modified Lorentz generators, 
we first show that all the null-vectors are really `constant' 
under our superspace covariant derivatives \nishone
$$ \li{& \nabla\low M n^- = \partial\low M n^- 
     + \half \phi\low M{}^{a b} ({\TildeM}_{b a}) ^{-+} n_+ = 0 ~~,\cr  
&\nabla\low M m^+ =  \partial\low M m^+ 
     + \half\phi\low M{}^{a b}({\TildeM}_{b a})^{+ -} m_- = 0 ~~.
&(2.1.8)\cr } $$
We next establish the action of the Lorentz generators on spinorial components
\nishone:
$$ \li{ 
&\TildeM_{a b} \Psi_\a = + \big( \TildeM_{a b}\big)\du\a\b \Psi_\b~~, 
~~~~\TildeM_{a b} \Psi^\a = - \Psi^\b \big( \TildeM_{a b}\big) \du\b\a ~~,  \cr 
&\TildeM_{a b} \Bar\Psi_{\Dot\a} = + \big( \TildeM_{a b}\big)\du{\Dot\a}{\Dot\b} 
     \Psi_{\Dot\b}~~,  ~~~~ 
    \TildeM_{a b} \Bar\Psi^{\Dot\a} = - \Bar\Psi^{\Dot\b} 
     \big( \TildeM_{a b}\big) \du{\Dot\b}{\Dot\a} ~~. 
& (2.1.9) \cr} $$ 
Accordingly, we get the commutators involving the charge conjugation matrices
\nishone:
$$ \li{ & \[\TildeM_{i j} , C_{\a\b} \] = +\half \left( \g_{i j}\Pup
       \right)_{\a\b} ~~, ~~~~ 
       \[\TildeM_{i j} , C^{\a\b} \] = - \half \left( \g_{i j}\Pup
       \right)^{\a\b} ~~, \cr 
& \[\TildeM_{i j} , C_{\Dot\a\Dot\b} \] = +\half \left( \g_{i j}\Pup
       \right)_{\a\b} ~~, ~~~~ 
       \[\TildeM_{i j} , C^{\Dot\a\Dot\b} \] = - \half \left( \g_{i j}\Pup
       \right)^{\Dot\a\Dot\b} ~~, \cr
& \[ \TildeM_{-i}, C_{\a\b} \] = \[ \TildeM_{-i}, C^{\a\b} \] =
       \[ \TildeM_{-i}, C_{\Dot\a\Dot\b} \] 
       = \[ \TildeM_{-i}, C^{\Dot\a\Dot\b} \] = 0 ~~,  
&(2.1.10) \cr } $$
which can be easily confirmed using our definitions.  

The significance of (2.1.10) is that the charge conjugation matrices
transform under our Lorentz generators, but are no longer
constants.  Even though this sounds disastrous, we can easily see that all 
the $~\g\-$matrices used in our superspace constraints in (2.3.4) 
are shown to be 
constant, as desired.  In order to see this, we give the important relations
$$ \li{ & \[ \TildeM_{a b} , \left( \g^c\right)\du\g{\Dot\d} \] 
     = \Hat\d\du{\[a}c \big( \g_{b\]}\big)\du\g{\Dot\d}  
     	+ \half \big(\g\low{a b}\Pup\g^c  \big) \du\g{\Dot\d} 
      -\half \big(\g\low{a b}\Pdown \g^c\big) \du\g{\Dot\d} ~~, \cr 
& \[ \TildeM_{a b} , \big( \g^c\big)\du{\Dot\g}\d \] 
     = \Hat\d\du{\[a}c \big( \g_{b\]}\big)\du{\Dot\g}\d  
     	+ \half \big(\g\low{a b}\Pup\g^c \big) \du{\Dot\g}\d  
      -\half \big(\g\low{a b}\Pdown \g^c\big)\du{\Dot\g}\d ~~, 
&(2.1.11)\cr }  $$ 
which can help us to prove that 
$$ \eqalign{& \[ \TildeM_{a b}, \left(\nsl\right)\du\g{\Dot\d} \]  =  
    \[ \TildeM_{a b}, \left(\nsl\right)\du{\Dot\g}\d \]  = 0 ~~, \cr 
& \[ \TildeM_{a b}, \left(\msl\right)\du\g{\Dot\d} \]  = 
     \[ \TildeM_{a b}, \left(\msl\right)\du{\Dot\g}\d \]  = 0 ~~. \cr }
\eqno(2.1.12) $$
These relations yield the desirable commutativity of $~\nabla_A$~ with $~\nsl$~
and $~ \msl$:
$$\li{ &\[ \nabla_A, \, (\nsl)\du\a{\Dot\b} \]  
			= \[ \nabla_A, \,(\msl)\du\a{\Dot\b} \]  
   =  \[ \nabla_A, \, (\Pup)\du\a\b  \]   
   =  \[ \nabla_A,  \,(\Pdown)\du\a\b \] = 0 {~~,~~~~~ ~~~~~}  
&(2.1.13\rm a)  \cr   
&\[ \nabla_A,  \,(\g\low{c d} \Pup)\du\a\b  \] 
   =  \[  \nabla_A, \,(\Pdown \g\low{c d})\du{\Dot\a}{\Dot\b} \] =0~~, 
&(2.1.13\rm b) \cr 
&\[ \nabla_A, \, (\nsl\g^a\Pdown)_{\a\b} \]
   = \[ \nabla_A, \, (\Pup\g^a\nsl)_{\a\b} \] = 0 {~~.~~~} 
&(2.1.13\rm c) \cr } $$ 
As will be seen, these relations can help us to see, {\it e.g.}, 
$$ \[ \nabla_A, T\du{\a\b} c \] = 0~~, 
\eqno(2.1.14) $$
as desired for the total consistency of our constraints.  See Appendix C for 
other notes.

\bigskip\bigskip

\leftline{\bf 2.2~~Scalar Superfields Intact under Supersymmetry}

In ref.~\nishone, we formulated our 12D supergravity using the null-vectors 
introduced above.  However, this formulation had a drawback of breaking
the manifest $~SO(10,2)$~ Lorentz covariance in 12D.  In ref.~\lag, we have 
improved this for global supersymmetry by introducing a scalar field $~\varphi$~
whose gradient replaces  the null-vector: $~n_a\equiv \nabla_a\varphi$.  In this
section, we use  this prescription to re-formulate our $~N=1$~ supergravity in
\nishone, avoiding the usage of null-vectors, to make the $~SO(10,2)$~ Lorentz
covariance  as manifestly as possible.  In our supergravity formulation, we 
need an additional scalar superfield  
$~\Tilde\varphi$~ in addition to $~\varphi$~ \lag, 
satisfying the constraints \lag
$$\li{&\nabla_a \nabla_b\varphi =0 ~~, 
    ~~~~ \nabla_a\nabla_b\Tilde\varphi = 0~~, 
    ~~~~ \nabla_{\un\a}\varphi = 0 ~~, 
    ~~~~ \nabla_{\un\a}\Tilde\varphi = 0 ~~, \cr 
&(\nabla_a\varphi)^2 = 0 ~~, ~~~~(\nabla_a\Tilde\varphi)^2 = 0 ~~,
    ~~~~ (\nabla_a\varphi) (\nabla^a\Tilde\varphi) = 1 
    {~~. ~~~~~} 
&(2.2.1)  \cr } $$                            
As is easily seen, a set of non-trivial solutions to these differential equations
is $$\nabla_a \varphi = n_a ~~, ~~~~\nabla_a\Tilde\varphi = m_a ~~, 
\eqno(2.2.2) $$
where $~n_a,~m_a$~ are the same null-vectors we have already introduced.  The
novel feature of the superfield equations (2.2.1) is their manifest $~SO(10,2)$~
Lorentz covariance in 12D with no usage of null-vectors any longer, unless we
consider their solutions.  These scalar superfields enable us to  re-formulate
the whole supergravity systems given in  \nishone\nishtwo.  Until we replace
these gradient superfields by the null-vectors, all the superfield equations
in our system are manifestly $~SO(10,2)$~ Lorentz covariant, except for the
modified  Lorentz generators. 
Note also that if we choose a set of
solutions different from (2.2.2), {\it e.g.,} $~\varphi = \Tilde\varphi=0$, 
then the vacuum of the system will collapse to non-supersymmetric vacuum, even
{\it without} the supersymmetry algebra $~\{ Q_\a, Q_\b\} = \big( \g^{c d}\big) P_c
n_d$.  To put it differently, this feature is also understood as 
choosing non-trivial solutions for BPS condition $~\det\{ Q_\a,Q_\b \}=0$~ for
these higher-dimensional algebra of supersymmetry \stheory\barsfourteen. 

In an ordinary supersymmetric theory, there will arise a problem for
introducing any non-constant field invariant under supersymmetry.  For
example, the commutator $~\{ \nabla_\a , \nabla_\b\}\varphi =
\big(\g^c\big)_{\a\b}\nabla_c\varphi$~ does not hold for a
non-constant but superinvariant scalar field $~\varphi(x)$, 
because the l.h.s.~is zero due
to $~\nabla_\a\varphi=0$, while the r.h.s.~is non-zero due to 
$~\nabla_c\varphi\neq0$.  In our formulation, however, thanks to (2.2.1) and
$~T\du{\a\b} c$~ in (2.3.4a), the r.h.s.~also vanishes due to $~T\du{\a\b}c
\nabla_c\varphi = 0$~ and $~T\du{\a\b} c \nabla_c\Tilde\varphi = 0$, as 
will be seen in (2.3.4a), 
consistently with the vanishing l.h.s.~both for $~\varphi$~ and
$~\Tilde\varphi$.  This is one of the most important features of our formulation
based on superinvariant scalar fields, replacing the original null-vectors.   

The prescription of replacing all the null-vectors in \nishone\ by the
gradients of scalars superfields is transparent in superspace formulation, 
where the superfield equations are directly associated with  Bianchi
identities.  For example, the projection operators (2.1.4) can be re-expressed
as  
$$ \Pup \equiv \half \big(\g^a\g^b\big) 
        (\nabla_a\varphi) (\nabla_b\Tilde\varphi)~~, 
        ~~~~\Pdown \equiv\half \big(\g^a\g^b\big)  
        (\nabla_a\Tilde\varphi) (\nabla_b\varphi)~~, 
        ~~~~\Pupdown \equiv \Pup - \Pdown~~.      
\eqno(2.2.3) $$        
This is because even though these gradients seem `non-constant' as they stand, 
once those additional superfield equations (2.2.1) are taken into account,  they
are effectively `constant', and these operators play exactly the same  role as
the projection operators in (2.1.4).  The same is also true for 
Lorentz covariant
derivatives $~\nabla_A$.  

In (2.1.8) we were concerned with the null-vectors.  We can repeat the 
same analysis with the gradients $~\nabla_A\varphi$~ and
$~\nabla_A\Tilde\varphi$, now with $~n_a$~ replaced by  $~\nabla_a\varphi$~
and $~m_a$~ replaced by $~\nabla_a\Tilde\varphi$:
$$ \li{&\nabla\low M \nabla^-\varphi = \partial\low M \nabla^-\varphi 
     + \half \phi\low M{}^{a b} ({\TildeM}_{b a}) ^{-+} \nabla_+\varphi = 0
     ~~,\cr   &\nabla\low M \nabla^+\Tilde\varphi =  \partial\low M
     \nabla^+\Tilde\varphi + \half\phi\low M{}^{a b}({\TildeM}_{b a})^{+ -}
     \nabla_-\Tilde\varphi = 0~~. 
&(2.2.4)\cr } $$  
In principle, we can use {\it unmodified} Lorentz generators $~{\cal M}_{a b}$~ 
everywhere in these equations.  However, as we will see in (2.3.10),  
a Bianchi identity at dimension one requires the modified form of  
Lorentz generators with $~\Pup$~ inserted like in (2.1.6).  Or to put it
differently, only a particular set of solutions for the scalar fields
$~\varphi,~\Tilde\varphi$~ achieves the satisfaction of all the Bianchi
identities in a non-trivial way.  In this sense, the original $~SO(10,2)$~
Lorentz covariance is broken at the level of solutions, or equivalently by the
particular choice of modified Lorentz generators (2.1.6).  

As shrewd readers may have already noticed, we can even make our modified 
Lorentz generators (2.1.6) themselves more `covariant', by  
$$ \li{&\Tilde \d\du a b \equiv \d\du a b - \big( \nabla_a \Tilde\varphi\big) 
     \big( \nabla^b\varphi\big)~~,   
&(2.2.5) \cr} $$ 
replacing (2.1.7).  As has been already mentioned, since the $~\Pup,~\Pdown$~ 
can be replaced by (2.2.3), this prescription removes all the non-covariant 
ingredients in our 12D supergravity formulation.    
However, since this sort of field-dependent Lorentz generators, which
becomes constant only on-shell, might
be controversial, we do not claim that this method makes the whole system 
totally $~SO(10,2)$~ covariant, leaving this note just as another important 
ingredient of our supergravity formulations.

\bigskip\bigskip

\leftline{\bf 2.3~~Bianchi Identities and Superspace Constraints}

We next study the Bianchi identities in our system to be satisfied, which are 
$~T\-$, $~G\-$ and $~R\-$Bianchi identities:
$$
\li{&\frac12\nabla_{\[A} T\du{B C)}D - \frac12T\du{\[A B|} E T_{E| C)} - 
\frac14 R\du{\[A B| e}f \big(\calM \du f e \big)\du{|C)}D\equiv0~~,   
&(2.3.1) \cr
& \frac16 \nabla_{\[A} G_{B C D)} - \frac1 4 T\du{\[A B|} E G_{E| C D)}
\equiv 0   ~~,  
&(2.3.2) \cr
& \frac12\nabla_{\[A} R\du{B C) d} e 
     - \frac12 T\du{\[A B|} E R\du{E| C) d} e\equiv 0~~. 
&(2.3.3) \cr} $$ 
In all the sections for superspace, we use the symbol 
$~{\scst \[ ~)}$~ for (anti)symmetrization {\it without}
normalization, {\it i.e.,} $~A_{\[a b)} \equiv A_{a b} \mp A_{b a}$.  
We sometimes call (2.3.1) - (2.3.3) 
respectively the $~(A B C, D)$, ~$(A B C D)$~, and 
$~(A B C, d e)\-$type Bianchi identities for convenience sake.  

One subtlety to be examined is the satisfaction of the $~R\-$Bianchi identities
(2.3.3).  In the usual supergravity system, the $~R\-$Bianchi identities are
automatically satisfied, once the $~T\-$Bianchi identities hold 
\ref\dragon{N.~Dragon, \zp{2}{79}{29}.}.  In ref.~\nishone, this was non-trivial
due to the modified Lorentz generators.  The point there was that 
all the $~R\-$Bianchi identities are still satisfied, if we follow the same 
proof in the manifestly covariant case in \dragon, even
though the whole $~SO(10,2)$~ Lorentz covariance was lost.   

With these preliminaries at hand, we now present our superspace constraints.  
Our field  content is formally the same as the $~N=1$~ supergravity in 10D 
\ref\bergshoeff{E.~Bergshoeff, M.~de Roo, B.~de Wit and P.~van Nieuwenhuizen, 
\np{95}{82}{97}.}, namely  $~(e\du m a , \psi\du m\a, B_{m n},
\Bar\chi_{\Dot\a}, \Phi; \varphi, \Tilde\varphi)$, where  $~e\du ma$~ is the
zw\"olfbein, $~\psi\du m\a$~ is the Majorana-Weyl gravitino,  $~B_{m n}$~ is a
real antisymmetric tensor, $~\Bar\chi_{\Dot\a}$~ is an anti-chiral Majorana-Weyl
dilatino, and $~\Phi$~ is a real dilaton.    Our results for constraints in
superspace are \nishone    
$$ \li{&T\du{\a\b} c = \big( \g^{c d}\big)_{\a\b} \phid
     + (\g^{d e})_{\a\b} \big(\nabla^c\varphi\big)\big(\phid\big)
     \big(\tilphie\big) = \big( \g^{c d} \big)_{\a\b} 
      \phid + \left( \Pupdown \right)_{\a\b} \nabla^c\varphi {~~,~~~~~ ~~~~~} 
&(2.3.4\rma) \cr 
&G_{\a\b c} = T_{\a\b c} ~~, 
&(2.3.4\rmb)  \cr 
& T\du{\a\b} \g = (\Pup)\du{(\a|}\g \left( \g^c\Bar\chi
\right)_{|\b)}\phic - \big( \g^{a b} \big)_{\a\b} \left( \Pdown\g_a\Bar\chi
\right)^\g \phib 
&(2.3.4\rmc) \cr 
& \nabla_\a \Phi = \left( \g^c\Bar\chi \right)_\a \phic ~~, 
&(2.3.4\rmd) \cr 
& \nabla_\a\Bar\chi_{\Dot\b} = -\fracm1{24} \big( \g^{c d e}\Pup
\big)_{\a\Dot\b} G_{c d e} +\half \big( \g^c \Pup \big)_{\a\Dot\b} 
\nabla_c \Phi - \big( \g^c\Bar\chi \big)_\a \Bar\chi_{\Dot\b} \phic ~~, 
&(2.3.4\rme)\cr 
& T\du{\a b} c = 0 ~~, ~~~~ T\du{\a b}\g = 0 ~~,  ~~~~ G_{\a b c} = 0 ~~, 
&(2.3.4\rmf) \cr 
& T\du{a b} c = - G\du{a b} c ~~, 
&(2.3.4\rmg)  \cr     
&R_{\a\b c d} = + \big(\g^{e f}\big)_{\a\b} G_{f c d}\phie  ~~, 
&(2.3.4\rmh) \cr 
& \nabla_\a G_{b c d} = +\half \big( \g^e \g_{\[b} T_{c d\]} \big)_\a \phie = 
  - \nabla_\a T_{b c d} ~~, 
&(2.3.4\rmi) \cr  
& R_{\a b c d} = +\big(\g^e \g_{\[c} T_{d\] b} \big)_\a \phie ~~,
&(2.3.4\rmj) \cr   
&\nabla_a T\du{b c}\d = - \fracm14 (\g^{d e}\Pup)\du\a\d R_{b c d e} 
   + T_{b c}{}^\d \big(\g^e\Bar\chi \big)_\a \phie 
   + \big( \Pup\big)\du\a\d \big( \Bar\chi\g^e T_{b c} \big) \phie \cr 
& ~~~~~ ~~~~~ ~~~~~ 
     + \big( \g^{d e}T_{b c} \big)_\a \big( \Pdown\g_d \Bar\chi \big)^\d
       \phie {~,~~~~~~~~~~~~~}
&(2.3.4\rmk)   \cr 
&\nabla_{\ul\a}\varphi = \nabla_{\ul\a}\Tilde\varphi = 0 ~~, 
      ~~~~~ \big(\nabla_a\varphi\big)^2 = 
      \big(\nabla_a\Tilde\varphi\big)^2 = 0 ~~, ~~~~
      \big(\nabla_a \varphi \big) \big(\nabla^a\Tilde\varphi\big) = 1 ~~, 
&(2.3.4\ell) \cr 
& \nabla_a\nabla_b\varphi = \nabla_a\nabla_b\Tilde\varphi = 0 ~~.  
&(2.3.4\rmm)  \cr }  $$ 
Here $~\Pup,~\Pdown,~\Pupdown$~ are exactly the same as in (2.2.3).     
Our short-hand notation, such as $~\big(\g^c\Bar\chi\big)_\b \equiv 
\big(\g^c\big)\du\b{\Dot\g}\Bar\chi_{\Dot\g}$, and $~\big(\g^e\g_b T_{c d}\big) 
\equiv \big(\g^e\big)\du\a{\Dot\b} \big(\g_b\big)\du{\Dot\b}\g T_{c d\g}$, 
is always taken for granted.  As they
stand, these equations are formally $~SO(10,2)$~ Lorentz covariant.

\vfill\eject 

As usual in higher-dimensional supergravity, we have  
extra constraints \nishone:
$$ \li{ & T\du{A B} c \nabla_c\varphi = 0 ~~, ~~~~ 
     G_{A B c} \nabla^c\varphi =
     0~~, ~~~~   T\du{a B}C \nabla^a\varphi = 0 ~~, 
&(2.3.5) \cr 
& R\du{A B c} d \nabla_d\varphi  = R\du{a B c} d \nabla^a\varphi = 0 ~~, 
&(2.3.6) \cr 
& (\nabla^a\varphi) \nabla_a\Phi = 0 ~~,  
     ~~~~ (\nabla^a\varphi) \nabla_a \Bar\chi_{\Dot\b}= 0 ~~,
&(2.3.7) \cr 
& (\g^c)\du\a{\Dot\b} \Bar\chi_{\Dot\b} \nabla_c\Tilde\varphi = 0 ~~, ~~~~ 
    T\du{a b} \g (\g^d)\du \g {\Dot\a}\nabla_d\Tilde\varphi = 0 ~~,  
&(2.3.8) \cr
& \phi\du{A b} c \phic = \phi\du{a b} c \nabla^a\varphi = 0 ~~. 
&(2.3.9) \cr } $$
Note that $~\nabla_c\Tilde\varphi$~ appears instead of 
$~\nabla_c\varphi$~ in (2.3.8).
Notice that {\it not all} the extra components in these fields are deleted by 
these constraints (2.3.5) - (2.3.9).  For example, if we had imposed also $~G_{A
B c} \nabla^c\Tilde\varphi = 0$, then there would be no extra component left over
for the superfield $~G_{a b c}$, and therefore the system is totally reduced to
the  conventional 10D theory \bergshoeff.  We stress that the non-trivial
feature of our  12D theory is that {\it not all} extra components are deleted by
these conditions, while all the Bianchi identities 
are satisfied.  

We give next some details in the derivation of these results.  We first mention 
the second term in $~T\du{\a\b} c$~ in (2.3.4a), which is additional compared 
with the globally supersymmetric result in \ns.  As will be seen, this additional
term will be also important, when we confirm $~\k\-$fermionic symmetry in the 
Green-Schwarz superstring coupled to our supergravity background.  In fact, 
the crucial relationship $~\Pi\du\pm a\tilphia=0$~ will hold, 
only when the second term in (2.3.4a) is present in the system.  
Another important ingredient to be mentioned is the modification of our 
Lorentz generators (2.1.6).  This was required by the $~(\a\b\g,
\d)\-$type Bianchi identity at dimension $~d=1$.  
We found that terms like 
$~(\s^{a b})_{(\a\b|} (\s^{f c}) \du{|\g)} \d G_{a c d} \big(\phif\big) 
\big(\phib\big) \big(\nabla^d\Tilde\varphi\big)$~ 
would be left over, if we did not have the modification of 
$~\TildeM_{a b}$, 
and these terms are completely cancelled, when 
the Lorentz generator are modified like (2.1.6) with $~\Pup$, {\it via} the term
$$ R_{(\a\b\g)} {}^\d = -\frac14 R\du{(\a\b|}{c d} 
     \big( \g_{c d}\Pup \big) \du{|\g)}\d~~, 
\eqno(2.3.10) $$
with $~\Pup$~ inserted in the Lorentz generator, instead of the original one 
$\big(\g^{c d}\big)\du\g\d$.    

We next describe the derivation of our other constraints in (2.3.4).  
As usual in supergravity theory, we put some unknown coefficients 
$~a_1, ~a_2, ~c_1, ~c_2,~g$, like
$$\li{& T\du{\a\b} \g = a_1 (\Pup)\du{(\a|}\g \big( \g^d\Bar\chi
    \big)_{|\b)}\nabla_b\varphi + a_2\big( \g^{a b} \big)_{\a\b} 
    \left( \Pdown\g_a\Bar\chi \right)^\g \phib~~, \cr 
& \nabla_\a\Bar\chi_{\Dot\b} = c_1 \big( \g^{c d e}\Pup
    \big)_{\a\Dot\b} G_{c d e} + c_2 \left( \g^c \Pup \right)_{\a\Dot\b} 
   \nabla_c \Phi +g \left( \g^c\Bar\chi \right)_\a
   \Bar\chi_{\Dot\b}\nabla_c\varphi ~~, 
&(2.3.11) \cr } $$ 
and require the satisfaction of all the Bianchi identities.  First of all, 
$~c_2$~ is fixed by the closure on $~\Phi$, 
$$\{ \nabla_\a, \nabla_\b \}\Phi = \nabla_{(\a|} 
     \big[ \, \big(\g^c \Bar\chi\big)_{|\b)} \phic \, \big]
      = +2 c_2 \big( \g^{c d} \big)_{\a\b} \big(\nabla_c\Phi\big)
     \big(\phid\big) ~~,  
\eqno(2.3.12) $$ 
as $~c_2 =+1/2$, compared with $~T\du{\a\b} c \nabla_c\Phi$.  
Next step is to go to dimensions $~d=0$~ Bianchi identity of 
$~(\a\b\g\d)\-$type, which is easily satisfied by help of relation (A.5) 
in Appendix, as well as the properties of null-vectors like
$~\big(\delsl\varphi\big)\big(\delsl\varphi\big)\equiv 0$.    

The less trivial sector arises at $~d=1/2$~ for the $~(\a\b\g d)$~ and 
$~(\a\b\g,d)$~ Bianchi identities.  The former yields only three sorts of 
terms with at least two $~\nabla_a\varphi$'s:  
If we denote the l.h.s.~of the $~(A B C D)\-$Bianchi identity 
by $~X_{A B C D}$, then some appropriate manipulations yield 
$$\li{X_{\a\b\g d} =  2 \big( a_1 + a_2 \big) 
      \bigg[&\! + 2 \big(\g\du d a\big)_{(\a\b|} 
      \big( \g^b\Bar\chi \big)_{|\g)} (\phia)(\phib)  \cr  
& \! -  \big( \g^{a b} \big)_{(\a\b|} 
     \left( \g^c\Bar\chi \right)_{|\g)} (\phib)(\phic) (\nabla_d\varphi) 
     (\tilphia) \, \bigg]~~.  
&(2.3.13)  \cr} $$
Since these two terms are independent, we get the condition 
$$ a_1=-a_2~~. 
\eqno(2.3.14) $$ 
Fortunately, the $~(\a\b\g,d)\-$type Bianchi identity at $~d=1/2$~ is 
automatically satisfied, once this $~(\a\b\g\d)\-$type Bianchi identity  
holds.  

Next final non-trivial Bianchi identities are at $~d=1$, which are 
of (i) $(a b \a\b)$,~(ii) $(\a\b c,d)$~ 
and (iii) $(\a\b\g,\d)\-$types.  
Among these the first one is straightforward, while (ii) gives the relation
(2.3.4h), that in turn is used in (iii),
which is now composed of three sorts of terms:   
$~\nabla\Phi\-$terms, ~$G\-$terms, and $~\chi^2\-$terms.  Here the 
$~\nabla \Phi\-$terms are arranged as 
$$\li{\left(\nabla\Phi\-\hbox{terms}\right) =  
     - c_2 \left(a_1+a_2\right)  
 \bigg[ & \big( \g^{a b}\big)_{(\a\b|} \big( \g^{c d} \big)\du{|\g)}\d
     \big(\nabla_a\Phi\big) \big(\nabla_b\varphi\big) 
     \big(\nabla_c\varphi\big) \big(\nabla_d\Tilde\varphi\big) \cr  
& + \big( \g^{a b} \big)_{(\a\b} \d\du{\g)}\d 
     \big(\nabla_a\Phi\big) \big(\nabla_b\varphi\big) \, \bigg] ~~. 
&(2.3.15) \cr} $$ 
This gives $~a_1 = - a_2$, consistently with (2.3.13).  The $~G\-$terms are
$$ \li{\left( G\-\hbox{terms}\right) = & \big( -3c_1a_2 + \frac18\big) \cr 
& \times\bigg[ \, \big(\g^{a b} \big)_{(\a\b|} 
               \big(\g^{c d}\big) \du{|\g)}\d G_{c d a} \big(\phib\big)  
-2 \big( \g^{a b} \big)_{(\a\b|} \big(\g^{f c} \big)
            \du{|\g)}\d G_{c d a} \big(\phif\big) \big(\phib\big) 
            \big(\tilphid\big) \cr 
&~~~~~ ~~~~~ + \big( \g^{a b}\big)_{(\a\b|} \big(\g^{f c d g} \big)
            \du{|\g)}\d G_{c d a} \big(\phif\big) \big(\phib\big) 
            \big(\tilphig\big) \,\bigg] ~~.    
&(2.3.16) \cr } $$ 
This yields the condition 
$$ c_1 a_2 = + \fracm1{24} ~~. 
\eqno(2.3.17) $$ 
The remaining terms in (iii) are the $~\chi^2\-$terms which are after 
appropriate manipulations:
$$\li{\left(\chi^2\-\hbox{terms}\right) 
   = \, a_2\left(g-a_1-2a_2 \right) \big(\g^{a b}\big)_{(\a\b|} \bigg[   
     & \left(\g^c\Bar\chi\right)_{|\g)} 
     \big(\g^d\Bar\chi\big)^\d \big(\phic\big) \big(\phib\big) 
     \big(\phid\big) \big(\tilphia\big) \cr 
& - \left(\g^c\Bar\chi\right)_{|\g)} 
     \left(\g_a\Bar\chi\right)^\d \big(\phic\big)\big(\phib\big) 
     \,\bigg] {~~,~~~~~ ~~~~~} 
&(2.3.18) \cr } $$ 
yielding 
$$ g= a_1 + 2a_2~~. 
\eqno(2.3.19) $$ 

We now collect all the conditions on the unknown coefficients:
$$ a_1=- a_2 ~~, ~~~~ c_1a_2 = + \frac1{24} ~~, ~~~~ g= a_1 + 2a_2~~, 
\eqno(2.3.20) $$ 
which fortunately have a set of consistent solutions 
$$ a_1= - a_2 ~~, ~~~~ c_1 = \frac1{24} a_2^{-1} ~~, ~~~~g= - a_1~~. 
\eqno(2.3.21) $$
We can choose $~a_1$~ to be $~a_1=+1$, in order also to accord with the 
10D result after the dimensional reduction to be performed later. 
This fix all 
the coefficients in (2.3.11), and therefore 
our constraints (2.3.4) have been confirmed.  
Due to the limited resource of the publisher, as well as the interest of 
the majority of readers who need few technical details, we are to 
skip further details here.  

Our superfield equations in our system are much similar to those in 10D
\ref\gnz{M.T.~Grisaru,
H.~Nishino and D.~Zanon, \pl{306}{88}{625}; \np{314}{89}{363}.}:   
$$ \li{&(\g^{b c})_{\a\b} T\du{a b}\b \phic - 2 (\g^c)\du{\a}{\Dot\b}
     \big(\nabla_a {\Bar\chi}_{\Dot\b}\big) \phic =0 ~~,  
& (2.3.22) \cr 
& R_{a \[ b|} \nabla_{|c \]}\varphi + 4 \big(\nabla_a \nabla_{\[b|} \Phi\big)
\nabla_{|c\]}
     \varphi 
     - 4 \big( \Bar\chi\g^d T_{a \[b|} \big) \big(\nabla_{|c\]}\varphi\big)\phid
     =0 ~~,   
& (2.3.23) \cr 
& R_{\[a b\]} = -  \nabla_c G\du{a b } c~~.   
& (2.3.24) \cr } $$ 
These are obtained from Bianchi identities at $~d\ge 3/2$~ \nishone.  
Since this procedure is similar to the usual procedure, 
and nothing essential is peculiar to our system, we skip the 
details, except for the results.  
First, at $~d=3/2$, the $~(\a b c d)$~ Bianchi identity gives (2.3.4i), 
while $~(\a b c , d)$~Bianchi identity gives (2.3.4j).  
Now out of $~(a\b\g,\d)~$ Bianchi identity $~X\du{a\b\g}\d = 0$, we 
take the contraction $~X\du{a\b\g}\b$, to get 
$$X\du{a\b\g}\b = - \frac72 \left[\, \g^{b c} T_{a b} \phic
     + 2a_1 \big(\g^c\nabla_a\Bar\chi\big)\phic \, \right]_\g ~~ = 0 ~~, 
\eqno(2.3.25) $$ 
for the gravitino field equation (2.3.22).   

At $~d=2$, we have (i) $(a b \g,\d)$~ and (ii) $(a b c , d)$~ and (iii) $
(a b c d)\-$type Bianchi identities.  The first one gives (2.3.4k), 
which in turn can be combined with the gravitino superfield equation (2.3.22), 
as
$$ \li{0 & = + \big(\g_{d e} \big)^{\b\g} \nabla_\b \left[\,  
      \big(\g^{b c} \big)\du\g\d T_{a b \d}\, \phic 
      + 2a_1 \big(\g^b \big)\du\g{\Dot\d} 
       \big(\nabla_a\Bar\chi_{\Dot\d}\big) \phib \, \right] \cr 
&= + 8 \left[\, R_{a\[d} n_{e \]} 
     + 8 a_1c_2 \big( \nabla_a\nabla_{\[d} \Phi\big) \nabla_{e \]}\varphi 
     - 4a_1 \big(\Bar\chi\g^b T_{a\[d} \big) \big(\nabla_{e\]}\varphi\big) 
   \phib\,\right] ~~,    
&(2.3.26) \cr} $$ 
yielding the gravitational superfield equation (2.3.23) 
much like that for $~N=1$~ supergravity in 10D \gnz.  
Now the $~(a b c , d)\-$type Bianchi identity gives (2.3.24) 
after the contraction of $~_c$~ and $~^d$~ indices:  
$$ 0 = X\du{a b c} c = - R_{\[a b\]} - \nabla_c G\du{a b } c~~.
\eqno(2.3.27) $$  
The $~(a b c d)\-$type Bianchi identity gives no information, as usual. This 
concludes the satisfaction of all the Bianchi identities in our superspace, and 
therefore the confirmation of the consistency.  

We do not repeat the same remark as in \gnz\ about the peculiar structure
of our constraint system with no separate field equations for the dilaton
or dilatino, but mixed up with the zw\"olfbein or gravitino field
equations (2.3.23) and (2.3.22):  There is no loss of degree of freedom for  all
the physical fields for the same reason as in 10D \gnz\ after the dimensional
reduction.  As a matter of fact, in 10D the equivalence between the constraint
set in \gnz\ to the canonical set \bergshoeff\ was easily confirmed by super-Weyl
rescalings \ref\superweyl{S.J.~Gates, Jr.~and S.~Vashadkitze,
\np{291}{87}{172}.}.  

Before concluding this subsection, we give the component transformation rule
that can be easily obtained from our superspace constraints, 
by the aid of the standard technique in pages 321 - 327 of 
\ref\ggrs{S.J.~Gates, Jr., M.T.~Grisaru, M.~Ro\v cek and W.~Siegel, {\it
`Superspace'}, Benjamin/Cummings, Reading, 
MA (1983).}:\footnotew{The supercovariant derivative $~\nabla_a$~ corresponds to
the component supercovariant derivative $~D_a$.} 
$$ \li{& \d\low Q e\du m a = +\big( \e \g^{a b}\psi_m \big) D_b\varphi +
  \left( \e\Pupdown \psi_m \right) D^a\varphi ~~,~~~~ 
  \d\low Q\Phi = - \left( \e\g^m\Bar\chi\right) \partial_m\varphi~~,   
&(2.3.28\rma) \cr
& \d\low Q \psi\du m\a = D_m \e^\a + \left( \Pdown\e \right)^\a 
     \left( \Bar\chi \g^n\psi_m\right) \partial_n\varphi
     + \left( \Pdown\psi_m \right)^\a \left( \e\g^n\Bar\chi \right)
     \partial_n\varphi \cr 
& ~~~~~ ~~~~~ ~~~~~ - \left( \Pdown\g_a \Bar\chi \right)^\a  
   \big( \e\g^{a n}\psi_m \big) \partial_n\varphi{~~,~~~~~ ~~~~~}  
&(2.3.28\rmb) \cr 
& \d\low Q B_{m n} = + \big( \e \g\du{\[m} r\psi_{n\]} \big)\partial_r\varphi 
   -\left( \e\Pupdown\psi_{\[m} \right) \partial_{n\]}\varphi ~~, 
&(2.3.28\rmc) \cr
& \d\low Q\Bar\chi_{\Dot\a} = + \fracm1{24}\left( \Pdown\g^{m n r}\e
     \right)_{\Dot\a} G_{m n r} + \half \left( \Pdown\g^m\e 
     \right)_{\Dot\a} \partial_m \Phi 
     - \Bar\chi_{\Dot\a}\left( \e\g^m\Bar\chi \right) \partial_m\varphi~~,~~~~~ 
     ~~~~~    
&(2.3.28\rmd) \cr 
& \d_Q\varphi=0~~.
&(2.3.28\rme) \cr } $$ 
As is easily seen, the second term in (2.3.28a) is not important in component 
formulation, because it can be interpreted as an extra transformation 
for $~e\du m a$~ proportional to $~\nabla^a\varphi$, like 
supersymmetric Yang-Mills in 12D \ns.  
Note also that the common factor $~\Pdown~$ 
in front of the last three terms in (2.3.28b) is consistent with the 
constraint (2.3.8) for the gravitino field strength.  The same is also 
true with the first two terms in (2.3.28d).

\bigskip\bigskip

\leftline{\bf 2.4~~Dimensional Reduction} 

As the first important confirmation of the validity of our result, we 
perform simple dimensional reduction \nishone\ into 10D \gnz.  
This process is the standard one, 
namely we require all the dependence of the superfields on the extra coordinates
to vanish, truncating {\it all} the extra components as well, except those 
for null-vectors.  To be more specific, our $~64\times 64$~ 
~$\g\-$matrices in 12D will be dimensionally reduced as  
$$ \Hat\g_{\hata} = \cases{\hbox{$\Hat\g_a = \g_a \otimes \t_3 ~~,$}
\cr \hbox{$\Hat\g\low{(11)} = I\otimes \t_1 ~~,$} \cr
\hbox{$\Hat\g\low{(12)} = - I \otimes i\t_2 ~~,$} \cr }
\eqno(2.4.1) $$
while our charge conjugation matrix is to be
$$ \Hat C = C \otimes \t_1 ~~,   ~~~~\Hat \g_{13} = \g_{11} \otimes \t_3~~.
\eqno(2.4.2) $$
Here ~$\g_a, ~I,~C$~ and $~\g_{11}$~ are all $~32\times 32$~ matrices for the
10D Clifford algebra, while $~\t_1,~\t_2$~ and $~\t_3$~ are the standard
$~2\times 2$~ Pauli matrices.  Only in the sections for dimensional reductions,
we use the  {\it hats} for the quantities and indices in 12D, distinguished from 
{\it non-hatted} quantities and indices are in 10D.  
We next replace all the gradients of $~\varphi$~ and $~\Tilde\varphi$~ by 
the null-vectors as in (2.2.2).  
Accordingly, we have the dimensional reductions for the null-vectors and 
projection operators:
$$ \li{&(\Hat\nsl)\du{\hat\a}{\hat{\Dot\b}} = (\Hat\g^+)
     \du{\hat\a}{\hat{\Dot\b}}
     = {\sqrt2} I\otimes \pmatrix{0 & 1 \cr 0 & 0 \cr } ~~, ~~~~
     (\Hat\msl)\du{\hat\a}{\hat{\Dot\b}} =(\Hat\g^-)\du{\hat\a}{\hat{\Dot\b}}
     = {\sqrt2} I\otimes \pmatrix{0 & 0 \cr 1 & 0 \cr } ~~, \cr
&\Hat\Pup = I \otimes \pmatrix{1 & 0\cr 0 & 0 \cr } ~~, ~~~~
     \Hat\Pdown = I \otimes \pmatrix{0 & 0\cr 0 & 1 \cr} ~~. 
&(2.4.3) \cr } $$
Similarly, the dimensional reduction for our spinorial superfield goes as
$$ \left( \Hat{\Bar\chi}_{\hat{\Dot\a}} \right) 
  = \pmatrix{\Hat{\Bar\chi}_{\a\uparrow} \cr
    \Hat{\Bar\chi}_{\a\downarrow} \cr}  
  = \pmatrix{0 \cr \chi_\a\cr } ~~,  ~~~~ 
\left(\Hat T_{\hata\hatb}{}^{\hat\g} \right) 
  = \left( \Hat T_{\hata\hatb}{}^{\g\uparrow} ,
    \Hat T_{\hata\hatb}{}^{\g\downarrow} \right)  
  = \left( T_{\hata\hatb}{}^\g , 0 \right)~~,   
\eqno(2.4.4) $$  
where conveniently, we use also $~{\scst\uparrow}$~ or $~{\scst\downarrow}$~
to  denote the upper or lower eigen-components of $~\Pup$~ or $~\Pdown$~ 
in the spinors in the dimensional reductions.  Note that the components in 
$~\Hat T_{\hata\hatb}{}^{\hat\g}$~ are given as a row vector to be 
multiplied by 12D $~\g\-$matrices from the right, in accordance  
with our multiplication rule.  This is crucial when it comes to the extra
constraint (2.3.8).  

A typical example illustrating our process is 
$$\li{ & \Hat T\du{\a\uparrow\b\uparrow} c
=  (\Hat\g^{c +})_{\a\uparrow\b\uparrow}  
=  (\Hat\g^c\Hat\nsl)_{\a\uparrow\b\uparrow} 
= (\Hat\g^c)\du{\a\uparrow}{\hat{\Dot\g}}(\Hat\nsl)\du{\hat{\Dot\g}}{\hat\d} 
\Hat C_{\hat\d\b\uparrow}  \cr 
& ~~~~~ ~~~ = \left[\, (\g^c)\du\a\g \otimes \pmatrix{ 1 & 0 \cr 0 & -1 \cr } 
\d\du\g\d \otimes \pmatrix{ 0 & {\sqrt2} \cr 0 & 0 \cr} C_{\d\b}\otimes
\pmatrix{ 0 & 1 \cr 1 & 0  \cr }\,\right]_{\uparrow\uparrow} \cr 
& ~~~~~ ~~~ = {\sqrt2} (\g^c)_{\a\b} \otimes\pmatrix{ 1 & 0 \cr 0 & 0
\cr}_{\uparrow\uparrow} = {\sqrt2} (\g^c)_{\a\b} \equiv T\du{\a\b} c~~,   
&(2.4.5)  \cr 
& \Hat T\du{\a\uparrow\b\uparrow}{\g\uparrow} \rightarrow T\du{\a\b}\g 
=  {\sqrt2} \left[\, \d\du{(\a}\g \chi_{\b)} - (\g^a)_{\a\b} (\g_a\chi)^\g 
\, \right] ~~. 
&(2.4.6) \cr } $$

The dimensional reduction for our superfield equations is straightforward.  
We start with the gravitino field equation (2.3.22) by rewriting it as
$$ \big( {\Hat\g}^{\hat\b} {\Hat\g}^+ \big) \Hat T\du{\hata\hatb}{\hat\b} 
        +2 \big( {\Hat\g}^+ \big)_{{\hat\a}{\hat{\Dot\b}}} {\Hat\nabla}_{\hata} 
         \Hat{\Bar\chi}{}^{\Dot\b} ~~  = 0 ~~.   
\eqno(2.4.7) $$ 
There are in total four options for the free indices $~{\scst \hat\a\hata}$: 
(i) ${\scst\hat\a\hata~=~\a\uparrow a}$,
(ii) ${\scst\hat\a\hata~=~\a\downarrow a}$,
(iii) ${\scst\hat\a\hata~=~\a\uparrow +}$,
(iv) ${\scst\hat\a\hata~=~\a\downarrow +}$.  This is because when $~{\scst 
\hata~=~-}$, (2.4.7) is trivially satisfied by $~\Hat T\du{-\hatb}{\hat\b} = 0,~
\Hat\nabla_-\Hat{\Bar\chi}{}^{\Dot\b} = 0$~ of extra constraints (2.3.5) and 
(2.3.7).  Among these four cases, the case (i) yields 
$$\g^b T_{a b} + \nabla_a \chi = 0 ~~, 
\eqno(2.4.8) $$
under our dimensional reduction rules (2.4.1) - (2.4.4), giving
nothing but the 10D gravitino
field equation in ref.~\gnz.  All the other cases (ii) - (iv) can be easily 
satisfied by our dimensional reduction rules, such as $~T\du{a b}{\b\downarrow}
= 0$.  

We next perform the dimensional reduction of our 
zw\"olfbein field equation (2.3.23).  There are in total six possibilities for 
the free indices $~{\scst \hata\[\hatb\hatc\]}$:  
(i) ${\scst a\[ b+\]}$, (ii) ${\scst a\[+ -\]}$, (iii) ${\scst +\[b +\]}$,
(iv) ${\scst +\[+ -\]}$, (v) ${\scst -\[b +\]}$, (vi) ${\scst -\[+ -\]}$.  
First, the case (i) yields 
$$ \li{ 0 = &+ \Hat R_{a b} \Hat n_+ + 4 \Hat\nabla_a
     \Hat\nabla_b\Hat\Phi\Hat n_+ 
     - 4 \big( \Hat{\Bar\chi}\Hat\g^+ \Hat T_{a b} \big) \Hat n_+ ~~\cr 
= & + R_{a b} + 4\nabla_a\nabla_b \Phi - 4\Hat T\du{a b}{\g\uparrow}
    \big( \Hat\g^+ \big)\du{\g\uparrow}{\d\downarrow} 
     \Hat\chi_{\d\downarrow}~~,
&(2.4.9) \cr } $$
which gives the zehnbein field equation  
$$ R_{a b} + 4 \nabla_a\nabla_b\Phi 
- 4 {\sqrt 2} \big( {\Bar T}_{a b} \chi \big)  = 0 ~~                    
\eqno(2.4.10) $$
in \gnz.  All other sectors (ii) - (vi) turn out to be satisfied easily 
by our dimensional reduction prescription and constraints, such as 
$~R_{a -} = 0, ~\nabla_- \Phi = 0$, {\it etc.}  
In a similar fashion, eq.~(2.3.24) is easily reduced to 10D, yielding  
$$ R_{\[a b\]} = - 2 \nabla_c G\du{a b } c~~,   
\eqno(2.4.11)$$ 
as in \gnz, which is formally of the same form as in 12D.  This concludes the 
description of our dimensional reduction, as an important confirmation 
of our original 12D supergravity.

\bigskip\bigskip
\bigskip\bigskip

\centerline{\bf 3.~~Superstring on Background of ~$D=12,\,N=1$~ Supergravity}

\bigskip

Once our superspace formulation is established for $~N=1$~ supergravity 
in 12D, then the next natural task is to put some probe for the background, 
such as superstring.  The existence of consistent superstring on such 
background is also naturally expected from the viewpoint of F-theory \ftheory,
namely all the superstring theories such as heterotic or type IIB string,
that are not directly from 11D M-theory \km\bfss\mtheoryrev, are from 12D
F-theory \ftheory.  The first natural trial is to put Green-Schwarz superstring
on our supergravity background.  

We start with the postulate for the total action for Green-Schwarz superstring
\nishone: 
$$ \li{& S \equiv S_\s + S_B  + S_\L ~~,  
&(3.1) \cr 
& S_\s \equiv \int d^2\s \, \left[ \, V^{-1} \eta_{a b} 
\Pi\du + a\Pi\du - b \, \right] ~~, 
&(3.2) \cr 
& S_B  \equiv \int d^2\s\, \left[ \, V^{-1} \Pi\du + A 
\Pi\du - B B_{B A} \, \right] ~~, 
&(3.3) \cr 
& S_\L \equiv \int d^2\s \, \bigg[ \, V^{-1} \L_{+ +}  
 \big(\Pi\du -a \phia \big) \left(\Pi\du -b \tilphib\right)  \cr 
& ~~~~~ ~~~~~ ~~~~~ ~~~~~ + V^{-1}
 {\Tilde\L}_{+ +} \left\{ \left(\Pi\du -a \phia\right)^2 
  + \left(\Pi\du -a \tilphia\right)^2 \right\} \, \bigg] {~,~~~~~~} 
&(3.4) \cr } $$ 
where $~V\equiv\det(V_\pm{}^i)$~ is the determinant 
of the zweibein $~V\du\pm i$, 
and the indices $~{\scst i,~j,~\cdots~=~0,~1}$~ are for the curved 2D 
coordinates $~\s^i$, while $~{\scst\pm}$~ are for the local Lorentz 
light-cone coordinates.\footnotew{We try to avoid the simultaneous usage
of the ${\scst\pm}$-indices for the 12D target space-time, and these 2D
light-cone indices.}      

Due to our extra coordinates, we expect some symmetry that will get rid of 
non-physical components associated with them.  In fact, we have not only
the usual $~\k\-$symmetry \gsw, but also an additional fermionic
$~\eta\-$symmetry  in our total action, dictated by \nishone 
$$ 
\li { & \d V\du + i = \Bar \k\du+{\Dot\a} (\g^c)\du{\Dot\a}\b \big(\phic\big)
\Pi_{+\b} V\du - i \equiv \left(\Bar\k_+ \g^c\Pi_+ \right) V\du-i\phic~~,
&(3.5\rm a) \cr 
&\left( \g^c\right)\du\a{\Dot\b} {\Bar\k}_{+\Dot\b} \phic \equiv \left(
\g^c\Bar\k_+\right)_\a \phic = 0 {~~,~~~~~~~~~~~~} 
&(3.5\rm b) \cr 
& \d V\du- i = 0 ~~, ~~~~ \d\big( V^{-1} \big) = 0 ~~, ~~~~
\d\Bar E{}^{\Dot\a} = \d E^a = 0 ~~, 
&(3.5\rm c) \cr 
&\d E^\a = \half\left( \g_a\right)^{\a\Dot\b}{\Bar\k}_{+\Dot\b}\Pi\du-a 
  + \left( \Pup \right)^{\a\b}\eta\low\b 
 \equiv\half\left( \Pisl_- \Bar\k_+\right)^\a + \left( \Pup\eta\right)^\a ~~, 
&(3.5\rm d) \cr 
&\d\L_{+ +}  = - 2 \left( {\Bar\k}_+ \g^c \Pi_+ \right)\phic ~~, ~~~~ 
   \d{\Tilde\L}_{+ +} = 0 ~~, ~~~~ \d\varphi = 0~~, ~~~~ \d\Tilde\varphi=0~~,  
&(3.5\rm e)  \cr } $$ 
where $~\k$~ and $~\eta$~ are infinitesimal arbitrary $~\s\-$dependent  
fermionic parameters.  Notice the important significance of (3.5a) through 
(3.5c)  that the effective 2D gravitational field is $~h_{++}\approx V^{-1}
g_{++}$~ in the light-cone coordinates, and $~\d h_{++} \approx
\big(\Bar\k_+\nablasl\varphi\Pi_+\big)$, so that the only non-trivial 
component for the energy-momentum tensor will be $~T_{--}$.  
This feature will be important, when we later study the contributions of 
the extra string coordinates to the conformal anomaly.  

We first confirm the $~\k\-$invariance of the total action.  To this end,  
we need the basic relations for variations in Green-Schwarz $~\s\-$model, such
as  $$ \li{\d_\k\Pi\du\pm A = \, & V\du\pm i D_i \big(\d_\k E^A \big)  
     + \big( \d_\k V\du\pm i \big) \Pi\du i A 
     - \Pi\du\pm D \big( \d_\k E^C \big)  
            \big( T\du{C D} A  - \phi\du{C D} A \big) ~~, 
&(3.6) \cr } $$
where the explicit Lorentz connection $~\phi$~ will
automatically disappear or will be absorbed into covariant derivatives, 
as in the usual 10D case.   Using this, we get that
$$ \li{\d_\k \left( S_\s + S_B \right) 
= \, & + \left(\Bar\k_+\g^c\Pi_+ \right) \left( \Pi\du-a\right)^2 \phic 
     - \frac12 \big(\Bar\k_+\g^d\g^a\g^e \g^{b c} \Pi_+ \big) \Pi_{-e} 
      \Pi_{-b}\big(\phic\big) \big(\phid\big) \big(\tilphia\big) \cr 
& - \frac12 \left(\Bar\k_+\g^c\g^a\g^e \Pupdown 
      \Pi_+ \right) \Pi_{-e}\Pi\du-b
      \big(\phib\big)\big(\phic\big)\big(\tilphia \big) \cr  
=  \, & +  2 V^{-1} \left(
      \Bar\k_+ \g^c\Pi_+ \right) 
      \left(\Pi\du- a \phia \right) \big(\Pi\du- b \tilphib \big)\phic ~~. 
& (3.7) \cr } $$ 
From the second toward the third line, we have performed $~\g\-$matrix
manipulations, such as $~\g^e\g^{b c} = \g^{e b c} + \eta^{e \[b} \g^{c\]}$, as
well as the  constraint (3.5b),  
so that $~\Bar\k_+ \g^a\g^b\big(\nabla_a\Tilde\varphi\big)\big(\phib\big) 
= + 2 \Bar\k_+$, under the null-vector conditions (2.3.4$\ell$).  We see that the 
first term in the second line is cancelled by other like terms, while we are 
left only with one term in the third line.  Similarly, we get 
$$ \d_\k S_\L = - 2 V^{-1} \big( \Bar\k_+ \g^c \Pi_+ \big)  
\big(\Pi\du- a \phia \big) \big(\Pi\du- b \tilphib \big) \phic  ~~,   
\eqno(3.8) $$  
by the use of the relations such as $~\d_\k \parenth{\Pi\du- a \phia} = 
\d_\k \parenth{\Pi\du- a \tilphia} = 0$, which are easily confirmed first by
showing that $~T\du{\a\b} c \tilphic= 0$.  Here technically we need also the 
feature of the Lorentz generators, such as 
$~\big(\Tilde M_{b c} \big) ^{- d} = 0$.    
After all, we have 
$$ \d_\k S = \d_\k ( S_\s + S_B + S_\L) = 0~~.  
\eqno(3.9) $$ 

As for the second fermionic symmetry, we can similarly confirm the invariance
\nishone:
$$ \d_\eta S = 0 ~~. 
\eqno(3.10) $$

The two fields $~\L_{++}$~ and $~\Tilde\L_{++}$~ are playing roles of 
lagrange multipliers, yielding the two field equations
$$ \left(\Pi\du- a \phia\right) \big(\Pi\du- b \tilphib\big) = 0~~, ~~~~
\left(\Pi\du - a \phia\right)^2 + \left(\Pi\du - a \tilphia\right)^2 = 0~~,    
\eqno(3.11)  $$ 
which in turn are equivalent to the two equations\footnotew{Some ideas similar 
to these constraints have been suggested in various contexts
\ref\preit{{\it See, e.g.,} E.~Bergshoeff, L.A.J.~London and P.K.~Townsend, 
\cqg{9}{92}{2545}; T.~Hurth, P.~van Nieuwenhuizen, A.~Waldron and
C.~Preitshopf, Nucl.~Phys.~Proc.~Suppl.~{\bf 56B} (1997) 310.}.}   
$$ \li{&\Pi\du- a \phia = 0~~, ~~~~\Pi\du - a \tilphia = 0~~.   
&(3.12) \cr } $$  

Note that (3.12) is a consequence of field equations
out of the Lagrange multiplier action $~S_\L$, but not imposed by hand.  
The reason is that for the invariance check of the total action, 
we should not impose by hand the constraint with the 
first-order derivative such as (3.12) which can be interpreted as
unidexterous field equations in 2D.  This is a particular caution needed for  
action invariance in 2D.  
Interestingly, the action $~S_\L$~ also cancels the unwanted terms in (3.7).  
As mentioned after eq.~(3.5), 
the only non-trivial component of the energy-momentum tensor 
coupled to 2D zweibein field is $~T_{--}$, therefore the deletion of the 
components $~\P_-{}^a \tilphia= \P_-{}^a \ma$~ and $~\P_-{}^a \phia =
\Pi_-{}^a \na$~ removes any 
additional contribution from the extra string variables 
to the conformal anomaly.  Accordingly, the usual 2D conformal anomaly
cancellation works in the same way as in the 10D Green-Schwarz superstring
\gsw.  

There is another crucial point related to eq.~(3.12).  Note that these
constraints effectively force the string variables $~X^\pm$~ to depend only on
$~\s^+$.  In other words, there are non-vanishing extra components
$~X^\pm(\s^+)$~ which distinguish our system from just a `rewriting' of the
conventional 10D superstring theory \gsw.  Due to these non-trivial components,
our Green-Schwarz superstring \gsw\ coupled to 12D supergravity is by no means
just  a rewriting of the conventional $~N=1$~ superstring coupled to 10D
supergravity `in disguise'.  To put it differently, our system cleverly
maintains the conformal anomaly cancellation of the  conventional 10D
superstring, while keeping new variables inherent in the theory.  

We mention that there is an alternative form of our $~\k\-$symmetry.  
This can be obtained by the replacement $~\Bar\k_{+\Dot\a} 
= \parenth{\g^c\l_+}_{\Dot\a}\phic$, with the constraint 
$~\g^c\l_+\phic = 0$.  This is merely a rewriting of the 
original $~\k\-$symmetry, with nothing significant, reflecting just 
the nilpotency of $~\g^a\tilphia$~ and $~\g^a\phia$.  

The counting of the physical degrees of freedom can be easily done, 
by considering the components deleted by these fermionic symmetries.  
First of all, the $~\eta\-$symmetry deletes half of the original 32 components 
of the fermionic chiral coordinates $~\theta^\m$~ in superspace in 12D, 
and thus at most 16 components can be physical.  
Subsequently, the usual $~\k\-$symmetry \gsw\ deletes further half of 16
components, and we are left with  the usual 8 components in accordance with the
light-cone gauge in Green-Schwarz superstring \gsw.      

We finally stress that the 
null-vector conditions in (2.3.4$\ell$) are also required 
by these fermionic invariances on
the Green-Schwarz superstring world-sheet.  Therefore these 
world-sheet fermionic symmetries provide an independent validity
confirmation of our 12D supergravity constraints in superspace.

\bigskip\bigskip
\bigskip\bigskip

\centerline{\bf 4.~~$N=2$~ Supergravity in $~D=10+2$}
\bigskip

\leftline{\bf 4.1~~Notations} 

We have so far worked on $~N=1$~ supergravity in $~D=10+2$~ and its related 
features.  We now turn to $~N=2$~ chiral supergravity which is supposed to be 
the strong coupling limit of F-theory \ftheory.  
Once we have understood how our peculiar Lorentz generators work for 
$~N=1$, it is easier to handle the $~N=2$~ theory 
in component language, where
we  can get directly the transformation rules and field equations.  

Our basic conventions are 
consistent with the notation in the preceding sections, except for 
minor differences peculiar to the component formulation.  
One of them is the index convention such as $~{\scst \m,~
\n,~\cdots~=~0,~1,~\cdots,~9,~11,~12}$~ used for curved indices, while 
$~{\scst m,~n,~ \cdots~=~(0),~(1),~\cdots,~(9),~(11),~(12)}$~ 
for local Lorentz indices.    
Another difference from the superspace notation is the normalized 
anti-symmetrization, such as $~A_{\[\m\n\]}\equiv (1/2)(A_{\m\n} 
- A_{\n\m})$, and the component covariant derivative $~D_\m$, {\it etc.} 
Other than these, we use the same null-vectors $~\delsl\varphi,~
\delsl\Tilde\varphi~$ or the operators $~\Pup,~\Pdown,~\Pupdown$, as in
(2.2.1) and (2.2.2).  Due to the chiral nature of our system, we need to
distinguish the chiralities for the $~N=2$~ case.  
The explicit representations for 
$~\Tilde{\cal M}_{m n}$~ is the exactly the same as (2.1.6).  

Similarly to the $~N=2$~ chiral supergravity in 10D \ref\schwarz{J.H.~Schwarz, 
\np{226}{83}{269}.}, our system also has 
the coset $~SU(1,1)/U(1)$~ parametrized by the scalar fields   
playing roles of coordinates on this manifold.  
The scalar fields $~V\du\pm\a$~ are $~SU(1,1)$~ group 
matrix-valued, transforming as
$$\d V\du\pm\a = m\ud\a\b V\du\pm\b \pm i \S V\du\pm \a ~~.   
\eqno(4.1.1) $$
Here the indices $~{\scst \a,~\b,~\cdots~=~1,~2}$~ should not
be confused with the 12D spinorial indices in (2.1.6), 
as long as they are clear from the context.   
The explicit matrix representations for $~V\du\pm\a$, 
such as 
$$ \pmatrix{V\du- 1 & V\du +1 \cr 
V\du-2 & V\du+2 \cr} = \exp\,\pmatrix{ i\varphi & A\cr A^* & - i \varphi\cr} 
  = \pmatrix{ \cosh\r + i\varphi\fracm{\sinh\r}\r & A\fracm{\sinh\r}\r \cr 
A^*\fracm{\sinh\r}\r & \cosh\r - i\varphi\fracm{\sinh\r}\r \cr } ~~, 
\eqno(4.1.2) $$ 
are sometimes useful, where $~\r^2 \equiv A^*A - \varphi^2$.  
The constant parameter 
$$\left( m\ud\a\b \right) = \pmatrix{i\g & \a\cr \a^* & - i\g\cr} ~~, 
\eqno(4.1.3) $$
is for the global $~SU(1,1)$~ group, while $~\S$~ is a real parameter for 
the $~U(1)$~ transformation.  

The $~V's$~ satisfy the relationships
$$\e_{\a\b} V\du - \a V\du + \b = \det \,V = 1 ~~, ~~~~ 
     V\du-\a V\du + \b - V\du+\a V\du- \b = \e^{\a\b} ~~, 
\eqno(4.1.4) $$
so that we do not need their inverse matrices.  
The composite $~U(1)$~ connection defined by  
$$Q_\m = - i \e_{\a\b} V\du - \a \partial_\m V\du + \b~~,  
\eqno(4.1.5) $$ 
transforms as $~\d Q_\m=\partial_\m \S$.  The
$~SU(1,1)$~ invariant field strength $~P_\m = 
- \e_{\a\b} V\du+\a \partial_\m V\du+\b $~
transforms as $~\d P_\m = 2 i \S P_\m $.  Among the fields in
our supergravity multiplet $~(e\du\m m, \psi_\m, A_{\m\n\r\s}, \l,
A_{\m\n}{}^\a,$ $V\du\pm\a; \varphi,\Tilde\varphi)$, the following fields
transform under $~SU(1,1) \otimes U(1)$:  
$$ \d A_{\m\n}{}^\a = m\ud\a\b A\du{\m\n}\b ~~, ~~~~ 
      \d\psi_\m = \fracm i 2 \S\psi_\m ~~, ~~~~ 
      \d\l = \fracm{3i} 2 \S\l ~~.  
\eqno(4.1.6)$$

Useful relation associated with (anti)self-duality are such as (A.20) in
Appendix A, and 
$$ \li{& \g^{\[6\]} \psi_+ S_{\[6\]} \equiv 0~~, ~~~~ 
   \g^{\[6\]} \psi_- A_{\[6\]} \equiv 0~~, ~~~~
    \g_{13} \psi_\pm \equiv \pm \psi_\pm~~, ~~~~ \cr 
& S_{\[6\]} = + \frac1{6!} \e\low{\[6\]}{}^{\[6\]'} S_{\[6\]'} ~~, ~~~~
  A_{\[6\]} = - \frac1{6!} \e\low{\[6\]}{}^{\[6\]'} A_{\[6\]'} ~~. ~~~~ 
& (4.1.7) \cr } $$ 
Here $~{\scst \[ n \]}$~ denotes the normalized antisymmetrization of $~n$~
indices.   
 
We finally mention the important relations with respect to 
inner products of spinors in our $~N=2$~ system, {\it e.g.}, for two 
Weyl spinors $~\psi_1$~ and $~\psi_2$, we have
$$ \li{& \left( \Bar\psi_1 \g^{\m_1\cdots\m\low N} \psi_2 \right) 
= + \left( \Bar\psi_2 \g^{\m\low N\cdots\m_1} \psi_1 \right) 
= (-)^{N(N-1)/2} \big( \Bar\psi_2 \g^{\m_1\cdots\m\low N} \psi_1 \big) ~~, 
&(4.1.8\rma) \cr    
& \left( \Bar\psi_1 \g^{\m_1\cdots\m\low N} \psi_2 \right) ^\dagger 
     = \psi_2^\dagger \left(\g^{\m_1\cdots\m\low N} \right)^\dagger
     \Bar\psi_1{}^\dagger 
     = + \left( \Bar\psi_2 \g^{\m\low N\cdots\m_1} 
     \psi_1 \right) \cr 
     & ~~ = (-1)^{N(N-1)/2} \left(\Bar\psi_2 \g^{\m_1\cdots\m\low N} 
     \psi_1 \right) = + \Big( \Bar\psi_1^{\,*} \g^{\m_1\cdots\m\low N} 
     \psi_2^* \Big)  ~~,  
& (4.1.8\rmb) \cr} $$ 
where the dagger $~\dagger$~ denote a hermitian conjugate, and 
$~^*\-$symbol is a complex conjugation of a Weyl spinor, such as 
$~\psi^* = \left(\psi^{(1)} + i \psi^{(2)}\right)^*  = \psi^{(1)} - i \psi
^{(2)} $~ for two Majorana-Weyl spinors $~\psi^{(1)}$~ and $~\psi^{(2)}$~ 
forming a Weyl spinor $~\psi$.

\bigskip\bigskip

\vbox{
\leftline{\bf 4.2~~$~N=2$~ Supergravity in $~D=10+2$} 

We first give our result for supersymmetry transformation for our 
multiplet of supergravity $~(e\du\m m, \psi_\m, A_{\m\n\r\s}, \l,
A_{\m\n}{}^\a, V\du\pm\a; \varphi,\Tilde\varphi)$~ \nishtwo:
}
$$ \li{
&\d\low Q e\du\m m = \left[\, \big( \Bar\e\g^{m n} \psi_\m \big) D_n \varphi
   + \big( \Bar\e\Pupdown\psi_\m \big) D^m\varphi \,\right] + \hbox{c.c.}~~, 
&(4.2.1\rma) \cr 
\noalign{\vskip0.1in}
&\d\low Q\psi_\m = \Hat D_\m\e - \fracm i{480} \big(\Pdown\g^{\[5\]} 
    \g\low\m\e \, 
    \big) \Hat F_{\[5\]}+\fracm1{96}\Pdown \left( \g\low\m{}^{\[3\]}  
    \Hat G_{\[3\]} - 9 \g^{\[2\]} \Hat G_{\m\[2\]} \right)
    \e^* {~~,~~~~~ ~~~~~}  
&(4.2.1\rmb) \cr  
\noalign{\vskip0.1in}
&\d\low Q A\du{\m\n} \a = V\du+\a \big( \Bar\e{}^{\,*} \g\low{\m\n}{}^\r \l^* 
    \big) \phiroh + V\du-\a\big( \Bar\e \g\low{\m\n}{}^\r \l \big)
     \phiroh \cr  
& ~~~~~ ~~~~~ ~~~~~ - 4 V\du+ \a \big(
   \Bar\e\g\low{\[\m|}{}^\r
    \psi_{|\n\]}^* \big) \phiroh - 4 V\du-\a \big( \Bar\e{}^{\,*} 
    \g\low{\[\m|}{}^\r \psi_{|\n\]} \big) \phiroh~~,
&(4.2.1\rmc) \cr 
\noalign{\vskip0.1in}
& \d\low Q A_{\m\n\r\s}  
    = i \big( \Bar\e\g\low{\[\m\n\r|}{}^\t \psi_{|\s\]} \big)
    \phitau - i \big( \Bar\e{}^{\,*} \g\low{\[\m\n\r|}{}^\t \psi_{|\s\]}^* 
    \big) \phitau -\fracm{3i}8 \e_{\a\b} A\du{\[\m\n}\a \d\low Q 
    A\du{\r\s\]} \b~~, 
&(4.2.1\rmd) \cr                                                 
\noalign{\vskip0.1in}
& \d\low Q\l = - \left( \Pdown \g^\m \e{}^* \right) \Hat P_\m  
     -\fracm1{24} \left( \Pdown \g^{\m\n\r} \e\right) \Hat G_{\m\n\r} ~~, 
&(4.2.1\rme) \cr 
\noalign{\vskip0.1in}
& \d\low Q V\du+\a = V\du-\a \big( \Bar\e{}^{\,*} \g^\m \l \big) \phimu~~, ~~~~
  \d\low Q V\du-\a = V\du+\a \big( \Bar\e \g^\m\l{}^* \big) \phimu~~,           
&(4.2.1\rmf) \cr 
&\d\low Q\varphi = 0 ~~, ~~~~ \d\low Q\Tilde\varphi = 0 ~~, ~~~~
&(4.2.1\rmg) \cr } $$  
where $~e\du\m m$~ is the zw\"olfbein, $~\psi_\m$~ is a pair of 
two Majorana-Weyl
spinors of the same chirality: $~\g_{13}\psi_\m=-\psi_\m$, or equivalently a Weyl
spinor for $~N=2$~  supersymmetry, $~A\du{\m\n}\a$~ is a pair of complex vector
fields,  $~A_{\m\n\r\s}$~ is a real fourth-rank antisymmetric tensor, $~\l$~ is
a Weyl spinor sometimes called gravitello satisfying $~\g_{13}\l=+\l$, and
$~V\du+\a$~ is a scalar field  parametrizing the coset $~SU(1,1)/U(1)$.    
The field strengths with the {\it hat}-symbols are meant to be 
supercovariantization \ref\pvn{P.~van Nieuwenhuizen,
\prep{68}{81}{189}.}\schwarz\ of the field strengths defined by
\nishtwo\footnotew{We will not confuse $~P_\m$~ with the projectors
$~\Pup,~\Pdown$~ of $~\Pupdown$, as long as we are careful about the context.} 
$$\li{  & ~~~G_{\m\n\r} \equiv - \e_{\a\b} V\du+\a F\du{\m\n\r}\b ~~, ~~~~
     P_\m \equiv -\e_{\a\b} V\du+\a \partial_\m V\du+\b ~~, ~~~~
     Q_\m \equiv - i \e_{\a\b} V\du -\a \partial_\m V\du +\b~~, \cr   
& ~~~F\du{\m\n\r}\a \equiv 3\partial_{\[\m} A\du{\n\r\]} \a ~~, ~~~~ 
       F_{\m\n\r\s\t} \equiv 5\partial_{\[\m} A_{\n\r\s\t\]}  
       + \fracm{5 i}8 \e_{\a\b} A\du{\[\m\n}\a F\du{\r\s\t\]}\b ~~,  
& (4.2.2) \cr } $$ 
which satisfy useful identities such as \nishtwo 
$$ \li{&D_{\[\m} P_{\n\]} = 0 ~~, ~~~~ 
     D_{\[\m} G_{\n\r\s\]} = + P_{\[\m} G^*_{\n\r\s\]} ~~, 
&(4.2.3\rma) \cr 
&\partial_{\[\m_1} F_{\m_2\cdots\m_6\]} \equiv \fracm{5 i}{12} 
     G_{\[ \m_1\m_2\m_3} G^*_{\m_4\m_5\m_6\]} ~~, 
&(4.2.3\rmb) \cr 
&\partial_{\[\m} Q_{\n\]} = - i P_{\[ \m} P^*_{\n \]} ~~
     {\hskip 6pt ,  ~~~~~ ~~~~~} 
&(4.2.3\rmc) \cr } $$
parallel to the 10D case \schwarz.  

As in the $~N=1$~ supergravity theory \nishone, 
we have the extra constraints imposed 
on the field strengths and spinors \nishtwo:\footnotew{We use 
$~D_m\varphi\equiv 
e\du m\m \partial_\m \varphi$~ in {\it e.g.} (4.2.4e), avoiding 
the misleading expression $~\partial_m \varphi$~ with the local Lorentz index
$~{\scst m}$.}  
$$ \li{& \Hat G\du{\m\n}\r \phiroh = 0 ~~,~~~~\Hat F\du{\m\n\r\s} \t
     \phitau = 0 
     ~~, ~~~~ \Hat R\du\m\n{}^{m n} \phinu  = 0 ~~, ~~~~
     \Hat R_{\m\n}{}^{m n} \phim = 0 ~~, ~~~~~ ~~~~~ ~~~~~ 
&(4.2.4\rma) \cr 
& \Hat P^\m \phimu = 0 ~~, ~~~~ Q^\m \phimu = 0 ~~, 
&(4.2.4\rmb) \cr 
& \Hat{\cal R}\du\m\n \phinu = 0 ~~, 
    ~~~~\Bar{\Hat{\cal R}}_{\m\n}\g^m D_m\Tilde\varphi  = 0~~,  
&(4.2.4\rmc) \cr 
& \g^\m \l \partial_\m\Tilde\varphi = 0 ~~, 
     ~~~~ \big(\Hat D_m \l\big) \big(D^m \varphi\big) = 0~~.   
&(4.2.4\rmd) \cr 
& D_m D_n \varphi = 0 ~~, ~~~~D_m D_n \Tilde\varphi = 0 ~~,~~~~ 
     \big(D_m\varphi \big) \big(D^m\Tilde\varphi \big) = 1 ~~, \cr 
& \big(D_m \varphi\big)^2  = 0 ~~, ~~~~
     \big(D_m\Tilde\varphi\big)^2  = 0 {~~.~~~~~}  
&(4.2.4\rme) \cr } $$ 
The conditions in (4.2.4e) are the same as (2.2.1) for our scalar fields, 
whose non-trivial solutions are (2.2.2).  Note the involvement of $~D_m\Tilde
\varphi$~ in (4.2.4c) and (4.2.4d).    

Our component fields undergo also extra transformations 
in addition to our supersymmetry, translation
and Lorentz rotations, or gauge transformations, dictated symbolically for 
a general component field $~\phi\low{\[m\]}{}^{\[n\]}$~ in our multiplet
by  
$$ \d\low{\rm E} \phi_{\m_1\cdots\m_m}{}
      ^{r_1\cdots r_n} = \Omega_{\[\m_1\cdots\m_{m-1}}{}^{r_1\cdots r_n} 
      \partial_{\m_m\]}\varphi 
      + \Omega'{}_{\m_1\cdots\m_m}{}^{\[ r_1\cdots r_{n-1}} 
      D^{r_n\]}\varphi  ~~. ~~~~
\eqno(4.2.5) $$ 
For example, for $~A_{\m\n\r\s}$~ 
with $~m=4,~n=0$:  
$$\d_{\rm E} A_{\m\n\r\s} = \Omega_{\[\m\n\r} \partial_{\s\]}\varphi ~~, 
\eqno(4.2.6)  $$
where $~\Omega_{\[3\]}$~ is a infinitesimal local parameter.  
Since each component 
field has different index structure, eq.~(2.4.5) expresses 
collectively all of these extra transformations.  
As will be shortly mentioned, these extra transformations are needed also 
for the closure of supersymmetries.   

We now list up our field equations
$$\vbox{
\li{& \Hat D_\m\Hat P^\m - \fracm1{24} \Hat G_{\m\n\r}^2 
   + \fermionsquare = 0 ~~, 
&(4.2.7\rma) \cr                                                              
\noalign{\vskip0.1in}
&\big(\Hat D_\m \Hat G\ud\m{\[\n\r} \big) \partial_{\s\]}\varphi 
     +\Hat P^\m \Hat G_{\m\[ \n\r}^* 
     \partial_{\s\]}\varphi + \fracm{2i}3 \Hat F\du{\[ \n\r| } {\t\omega\l} 
     \Hat G_{\t\omega\l} \partial_{| \s\]}\varphi + \fermionsquare = 0~~, 
&(4.2.7\rmb) \cr  
\noalign{\vskip0.1in}
&\bigg( \Hat R_{\r\[\m |} - \Hat P_\r \Hat P_{\[\m|}^* - \Hat P_{\[\m|} 
     \Hat P_\r^* - \fracm16 \Hat F_{\[4\]\r} \Hat F\ud{\[4\]}{\[\m |} \cr 
& ~~ - \fracm18 \Hat G_\r{}^{\s\t}\Hat G_{\s\t\[\m|}^* - \fracm18
   \Hat G_{\s\t\[\m|} \Hat G_\r^*{}^{\s\t} 
   + \fracm1{48} g\low{\r\[\m|} \Hat G^{\[3\]} \Hat G_{\[3\]}^*
   \bigg) \partial_{|\n\]}\varphi + \fermionsquare = 0
   {~~,~~~~}                        
&(4.2.7\rmc) \cr                                                              
\noalign{\vskip0.1in} & \Hat F_{\[\m_1\cdots \m_5}\partial_{\m_6\]}\varphi 
    = - \fracm 1{6!} e^{-1} \e_{\m_1\cdots \m_6}{}^{\n_1\cdots\n_6} 
     \Hat F_{\n_1\cdots \n_5}\partial_{\n_6}\varphi ~~, 
&(4.2.7\rmd) \cr 
\noalign{\vskip0.1in}
& \g^\s\bigg( \g^\r \Hat{\cal R}_{\r\[\m|} + \l^*
     \Hat P_{\[\m|} - \fracm1{48} \g^{\[3\]} \g\low{\[\m|} \l \, 
     \Hat G_{\[3\]}^* 
     - \fracm1{96} \g\low{\[\m|} \g^{\[3\]} \l \, 
     \Hat G_{\[3\]}^* \bigg) \big(\partial_{|\n\]}\varphi\big) 
     \big(\phisigma\big) =0{~~,~~~~~ ~~~~~}  
&(4.2.7\rme) \cr  
\noalign{\vskip0.1in}
& \g^\s\left( \g^\m\Hat D_\m\l - \fracm i{240} \g^{\[5\]} \l \,\Hat F_{\[5\]} 
   \right) \phisigma = 0 ~~.  
&(4.2.7\rmf) \cr } 
} $$ 
The terms denoted by $~{\cal O}(\psi^2)$~ 
are fermionic terms 
in bosonic field equations, that are skipped in this paper 
as in ref.~\schwarz.  Note that the usual self-duality condition 
on $~F_{\[ 5  \]}$~ in 10D \schwarz\ corresponds to 
anti-self-duality condition (4.2.7\rmd) in                                
12D.  This is merely due to our notation related to the $~\e\-$tensor (A.20). 

We now give the detailed derivation of our transformation rule and 
field equations.  We first confirm the transformation rule (4.2.1), by 
taking a closure of two supersymmetry transformations on all the bosonic 
component fields, relying on the useful relationships in (4.1.8).   
As a typical example, we give the case on $~A\du{\m\n}\a$:  Using
(4.1.8), we get 
$$ \li{ \[ \d\low Q(\e_1) , \d\low Q(\e_2) \] A\du{\m\n}\a = & \, 
     \bigg[ - \frac3 4 V\du+\a \left( {\Bar\e}_2 \g^{\s\r} \e_1 \right) 
         G^*_{\s\m\n} \phiroh 
        +\frac14 V\du-\a\left({\Bar\e}_2 \g^{\t\r}\e_1\right) 
          G_{\r\m\n} \phitau \cr 
& ~~~~~ ~~~ +\frac1{24} V\du-\a \left( {\Bar\e}_2\g\du{\m\n}{\r\s\t\o}\e_1 
     \right) G_{\r\s\t} \phiomega \cr 
& ~~~~~ ~~~ - \frac1{24} V\du+\a \left({\Bar\e}_2 
      \g\du{\m\n}{\r\s\t\o} \e_1 \right) G^*_{\r\s\t}\partial_\o\varphi  
     \, \bigg] - {\scst (1\leftrightarrow 2)} + \hbox{c.c.} \cr 
= & \, \left({\Bar\e}_1 \g^{\r\s} \e_2 \right) F_{\r\m\n}{}^\a \phisigma   
     = \xi^\r F\du{\r\m\n}\a ~~, 
&(4.2.8) \cr } $$ 
as the $~G\-$linear terms,  
where $~\xi^\r \equiv \left({\Bar\e}_2 \g^{\r\s} \e_1 \right) \phisigma$, 
and we have used the relationship 
$$ \bigg[ \, \frac14 V\du-\a \left( {\Bar\e}_2\g^{\t\r} \e_1 \right) 
      G_{\r\m\n} \phitau - \frac3 4 V\du+\a \left({\Bar\e}_2\g^{\s\r} 
      \e_1 \right) G_{\s\m\n}^* \phiroh \, \bigg]  
      - {\scst (1\leftrightarrow 2)} + \hbox{c.c.}  
= + \xi^\r F_{\r\m\n}{}^\a  ~~.     
\eqno(4.2.9) $$
All other terms like those proportional to $~F_{\m\n\r\s\t}$~ cancel 
themselves.  Needless to say, this closure is up to the terms 
understood as extra transformations.    

We next outline the derivation of our field equations.  The main
ingredient  in this process is much like the $~N=2$~ chiral supergravity in 
10D \schwarz,  {\it except for} the involvement of the gradients
$~\partial\varphi$~ and $~\partial\Tilde\varphi$~ which are sometimes subtle.  
We first postulate our gravitello $~\l\-$field equation as  
$$ \Pdown \g^\m {\Hat D}_\m \l - i a_1 \Pdown\g^{\[5\]} \l \,
     \Hat F_{\[5\]} = 0 ~~, 
\eqno(4.2.10) $$
and take its variation under supersymmetry as in 10D \schwarz.  Note that the
supercovariantization of the derivative and field strength is also crucial.  
After the variation, all the terms
are categorized either into $~\e\-$terms or $~\e^*\-$terms.  The former is 
further composed of three sorts of terms: ~(i) $F G\-$terms, 
(ii) $D G\-$terms, and (iii) $G^* P\-$terms, where $~D$~ in $~D G$~ denote
derivative acting on $~G$.  
After appropriate algebra, the (i) $F G\-$terms are arranged as 
$$ \li{\left( F G\-\hbox{terms}\right) = & \, + i \left( \frac1{320} 
     - \frac3 4 a_1 \right) \big(\delsl\Tilde\varphi\big) \g^{\r\s\n_1\cdots\n_5}
\e 
     F_{\[\n_1\cdots\n_5|} G\du{\r\s}\t \partial_{|\t\]}\varphi \cr 
&\, - i \left( \frac1{16} + 5 a_1 \right) \Pdown 
     \g^{\m\n}\e F\du{\m\n}{\[3\]} G_{\[3\]} ~~.   
&(4.2.11) \cr}  $$ 
As in the 10D case \schwarz, we require the first line to vanish, getting 
the condition 
$$  a_1 = + \frac1{240} ~~, 
\eqno(4.2.12) $$ 
while the second line contributes to the $~A_{\m\n}{}^\a\-$field equation, as 
will be seen later.  The (ii) $D G\-$terms and the (iii) $G^* P\-$terms
talk to each other under the Bianchi identity (4.2.3a).  After all, we get  
$$ \li{ \left( \e\-\hbox{terms}\right) & =    
     - \frac18 \Pdown \g^{\m\n} \e \left[ \, D_\t G\du{\m\n} \t + 
       P_\t G^*\du{\m\n}\t  
        + \frac{2i} 3 F\du{\m\n} {\[3\]} G_{\[3\]} \right]  
&(4.2.13\rma)  \cr    
& = - \frac14 \g^\s \g^{\r\m\n} \e \Big[ \,  D_\t G\ud\t{\[\m\n}
     \big(\partial_{\r\]}\varphi\big) 
     + P^\t G^*_{\t\[\m\n} \big(\partial_{\r\]}\varphi\big) \cr 
& ~~~~~ ~~~~~ ~~~~~ ~~~~~ ~~~~~ ~~~~~ ~~~~~ ~~~~~  + \frac{2i}3 F\du{\[\m\n|}
     {\[3\]} G_{\[3\]} \big(\partial_{|\r\]}\varphi\big) \, \Big] \phisigma = 0 
     {~~,~~~~~ ~~~~~}   
&(4.2.13\rmb) \cr } $$  
which yields our $~A_{\m\n}{}^\a\-$field equation (4.2.7b).  
Note that it is too strong to require the vanishing of the square bracket in 
(4.2.13a), because of the multiplication of $~\Pdown$~ in front.  

We next look into the $~\e^*\-$terms.  They consist of three sectors: (i) 
$P F\-$terms, ~(ii) $D P\-$terms, and (iii) $G^2\-$terms.  Here the 
(i) $P F\-$terms are arranged as
$$ \left( P F\-\hbox{terms}\right) =i \left( \frac1{24} - 10a_1 \right) 
     \Pdown \g^{\[4\]} \e^* P^\m F_{\[4\]\m} ~~, 
\eqno(4.2.14) $$ 
yielding the condition 
$$ a_1 = +\frac1{240} ~~,  
\eqno(4.2.15) $$
consistently with (4.2.12).  Now the remaining (ii) $D P\-$ and 
(iii) $G^2\-$terms are arranged under (4.2.3a) to give  
$$ \left( \e^*\-\hbox{terms}\right)  
     = \Pdown \e^* \left( D_\m P^\m - \frac1{24} G_{\r\s\t}{}^2\right) ~~, 
\eqno(4.2.16) $$
resulting in the scalar field equation (4.2.7a).  This concludes all the 
variation of the gravitello $~\l\-$field equation.   

We next postulate the gravitino field equation as
$$ \li{ \g^\l \bigg[ & \g^\r \Hat{\cal R}_{\r\[\m|} 
     + b_2 \left(\g^\r\g_{\[\m|} \l^* \right) \Hat P_\r + b_3 \left( \g_{\[\m|} 
     \g^\r \l^* \right) \Hat P_\r \cr 
& + b_4 \left( \g^{\r\s\t} \g_{\[\m|} \l \right) \Hat G^*_{\r\s\t} 
    + b_5 \left( \g_{\[\m|} \g^{\r\s\t} \l \right) \Hat G^*_{\r\s\t} \, \bigg ] 
       \big(\partial_{|\n\]}\varphi\big) \big(\philambda\big)= 0 ~~, 
&(4.2.17) \cr } $$ 
with the constants $~b_2, ~\cdots, ~b_5$~ to be fixed.  
Note that the supercovariantization of the gravitino field strength is 
crucial, while that of $~P_\m$~ is not, due to the higher dimensions of the
latter, affecting only fermionic terms in bosonic field equations that 
we skip in this paper.  
The basic structure of this form is fixed after some trial and error process 
we performed in order
to produce the $~e\du\m m$~ and $~A_{\m\n\r\s\t}\-$field equations, as in 
the $~N=2$~ chiral supergravity  in 10D \schwarz.  To be more specific, our
first guiding principle was to rely on the anti-self-duality equation (4.2.7d), 
and we take its variation under supersymmetry.  It basically yields the equation
$$ \g_{\[ \m_1\m_2\m_3 }{}^\n {\cal R}_{\m_4\m_5}
     \big(\partial_{\m_6\]}\varphi\big) \big(\phinu\big)  
    + \frac1{720} \e\du{\m_1\cdots\m_6}{\n_1\cdots\n_6} \g\du{\n_1\n_2\n_3}\r 
     {\cal R}_{\n_4\n_5} \big(\partial_{\n_6}\varphi\big) \big(\phiroh\big) 
     + {\cal O}(\phi^2) = 0~~, 
\eqno(4.2.18)  $$
which is arranged after some manipulations of $~\g\-$matrices to be
$~\big(\delsl\varphi\big)\g^\r{\cal R}_{\r\[\m} \partial_{\n\]}\varphi = {\cal
O}(\phi^2)$,  leading us to the postulate (4.2.17).  Here $~{\cal O}(\phi^2)$~ 
denotes the bilinear or higher-order terms in {\it physical} fields, {\it e.g.,} 
the $~\varphi\-$field is {\it not} physical.   

The variation of 
the gravitino field equation consists of two parts: The $~\e\-$terms and 
$~\e^*\-$terms.  Now we first see that the $~\e^*\-$terms consists further of 
three parts:  (i) $D G\-$terms, (ii) $P G^*\-$terms, and (iii) $F G\-$terms.
Due to the Bianchi identity (4.2.3a), the first two sectors talk to each 
other, yielding  
$$ \li{ \big( D G\-\hbox{terms}\big)  
     + \left( P G^* \-\hbox{terms}\right) 
= \,&+ \left( -\frac1{96} - b_5 \right) \g^\l \g\du{\[\m|} {\r\s\t\o} \e^*      
     P_\r G^*_{\s\t\o} \big(\partial_{|\n\]}\varphi\big)
     \big(\philambda\big) \cr 
& + \left( \frac1{32} - \frac1{12} b_2 -b_5 \right) \g^\l\g^{\r\s\t} \e^* 
     P_{\[\m|}G^*_{\r\s\t}\big(\partial_{|\n\]}\varphi)
     \big(\philambda\big)\cr   
& + \left( - \frac3{32} - 9 b_5 \right) 
     \g^\l \g^{\r\s\t} \e^* P_\r G^*_{\s\t\[\m}
      \big(\partial_{\n\]}\varphi\big)\big(\philambda\big)  \cr 
& + \left( + \frac3{16} + 18b_5 \right) 
      \g^\l\g^\r \e^* P^\r G^*_{\s\r\[\m} \big(\partial_{\n\]}\varphi\big)
      \big(\philambda\big) \cr 
& + \left(+\frac1{32} + 3b_5  \right)\g^\l\g_{\[\m|}{}^{\r\s} 
     \e^* P^\t G^*_{\r\s\t}
     \big(\partial_{|\n\]}\varphi\big)\big(\philambda\big) \cr  & - \frac i8
     \g^\l\g^\r\e^* F\du{\r\[\m|}{\[3\]} G_{\[3\]} \big(\partial_{|\n\]}
     \varphi\big) \big(\philambda \big) \cr 
& + \frac i{48} \g^\l \g\du{\[\m|} {\r\s} \e^* F\du{\r\s}{ \[3\]} 
      G_{\[3\]} \big(\partial_{|\n\]}\varphi\big)\big(\philambda\big) \cr 
& + \left[ \, (b_2-b_3)\-\hbox{terms} \, \right] 
     + \left[ \, (b_4-2b_5)\-\hbox{terms} \, \right] {~~.~~~~~ ~~~~~ ~~~~~} 
&(4.2.19) \cr } $$ 
Even though we skip the explicit structure of the last line, 
the important point 
is that these two sorts of terms are independent, yielding two conditions
$~b_2-b_3=0$~ and $~b_4-2b_5=0$.  Now requiring the vanishing of each 
$~P G^*\-$terms in (4.2.19), we can fix the values 
$$ b_2 = + \frac12~~, ~~~~ b_3 = + \frac12~~, ~~~~ b_4 = - \frac1{48} ~~, 
    ~~~~ b_5 = - \frac1{96} ~~. 
\eqno(4.2.20) $$
Even though we do not give the details here, we stress the usage of 
various relationships 
based on the properties of $~\delsl\varphi,~\delsl\Tilde\varphi$~ together with
$~\Pup,~\Pdown$,  in addition to our extra constraints (4.2.4).  For 
instance, we can show the relationship
$$ \li{ \g^\l \g^{\r\s\t} & \g_{\[\m|} \Pdown \g^\o \e^* P_\o G_{\r\s\t}^* 
     \big(\partial_{|\n\]}\varphi\big) \big(\partial_\l\varphi\big) \cr 
= \, & - \g^\l \g\du{\[\m|}{\r\s\t\o} \e^* P_\o G_{\r\s\t}^* 
     \big(\partial_{|\n\]}\varphi\big)\big(\partial_\l\varphi\big) 
     - 3\g^\l \g\du{\[\m|}{\r\s} \e^* P^\t G_{\r\s\t}^*
     \big(\partial_{|\n\]} \varphi\big)  \big(\partial_\l\varphi\big) \cr 
& + \g^\l \g^{\r\s\t}\e^* P_{\[\m|} G_{\r\s\t}^* 
     \big( \partial_{|\n\]} \varphi \big) \big(\partial_\l\varphi\big)\cr 
& + 3\g^\l\g^{\r\s\t} \e^* P_\t G_{\r\s\[\m|}^* 
     \big(\partial_{|\n\]} \varphi \big) \big( \partial_\l\varphi \big) 
     + 6\g^\l\g^\r \e^* P^\s G_{\r\s\[\m|}^*  
     \big( \partial_{|\n\]} \varphi \big) \big( \partial_\l  \varphi \big) 
      {~~.~~~~~ ~~~~~}
&(4.2.21) \cr } $$
For example, in the left hand side, we can push the particular combination
$~\g^\l\philambda$~ all the way to 
the left side of $~\Pdown$, making this projection operator redundant.   
This is because of the constraint $~G_{\r\s}^{*~~\t}\partial_\t\varphi = 0$~ 
and the antisymmetrization $~_{\[\m~\n\]}$.  Once the projection operator
$~\Pdown$~ is deleted, we see that the $~\g\-$matrix algebra is parallel to
the $~N=2$~ chiral supergravity in 10D \schwarz, and even the coefficient turns
out to be the same, except for $~\delsl\varphi$~ 
in front as an overall factor.  In our 12D computation, we frequently
use this technique of moving around the combination $~\delsl\varphi$, 
using the constraints (4.2.4) on the fields.   

We now look into the remaining (iii) $F G\-$terms in the $~\e^*\-$sector.  This 
sector is actually the most involved, as in 10D case \schwarz.  However, we can 
categorize all the terms with respect to the number of $~\g\-$matrices involved, 
so in total we have (1) $\delsl\varphi\g^{\[1\]}\e^*\-$terms, 
(2) $\delsl\varphi\g^{\[3\]}\e^*\-$terms,
(3) $\delsl\varphi\g^{\[5\]}\e^*\-$terms, (4) $\delsl\varphi\g^{\[7\]}\e^*\-$terms.  
After some manipulations, we soon notice that  (1) $~\delsl\varphi\g^{\[1\]}\e^*\-$terms
cancel themselves, while (2) $\delsl\varphi\g^{\[3\]}\e^*\-$terms and 
(4) $~\delsl\varphi\g^{\[7\]}\e^*\-$terms cancel each other, due to the self-duality
of the fifth-rank antisymmetric field strength (4.2.7d), 
expressed more symmetrically as 
$$ {\cal F}_{\[ 6\]} = - \frac1{6!} \e\low{\[6\]}{}^{\[6\]'} {\cal F}_{\[ 6\]'}
~~,~~~~{\cal F}_{\m_1\cdots\m_6} \equiv  F_{\[\m_1\cdots\m_5}
     \partial_{\m_6\]}\varphi ~~,    
\eqno(4.2.22) $$ 
combined with the technical relations associated with the $~\e\-$tensor 
in (A.20).  Finally, 
we see that the (3) $\delsl\varphi\g^{\[5\]}\e^*\-$terms cancel themselves, 
by help of the relations such as $~F_{\[3\] \r\[\m } \partial_{\n\]}\varphi 
= + 3 {\cal F}_{\[3\] \r\m\n}$.  After all, all the 
(iii) $F G\-$terms cancel themselves consistently, and they do not yield any 
field equations, as expected also from the experience of $~N=2$~ supergravity 
in 10D \schwarz.  This concludes all the $~\e^*\-$terms in the variation of 
(4.2.17).    

We next compute the $~\e\-$terms in the supersymmetric variation 
of the gravitino
field equation.  They are categorized as (i) $R\-$terms, (ii) $P P^*\-$terms, 
(iii) $D F\-$terms, (iv) $F^2 \-$terms, (v) $G G^* \-$terms.  Here the 
(i) $R\-$terms contain the Riemann tensor, 
coming from the commutator $~\[ D_\m, D_\n \] \e$.  
These $~R\-$terms give the leading Ricci tensor term in the zw\"olfbein 
$~e\du\m m\-$field equation, as will be in (4.2.34).  The (ii) $P P^*\-$terms are
arranged as $$ \li{ \left( P P^*\-\hbox{terms}\right)  = \, & + \g^\l\g^\r\e
     \left[ \, -\frac12 P_\r P^*_{\[\m|} +\left( \fracm12 - 2b_2 \right) 
      P_{\[\m|} P_\r^* \, \right] \big(\partial_{|\n\]}\varphi\big)
     \big(\philambda\big)\cr  & + (b_2 - b_3) \g^\l\g_{\[\m|} \g^\r 
     \Pdown \g^\s \e P_\r P^*_\s \big(\partial_{|\n\]}
     \varphi\big)\big(\philambda\big) ~~. 
&(4.2.23) \cr  } $$
Here the first two terms will contribute to the energy-momentum tensor, while
the last line is to vanish, yielding the result
$$ b_2 = b_3 = +\frac12~~, 
\eqno(4.2.24) $$
consistently with (4.2.20).  Now the (iii) $D F\-$terms turn out to be 
equivalent to the (v) $~G G^* \-$terms by the use of the Bianchi identity 
(4.2.3b).  In fact, we get
$$ \li{ \left( D F\-\hbox{terms} \right)  = \, &  
     - \frac1{192} \g^\l\g^{\[ 3 \] \[ 2 \]} \e \left[\, G_{\[ 3 \]} 
     G^*_{\[ 2 \]\[\m|} 
     \big(\partial_{|\n\]}\varphi\big) 
     - G_{\[ 2 \] \[ \m|} G^*_{\[ 3 \]} \big(\partial_{|\n\]}\varphi\big)
     \,\right]
     \big(\philambda\big)  \cr 
& - \frac1{576} \g^\l\g_{\[\m|}{}^{\[3\] \[ 3 \]'} \e\, G_{\[3\]} G^*_{\[3\]'} 
     \big(\partial_{|\n \]}\varphi\big)\big(\philambda\big) ~~,  
&(4.2.25) \cr } $$
which will be combined with the explicit (v) $G G^*\-$terms below.  
We will combine these terms with the explicit 
~$G G^* \-$terms of the category
(v) later.  Now (iv) $F^2 \-$terms are arranged to be
$$ \big( F^2\-\hbox{terms} \big) = -\fracm1{12} \g^\l\g^\r \e F_{\[4\]\[\m|} 
     F^{\[4\]}{}_\r \big(\partial_{|\n\]}\varphi\big)\big(\philambda\big) ~~,  
\eqno(4.2.26) $$ 
which contributes to the energy-momentum tensor in the $~e\du\m m\-$field 
equation.  For these complicated $~F^2\-$terms, we have used the 
following important technique.  Notice, that {\it e.g.,} in (4.2.26) all the
indices on $~F$~ including the contracted ones take {\it purely} 10D values. 
This is because the $~{\scst\r}\-$index can take only the 10D value, because of
$~\delsl\varphi$~ in front, while the contracted ones $~{\scst \[4\]}$~ stay
also within 10D,  due to the constraint (4.2.4a).  This implies that we can
basically use the  purely 10D relations for simplifications of these
$~F^2\-$terms.  In fact, the anti-self-duality (4.2.22) in 12D implies, as
desired, the self-duality in 10D \schwarz:   
$$ F_{i_1\cdots i_5} = +\frac1{5!} \e\du{i_1\cdots i_5}{j_1\cdots j_5} 
        F_{j_1\cdots j_5} ~~, 
\eqno(4.2.27) $$ 
where $~{\scst i_1,~i_2,~\cdots}$~ are purely 10D indices.       
This relation in turn leads to other identities, such as  
$$ \g^{i\[2\]} \e F\ud{\[3\]}{i j} F_{\[2\]\[3\]} \equiv 0 ~~,  
~~~~ \g^{\[3\]\[4\]} \e F_{\[3\] i j} F\du{\[4\]}j \equiv 0 ~~.   
\eqno(4.2.28) $$   
Fortunately, we found that all of these relevant $~F^2\-$terms always have  
purely 10D indices on $~F$'s, and we can keep using this technique in this
sector.    

We finally arrange all the $~G G^*\-$terms which are explicitly from 
the (v) $~G G^* \-$terms and from the (iii) $D F\-$ terms {\it via}
(4.2.25).  These terms are rather involved, but we can further 
categorize them by the number of $~\g\-$matrices, as
(1) $\delsl\varphi\g^{\[7\]} \e\-$terms, (2) $\delsl\varphi\g^{\[5\]} \e\-$terms, 
(3) $\delsl\varphi\g^{\[3\]} \e\-$terms, (4) $\delsl\varphi\g^{\[1\]} \e\-$terms.  
Here (1) $\delsl\varphi\g^{\[7\]} \e\-$terms turn out to be 
$$ \li{ & \left( \delsl\varphi\g^{\[7\]} \e\-\hbox{terms} \right)   
=  \frac1{768} \left[ \, 32(-b_4 +b_5 ) + \frac1 3\, \right] 
     \g^\l\g_{\[\m|}{}^{\[3\]\[3\]'} \e \, G_{\[3\]} G_{\[3\]'} ^*
     \big(\partial_{|\n\]}
     \varphi\big) \big(\philambda\big) {~~.~~~~~ ~~~~~}
&(4.2.29)  \cr }
$$ 
The $~\g\-$matrix structure here is different from the leading Ricci-tensor 
term in the $~e\du\m m\-$field equation, so that these terms should vanish, 
yielding the condition 
$$ b_4 - b_5 = - \frac1{96}~~,  
\eqno(4.2.30) $$
consistent with all our previous values.  Next the 
(2) $\delsl\varphi\g^{\[5\]} \e\-$terms can be arranged after some algebra into
$$ \li{\left( \delsl\varphi\g^{\[5\]} \e\-\hbox{terms} \right) 
      = & + \frac1{256} \left[ \, 96(b_4 - b_5 ) + 1 \, \right] 
     \g^\psi\g_{\[\m|}{}^{\r\s\l\o} \e \, G_{\r\s\t} 
      G^*\du{\l\o}\t \big(\partial_{|\n\]}\varphi\big)\big(\phipsi\big) \cr 
& + \frac1{3072} \left[ \, -12 -384(b_4 + b_5) \, \right] 
     \g^\l \g^{\[3\]\[2\]} \e\,G_{\[3\]} G_{\[2\] \[\m}^*
     \big(\partial_{\n\]}\varphi\big) \big(\philambda\big)  \cr 
& + \frac1{3072} \left[ \, - 4 - 384 (b_4 - b_5) \, \right] 
     \g^\o\g^{\r\s\t\l\o} \e\, G_{\l\o\[\m|} G^*_{\r\s\t}
     \big(\partial_{|\n\]}\varphi\big)\big(\phiomega\big) 
     {~~. ~~~~~ ~~~~~ ~~~~~}
&(4.2.31)  \cr } $$
Fortunately, each line vanishes, for the same values of 
$~b_4$~ and $~b_5$~ as in (4.2.20).  In a similar fashion, the
(3) $~\delsl\varphi\g^{\[3\]} \e\-$terms are arranged to
$$ \li{ \left( \delsl\varphi \g^{\[3\]} \e\-\hbox{terms} \right) 
       = & +\frac1{512} \left[ \, 4 + 384(b_4 - b_5) \, \right] 
      \g^\psi\g_{\[\m|}{}^{\r\o}  \e \, G_{\r\s\t} G^*\du\o{\s\t}
      \big(\partial_{|\n\]}\varphi\big) \big(\phipsi\big) \cr 
& + \frac1{512} \left[ \, -12-384(b_4+b_5) \, \right]
      \g^\psi\g^{\s\t\l} \e\, G_{\s\t\o} G_\l^{*~\o}{}_{\[\m}
     \big(\partial_{\n\]}\varphi\big)\big(\phipsi\big) \cr   
& + \frac1{512} \left[ \, + 4 + 384(b_4- b_5 )  \, \right] 
     \g^\o\g^{\r\s\l}
     \e\, G_{\l\t\[\m|} G^*\du{\r\s}\t \big(\partial_{|\n\]}\varphi\big)
     \big(\phiomega\big) {~~.~~~~~ ~~~~~ ~~~~~} 
&(4.2.32) \cr } $$ 
Fortunately, we see that each of these line vanish consistently with the 
values of $~b_4,~b_5$, as in (4.2.20).  Finally we look into the 
(4) $~\delsl\varphi\g^{\[1\]} \e\-$terms which contribute to the energy-momentum tensor:
$$\li{& \big( \delsl\varphi\g^{\[1\]} \e\-\hbox{terms} \big)  \cr 
& ~~~~~ ~~~~~ = \frac1{16} \g^\l \g^\t \e \bigg[ \, -G_{\r\s\t}
      G^*_{\r\s\[\m |} - G_{\r\s\[\m|} G^*_{\r\s\t}  
+ \frac16 g\low{\t\[\m|} \left| G_{\r\s\o} \right|^2 \, \bigg]
     \big(\partial_{|\n\]}\varphi\big) \big(\philambda\big) {~~. ~~~~~ ~~~~~} 
&(4.2.33) \cr } $$ 
This completes all the $~G G^*\-$terms.  

We now collect all the terms to contributing to the 
$~e\du\m m \-$field equation. 
They are all with $~\delsl\varphi\g^{\[1\]}\e$, arranged as 
$$ \li{ \fracm12 \g^\l \g^\r \e \bigg[ & R_{\[\m|\r} 
     - P_\r P_{\[\m|}^* - P_{\[\m|} P_\r^*  - \frac1 6 F_{\[4\]\r} 
        F\ud{\[4\]}{\[\m|} \cr 
& - \frac18 G_{\[2\] \r} G^*_{\[2\] \[\m|} 
         - \frac18 G_{\[2\] \[\m|} G^*_{\[2\]\r} 
         + \frac1{48}g\low{\r\[\m|} G_{\s\t\o} G^{*\s\t\o} \, \bigg] 
     \big(\partial_{|\n\]}\varphi\big) \big(\philambda\big) = 0  
     {~~. ~~~~~ ~~~~~ ~~~~~}
&(4.2.34) \cr} $$
yielding nothing but (4.2.7c).   This concludes the internal consistency check,
as well  as the derivation of all of our field equations, after the
supersymmetric  variation of our fermionic field equations.  

\bigskip\bigskip

\leftline{\bf 4.3~~Dimensional Reduction into $~D=9+1$}

As before, we can perform dimensional reduction into $~D=9+1$~ in order to 
see the validity of our $D=12,\,~N=2$~ supergravity.  The most 
frequently-used relationships are the same as (2.4.1) - (2.4.4), and the 
only new ones for $~N=2$~ case are 
$$ \Bar{\Hat{\cal R}}_{\hat\m\hat\n} = \big(\Bar{\cal R}
     _{\hat\m\hat\n}, 0 \big) ~~, ~~~~ 
     \Hat\l = \pmatrix{0\cr \l} ~~.  
\eqno(4.3.1)  $$ 
Since most of the dimensional reduction prescription is the same as in the 
$~N=1$~ case, we skip the details to get 10D field equations in agreement 
with ref.~\schwarz.  As a matter of fact, their forms can be easily figured out,
due to our 12D field equations which already resemble those in 10D.

\bigskip\bigskip
\bigskip\bigskip

\centerline{\bf 5.~~Super $~(2+2)\-$Brane on 
Background of ~$D=12,\,N=2$~ Supergravity}
\bigskip

As we have studied the Green-Schwarz superstring on the background of 
$~D=12,\,N=1$~ supergravity to see the validity of our theory as a probe, 
we can try to put super $~(2+2)\-$brane on the 
$~N=2$~ supergravity in 12D \nishtwo.  Here we choose the super $~(2+2)\-$brane 
\ref\hp{S.~Hewson and M.~Perry, \np{492}{97}{249}.}, because 
the super $~(2+2)\-$brane is the right p-brane action \ref\pbrane{A.~Achucarro,
J.~Evans, P.~Townsend and D.~Wiltshire, \pl{198}{87}{441}.}  
on such a background which 
is supposed to be the strong coupling limit of F-theory \ftheory\ with 12D
space-time  with two time coordinates.  The existence of the forth-rank
antisymmetric tensor $~A_{\[4\]}$~ also suggests this is the natural super
p-brane \pbrane.  

Our postulate for the total action for the super $~(2+2)\-$brane is \nishtwo
$$\li{&S = S_\s + S_A ~~,  
&(5.1\rma) \cr 
&S_\s \equiv \int d^4 \s \, \left( \half {\sqrt g} g^{i j} 
   \eta_{a b} \Pi\du i a \Pi\du j b - {\sqrt g} \right)~~,  
&(5.1\rmb) \cr  
&S_A \equiv \int d^4 \s \, \left( - \fracm i 6 \e^{i_1\cdots i_4} 
    \Pi\du{i_1}{B_1} \cdots \Pi\du{i_4}{B_4} A_{B_4\cdots B_1} \right)~~.
&(5.1\rmc) \cr } $$ 
Here the indices $~{\scst i,~j,~\cdots~=~1,~2,~3,~4}$~ stand for curved
coordinates in $~D=2+2$, and $~\Pi\du i A\equiv \big(\partial_i Z^M \big)
E\du M A$, {\it etc.}, similarly to section 3.  
Note the extra imaginary unit factor for $~S_A$~ understood
as  Wick rotation from Minkowskian $~D=1+3$~ to our $~D=2+2$.  
This feature is similar to the 
case for the topological $~F\Tilde F\-$terms in Euclidean $~D=4+0$~ from 
the usual Minkowskian $~D=3+1$~ 
\ref\vw{{\it See, e.g.}, C.~Vafa and E.~Witten, \prl{53}{84}{535}.}. 
Accordingly, $~g\equiv\det(g_{i j})$~ is positive definite, so 
we use $~\sqrt g$~ instead of $~\sqrt{-g}$.   

Our relevant superspace constraints are \nishtwo
$$ \li{&T\du{\a\b}c = (\g^{c d})\low{\a\b} \phid + (\Pupdown)\low{\a\b} 
     \nabla^c\varphi
     ~~, ~~~~ T\low{\Bar\a\Bar\b}{}^c = (\g^{c d})\low{\a\b} \phid 
     + (\Pupdown)\low{\a\b} \nabla^c\varphi ~~,    
&(5.2\rma) \cr  
&F_{\a\b c d e} = -\fracm i 4 (\g_{c d e}{}^f)\low{\a\b} \nabla\low f
     \varphi~~, ~~~~
    F_{\Bar\a\Bar\b c d e} = + \fracm i 4(\g_{c d e}{}^f)\low{\a\b}
    \nabla\low f\varphi~~,
&(5.2\rmb) \cr
& \nabla_a\nabla_b\varphi = 0 ~~, ~~~~\nabla_a\nabla_b\Tilde\varphi = 0 ~~,
    ~~~~\big(\nabla_a\varphi\big)\big(\nabla^a\Tilde\varphi \big) = 1~~,\cr 
& \big( \nabla_a\varphi\big)^2= 0 ~~, ~~~~ 
        \big( \nabla_a\Tilde\varphi\big)^2 = 0 ~~, ~~~~ \nabla_\a\varphi
        =0~~, ~~~~ \nabla_\a\Tilde\varphi=0 {~~.~~~~~ ~~~~~} 
&(5.2\rmc) \cr } $$  
These expressions are easily obtained from the component
results using the standard technique in \ggrs.  The {\it barred}
spinorial indices $~{\scst \Bar\a,~\Bar\b,~\cdots}$~
denote the complex conjugations in superspace, corresponding to
the $~{\it star}\-$operations in component in (4.1.8b).                 

As usual in general $~p\-$brane formulation \pbrane, 
we have the fermionic symmetries \nishtwo
$$\li{
&\d E^\a = \left(I + \G \right)\ud\a\b \k^\b 
   + \half \left[\,\big(\nablasl\varphi\big)\big(\nablasl\Tilde\varphi\big) 
    \,\right]^{\a\b}
    \eta\low\b 
   \equiv - \left[\,\left(I+\G \right)\k \,\right]^\a
   + \left( \Pup\eta \right)^\a  ~~,  
&(5.3\rma) \cr 
&\d \Bar E^{\Bar\a}=\left(I + \G\right)\ud\a\b{\Bar\k}{}^{\Bar\b}  
   + \half \left[\, \big(\nablasl\varphi\big) \big(\nablasl\Tilde\varphi\big) 
   \,\right]^{\a\b} 
   {\Bar\eta} \low{\Bar\b} 
   \equiv - \left[\,\left( I+\G \right) \Bar\k\, \right]^{\Bar\a} 
   + \left( \Pup\Bar\eta \right)^{\Bar\a} ~~, ~~~~
&(5.3\rmb) \cr 
&\d E^a = 0 ~~, ~~~~\d \varphi = 0 ~~, ~~~~\d\Tilde\varphi = 0 ~~, 
&(5.3\rmc) \cr } $$
with the fermionic parameters $~\k$~ and $~\eta$, 
where $~\nablasl\varphi \equiv \g^a\nabla_a\varphi$, 
and $~\G~$ defined by \hp\  
$$ \G\equiv \fracm1{24{\sqrt g}} \e^{i j k l} \Pi\du i a \Pi\du j b\Pi\du k c
\Pi\du l d \left( \g_{a b c d} \right) ~~,    
\eqno(5.4) $$ 
satisfies relations such as
$$ \li{& \G^2 = I~~,  
&(5.5\rma) \cr  
& \e_i{}^{j k l} \Pi\du j a \Pi\du k b \Pi\du l c \g_{a b c} 
   \G = + 6 {\sqrt g} \Pi\du i a \g_a ~~, 
&(5.5\rmb) \cr } $$
under the algebraic $~g_{i j}\-$field equation 
$~g_{i j} = \Pi\du i a\Pi_{j a}$,   
which are useful for the invariance check of our total action $~S$.  
In fact, the variation of $~S$~ under (5.3) takes the form
$$\eqalign{\d \left( S_\s + S_A \right) = \Big[ & + {\sqrt g} 
      g^{i j} \Pi_i{}^\g \big(\g_a \g^b \big)_{\g\b} \big(\phib\big)
       \big(\d E^\b \big) \Pi\du j a \cr   
      & + \fracm 1 6 \e^{i j k l} \Pi_i{}^\g\big( \g\low{b c d} 
     \g^a\big)_{\g\a} \big(\phia\big) \left( \d E^\a\right) 
      \Pi\du j b\Pi\du k c\Pi\du l d\,\Big]+ 
     ( \d E^\a \rightarrow \d{\Bar E}{}^{\,\Bar\a} ) {~~.~~~~~ ~~~~~ }  \cr} 
\eqno(5.6) $$   
For the $~\eta\-$transformation in (5.3), by the aid of (5.4) we see that two
sorts of terms cancel themselves, if  we impose the extra condition
$$ \Pi\du i a \phia = 0 ~~, 
\eqno(5.7) $$ 
together with the null-ness condition $~\big( \nabla_a\varphi\big)^2=0$~ in
(5.2c).   The extra condition (5.7) is formally the same as the first equation
in  (3.12) for the Green-Schwarz superstring 
on $~N=1$~ supergravity background.  However, the difference is that for the 
Green-Schwarz superstring, we could not impose such condition from outside, 
because these conditions are of the first order, interpreted 
as unidexterous field equations on 2D world-sheet.  On the other hand, 
in the present case of super $~(2+2)\-$brane, since the world-supervolume 
is 4D, the condition (5.7)
can  be imposed as constraint from outside, for the invariance check
of our total action.  As for the $~\k\-$transformation in (5.3) applied to
(5.6), we see that eq.~(5.5) helps us to rearrange two sorts of terms
cancelling each other under (5.7), but now without (5.2c).  This is more
natural than the $~N=1$~ case, because the
$~\eta\-$symmetry is associated with the extra dimensions governing 
the null-vector condition, while the $~\k\-$symmetry governs the 
physical freedom within 10D.  In other words, our null-ness condition (5.2c) is
not artificially put by hand, but required by one of the fermionic symmetries of
the super $~(2+2)\-$brane action.     

As in the case of Green-Schwarz superstring for the $~D=12,\,N=1$~
supergravity, we can easily understand the ordinary 16+16 degrees of freedom
come out  of the super $~(2+2)\-$brane by these fermionic symmetries:  First,
the $~\eta\-$symmetry deletes half of the original 64 components  of the
fermionic coordinates $~\theta^\m$~ in $~N=2$~ superspace in 12D, and thus at
most 32 components can be physical.   Next, the $~\k\-$symmetry deletes further
half of 32 components, leaving the usual 16 components in the
light-cone coordinates in $~N=2$~ Green-Schwarz superstring \gsw.

\bigskip\bigskip
\bigskip\bigskip


\centerline{\bf 6.~~$N=1$~ Supergravity in $~D=11+2$} 
\bigskip

\leftline{\bf 6.1~~Notations} 

Our metric in $~D=11+2$~ is $~\big(\eta_{m n}\big) = \hbox{diag.}~\big( -, +,
\cdots ,+, +, +, - \big)$, with the local Lorentz indices $~{\scst
m,~n,~\cdots,~=~(0),~(1), ~\cdots,~(9),~(10),~(12),~(13)}$, 
and our ~$\e\-$tensor
is defined by ~$ \e^{0 1 2 \cdots 10\,12\,13} = + 1$, and accordingly $~
\g\low{(13)} \equiv \g\low{(0)} \g\low{(1)} \g\low{(2)}\cdots\g\low{(9)}
\g\low{(10)}\g\low{(12)}$.  
Our null-vectors are defined in the same way as in 12D \nishone\nishtwo:
$$ \li{&(n_m)=\left(0, 0,\cdots, 0, +\frac1{\sqrt2},+\frac1{\sqrt2}\right)~~, 
     ~~~~(m_m) = \left(0, 0, \cdots, 0, +\frac1{\sqrt2}, -\frac1{\sqrt2}\right)
     {~~. ~~~~~ ~~~~~}
& (6.1.1) \cr} $$ 
As in 12D, we define the $~{\scst \pm}\-$indices by 
$$ \li{ & V_\pm \equiv \frac1{\sqrt 2} \big( V_{(12)} \pm V_{(13)} \big)~~,  
&(6.1.2) \cr } $$ 
so that we have $~n_+ = m^+ = + 1, ~n_- = m^- = 0,~ n_+ m^+ = n^-
m_- = + 1$~ as in 12D.  Other important relations such as (2.1.5) are formally
the same as in 12D \nishone\nishtwo.  Therefore we can identify these 
null-vectors with the gradients of superinvariant scalars $~\varphi,
~\Tilde\varphi$, as in (2.2.2): 
$$ \li{ & n_\m \equiv \partial_\m\varphi~~, ~~~~
        m_\m \equiv \partial_\m\Tilde\varphi~~, ~~~~ 
        D_m D_n\varphi = 0 ~~, ~~~~ D_m D_n\Tilde\varphi = 0~~. 
&(6.1.3) \cr} $$ 

As for the modified Lorentz generators in 13D, since its structure is to be  
exactly parallel to the 12D case in subsection 2.1, we do not repeat 
the details here.  We also skip other relations that are exactly and formally
parallel to 12D.

\bigskip\bigskip\bigskip

\leftline{\bf 6.2.~~$~N=1$~ Supergravity in $~D=11+2$}          
\bigskip

Based on our accumulated experience with 12D supergravity, the
construction of 13D supergravity is now straightforward. 
The field content of
our 13D supergravity is the same as in 11D supergravity 
\cjs, namely $~\big(e\du\m m, \psi_\m, A_{\m\n\r};
\varphi, \Tilde\varphi \big)$, with the dreizehnbein, a gravitino and the 
third-rank antisymmetric tensor, in addition to the supercovariant scalars.  
The supersymmetry transformation 
rule is fixed by the requirement of closure of gauge algebra  
\pvn.  

We first present our result of supersymmetry transformation rule:      
$$\li{&\d_Q e\du\m m = \big(\Bar\e \g^{m\n} \psi_\m \big) \partial_\n\varphi ~~, 
&(6.2.1\rma) \cr  
& \d_Q\psi_\m = D_\m\e + \frac1{144} \Pdown \big(\g\du\m{\[4\]} \e \, 
     \HatF_{\[4\]} - 8 \g^{\[3\]} \e \, \HatF_{\m\[3\]} \big)~~, 
&(6.2.1\rmb) \cr 
& \d_Q A_{\m\n\r} = +\frac32 \big(\Bar\e \g\du{\[\m\n} \s \psi_{\r\]} \big)
    \partial_\s\varphi~~,  
&(6.2.1\rmc) \cr 
&\d_Q\varphi = 0 ~~, ~~~~\d_Q\Tilde\varphi = 0 ~~,
&(6.2.1\rmd) \cr } $$ 
where as usual all the {\it hatted} field strengths are supercovariantized \pvn.
Our extra constraints are similar to the 12D case (4.2.4):
$$ \li{& \Hat F\du{\m\n\r\s} \t \phitau = 0 
     ~~, ~~~~ \Hat R\du\m\n{}^{m n} \phinu  = 0 ~~, ~~~~
     \Hat R_{\m\n}{}^{m n} \phim = 0 ~~, ~~~~~ 
&(6.2.2\rma) \cr 
& \Hat{\cal R}\du\m\n \phinu = 0 ~~, 
    ~~~~\Bar{\Hat{\cal R}}_{\m\n}\g^m D_m\Tilde\varphi  = 0~~,  
&(6.2.2\rmb) \cr 
& \big(D_m \varphi\big)^2  = 0 ~~, ~~~~
     \big(D_m\Tilde\varphi\big)^2  = 0~~, ~~~~
     \big(D_m\varphi \big) \big(D^m\Tilde\varphi \big) = 1 {~~.~~~~~}  
&(6.2.2\rmc) \cr } $$ 
The $~\calR_{\m\n}$~ is the field strength for the gravitino.            

We outline the derivations of this rule.  Based on the experience with the 12D
theories, we first postulate the form of the transformation rule as (6.2.1b)
and (6.2.1c) with three unknown coefficients $~a_1,~a_2$~ and $~a_3$, as  
$$\li{& \d_Q\psi_\m = \Pdown \big(a_1 \g\du\m{\[4\]} \e F_{\[4\]} 
     +a_2 \g^{\[3\]} \e F_{\m\[3\]} \big) ~~, 
&(6.2.3\rma) \cr 
& \d_Q A_{\m\n\r} = a_3\big(\Bar\e \g\du{\[\m\n} \s \psi_{\r\]} \big)
    \partial_\s\varphi ~~.    
&(6.2.3\rmb) \cr } $$ 
This structure is similar to 12D supergravities 
\nishone\nishtwo, where 
the null-vector is involved in the zw\"olfbein transformation (6.2.1a), 
while the projector $~\Pdown$~ is needed in (6.2.1b).  The
involvement of the null-vector in (6.2.1c) is expected
also from superspace as will be mentioned shortly, or as an analog of the
second-rank tensor transformation rule (2.3.28c) 
for the $~D=12,\,N=1$~ supergravity.   
Similarly to globally supersymmetric Yang-Mills \ns, our system has
also local extra symmetries:  
$$ \li{ & \d_{\rm E} e\du \m m = \a_\m
\big(D^m\varphi \big)  + \Tilde\a^m \partial_\m\varphi~~,   
&(6.2.4\rma) \cr 
& \d_{\rm E} A_{\m\n\r} = \b_{\[\m\n} \partial_{\r\]} \varphi ~~. 
&(6.2.4\rmb) \cr } $$   

The requirement of closure of supersymmetry  
on all the bosonic fields, up to the gauge, Lorentz transformations, or the
above-mentioned extra symmetries, fixes the unknown coefficients.  However, we do
not need to  confirm the closure on the gravitino, as long as we confirm the 
consistency among field equations.  

First, the closure on the dreizehnbein fixes the parameter of 
translation $~\xi^m\equiv \big(\Bar\e_2\g^{m\n}\e_1\big)\partial_\n\varphi$, 
up to local Lorentz transformation.  
Second, the closure on $~A_{\m\n\r}$~ yields 
$$ \li{ \[ \d_1, \d_2 \] A_{\m\n\r} 
   = & -12a_2 a_3\xi^\s F_{\s\m\n\r} + \partial_{\[\m} \L_{\n\r\]} \cr 
& - 2a_3 \big( 8a_1 + a_2 \big) \big( \Bar\e_2 
       \g\du{\[\m\n|}{\[3\] m } \e_1\big) F_{|\r\] \[3\]} D_m\varphi \cr 
& + a_1\Big[\, - 2\big(\Bar\e_2 \g_{\[\m|} \Pdown \g_{|\n|}{}^{\[4\]} 
      \e_1 \big) F_{\[4\]} 
      - \big( \Bar\e_2 \g_{\[\m\n|} \g^{\[4\]} \e_1 \big)F_{\[4\]} \,\Big] 
      \partial_{|\r\]} \varphi - {\scst (1\leftrightarrow 2)} \cr  
& + a_2 \left[ \, - 2\big(\Bar\e_2 \g_{\[\m|} \Pdown \g^{\[3\]} \e_1\big)
    F_{|\n|\[3\]} 
    + 6 \big( \Bar\e_2\g^{\[2\]} \e_1 \big)F_{\[\m\n|\[2\]} \, \right] 
    \partial_{|\r\]} \varphi - {\scst (1\leftrightarrow 2)}  
    {~~,~~~~~ ~~~~~ ~~~~~}  
&(6.2.5)  \cr } $$ 
where $~\L_{\m\n}$~ is the parameter of gauge
transformation, while the last three lines with $~\partial_\r\varphi$~ 
at the end are 
interpreted as the extra transformation (6.2.4b).  
The normalization of translation parameter in the
first term and the vanishing of the last term 
require the conditions 
$$ \li{ & a_2 a_3 = - \frac1{12} ~~, 
&(6.2.6\rma) \cr
& 8a_1 + a_2 = 0 ~~.
&(6.2.6\rmb) \cr  } $$
We can choose $~a_3= + 3/2$~ as the normalization to get the  
solutions 
$$ \li{& a_1 = + \frac1{144}~~, ~~~~a_2 = - \frac1{18} ~~, ~~~~
       a_3 = + \frac 3 2~~,  
&(6.2.7) \cr } $$
yielding (6.2.1), also in agreement with their corresponding terms in 
the 11D supergravity \cjs, as desired.   

We next give the list of our field equations: 
$$\li{ & \left[ \, \Hat R_{\r\[\m|} 
     + \frac 1 3 \Hat F_{\r\[3\]} \Hat F\du{\[\m|}{\[3\]} 
     - \frac1{36} g_{\r\[\m|} \Hat F_{\[4\]}{}^2  
       \, \right]\, \partial_{|\n\]}\varphi = {\cal O} (\psi^2) ~~, 
&(6.2.8) \cr 
& \big(\Hat D_\m \Hat F\ud\m{\[\n\r\s} \big) \partial_{\t\]}\varphi = +
     \frac1{2304} \e\du{\n\r\s\t} {\[4\]\[4\]' \m} 
      \Hat F_{\[4\]} \Hat F_{\[4\]'} \, \partial_\m\varphi  
     ~~, 
&(6.2.9) \cr 
& \g^\s\g^\r\Hat{\calR}_{\r\[\m} \big(\partial_{\n\]}\varphi\big)
     \big(\partial_\s\varphi) = 0 ~~.  
&(6.2.10) \cr } $$
These field equations are derived in the same way as in $~D=12,\,N=2$~
supergravity \nishtwo:  We first postulate the gravitino field
equation as in (6.2.10), and vary it under supersymmetry.  There 
are three types of terms generated: (i) $~R\-$terms, ~(ii) $D
F\-$terms, ~and (iii) $F^2\-$terms.  The (i) $~R\-$terms are going to give
the leading Ricci tensor term in the dreizehnbein (gravitational) field equation 
(6.2.8), while the (iii) $F^2\-$terms correspond to its energy-momentum tensor
terms, and (ii) $D F\-$terms give the $~F\-$field equation (6.2.9). 
However, these original terms talk to each other after the use of the $~F\-$~ 
and dreizehnbein field equations.  To clarify this point, we use the unknown
coefficients  $~\a,~\b$~ and $~\d$~ also for the $~F\-$ and the gravitational
field equations:
$$ \li{& \big( D_\m \HatF \ud\m{\[\n\r\s}\big) \partial_{\t\]}\varphi
    = \a e^{-1} \e\du{\n\r\s\t} {\[4\]\[4\]'\,\m} \HatF_{\[4\]} \HatF_{\[4\]'} 
    \, \partial_\m \varphi  ~~, 
&(6.2.11) \cr
& \left[\, \HatR_{\r\[\m|} + \d F_{\r\[3\]} F\du{\[\m|}{\[3\]}
     + \b g_{\r\[\m|} F_{\[4\]}{}^2 \, \right] 
     \partial_{|\n\]}\varphi = 0 ~~. 
&(6.2.12) \cr } $$   
For example, a typical arrangement by the use of the $~F\-$field equation is
$$ \li{ - 3a_2 \big(\delsl\varphi\big) 
      \g^{\[2\]}\e &\!\big( D_\r F\ud\r{\[2\] \[\m} \big) 
      \partial_{\n\]}\varphi 
     = - 24a_2 \a \big(\delsl\varphi\big) \g\ud{\[4\]\[4\]'}{\[\m|} 
     \e F_{\[4\]} F_{\[4\]'} 
     \partial_{|\n\]}\varphi \cr 
& - 6a_2 \big(\delsl\varphi \big) \g^{\r\s} \e  
     \left[ \, \big(D_\t F\ud \t {\[\r\s\m} \big)
     \partial_{\n\]} \varphi  
     -\a \e\du{\r\s\m\n}{\[4\] \[4\]'\t} F_{\[4\]} F_{\[4\]'} 
     \partial_\t\varphi \,\right]{\, .~~~~~ ~~~~~ ~}   
&(6.2.13) \cr} $$ 
where the last line vanishes by the $~F\-$field equation (6.2.9).  
Similarly, we have 
$$ \li{ -4a_1 &\! \big(\delsl\varphi\big)
       \g\du{\[\m|}{\[3\]} \e \big( D_\r F\ud\r{\[3\]} \big)
         \partial_{|\n\]}\varphi  
= + 768 a_1 \a \big(\delsl\varphi\big) 
     \g^{\[4\]\[3\]} \e F_{\[4\]} F_{\[3\] \[\m} 
     \partial_{\n\]}\varphi  \cr 
& - 8 a_1 \big(\delsl\varphi\big) 
      \g\du\m{\s\t\o} \e \left[\, \big(D_\r F\ud\r{\[\s\t\o}\big) 
     \partial_{\n\]}\varphi   
      - \a \e\du{\s\t\o\n}{\[4\]\[4\]' \l} F_{\[4\]} F_{\[4\]'} 
     \partial_\l\varphi \, \right] - {\scst(\m\leftrightarrow\n)} 
     {~~, ~~~~~ ~~~~~}
&(6.2.14) \cr } $$
with the vanishing second line.  Similarly,  
by the use of the gravitational field equation (6.2.8), we have 
$$ \li{ \frac12 \big(\delsl\varphi\big)
      \g^\r \e R_{\r\[\m} \partial_{\n\]}\varphi  
& = + \frac12 \big( \delsl\varphi \big) \g^\r \e 
    \left( \d F_{\r\[3\]} F\du{\[\m|}{\[3\]} \partial_{|\n\]}\varphi 
     - \b g_{\r\[\m|} F_{\[4\]}{}^2 \partial_{|\n\]}\varphi \right)\cr 
& + \frac12 \big(\delsl\varphi\big) 
     \g^\r \e \left( R_{\r\[\m} \partial_{\n\]}\varphi 
     + \d F_{\r\[3\]} F\du{\[\m|}{\[3\]} \partial_{|\n\]}\varphi 
     + \b g_{\r\[\m|} F_{\[4\]}{}^2 \partial_{|\n\]}\varphi \right)  
     {~.~~~~~ ~~~~~ ~~}   
&(6.2.15) \cr } $$
where the second line vanishes on-shell.    
In order to see the consistency of our transformation rule and field equations,
we keep the coefficients $~a_1, ~ a_2,~a_3$.  After these manipulations we get 
$$\li{ \d &\!\!\left[\big(\delsl \varphi\big)
      \g^\r {\Hat\calR}_{\r\[\m|} \partial_{|\n\]} \varphi \,\right]  
= \big( 768 \a a_1 + 32a_1^2 - 6a_1 a_2 - 2a_2^2 \big) N_{\m\n} \cr  
&~~~~~ + \big(-24a_2 \a +4a_1^2 + 2a_1 a_2 \big) W_{\m\n}  
      + \big( \frac12 \b + 288 a_1^2 \big) S_{\m\n} 
      + \big( - \frac12 \d + 1152a_1^2 + 36a_2^2 \big) P_{\m\n} \cr 
&~~~~~ - 36\big( 8a_1 + a_2 \big) \big( 2a_1- a_2 \big) Q_{\m\n} 
      - 6 \big( 8a_1+ a_2\big)^2 T_{\m\n} 
      - 72 a_1\big( 8a_1 + a_2 \big) U_{\m\n} {~.~~~~~ ~~} 
&(6.2.16) \cr }  $$  
Here $~N,~P,~Q,~S,~T,~U,~W$~ stand for different structures for 
the $~F^2\-$terms:  
$$\li{& N_{\m\n}\equiv \big(\delsl\varphi\big)
     \g^{\[4\]\[3\]} \e\,  
        F_{\[4\]} F_{\[3\] \[ \m} \partial_{\n\]}\varphi ~~, 
     ~~~~ P_{\m\n} \equiv \big(\delsl\varphi\big)
     \g^\r \e\, F\du\r{\[3\]} F_{\[3\]\[\m} 
    \partial_{\n\]}\varphi 
      ~~, \cr 
& Q_{\m\n} \equiv \big(\delsl\varphi\big) 
     \g^{\[2\]\r} \e\, F\du{\[2\]}{\[2\]'} 
      F_{\[2\]'\r\[\m} \partial_{\n\]}\varphi~~, ~~~~ 
      S_{\m\n} \equiv \big(\delsl\varphi\big)
      \g_{\[\m|} \e\, F_{\[4\]}{}^2  
     \partial_{|\n\]}\varphi ~~, \cr 
& T_{\m\n} \equiv \big(\delsl\varphi\big) 
      \g^{\[3\]\[2\]} \e\,  
      F\du{\[3\]} \r F_{\r\[2\]\[\m} \partial_{\n\]}\varphi~~, ~~~~ 
      U_{\m\n} \equiv \big(\delsl\varphi\big)      
      \g\ud{\[2\]\[2\]'}{\[\m|} \e\, 
     F_{\[2\]\[2\]''} F\du{\[2\]'} {\[2\]''} \partial_{|\n\]}\varphi~~, \cr 
& W_{\m\n} \equiv \big(\delsl\varphi\big) 
    \g\ud{\[4\]\[4\]'}{\[\m|} \e\, 
      F_{\[4\]} F_{\[4\]'} \partial_{|\n\]}\varphi~~. 
&(6.2.17) \cr } $$
The $~\a\-$dependent coefficients in $~N\-$~ and $~W\-$terms 
are the result of using the $~F\-$field equation (6.2.9), while the 
$~\b$~ and $~\d\-$ dependent coefficients in the $~S\-$~ 
and $~P\-$terms are from the dreizehnbein field equation (6.2.8).   
The vanishing of these $~N$~ and $~W\-$terms fixes the coefficient
$$ \li{&\a = + \frac1{2304} ~~, 
&(6.2.18) \cr }$$
after the use of (6.2.7).  The vanishing of the $~S$~ and $~P\-$terms 
fix the coefficients  
$$\li{&\b = - \frac1{36} ~~, ~~~~ \d = + \frac1 3 
&(6.2.19) \cr } $$ 
after the use of (6.2.7).  
Finally in (6.2.16), the ~$Q,~T$~ and $~U\-$terms vanish under (6.2.7).  
At this stage, all the coefficients $~\a,~\b$~ and
$~\d$~ in the field equations are fixed, consistently with the values for 
$~a_1, ~a_2$~ and $~a_3$~ in (6.2.7).  

\bigskip\bigskip\bigskip

\leftline{\bf 6.3.~~Superspace for $~N=1$~ Supergravity in $~D=11+2$}          
\bigskip

In this subsection, we briefly provide helpful information for 
the superspace formulation of our $~D=13,\,N=1$~ supergravity. 
The structures for constraints
in superspace are easily read from the component 
transformation rules (6.2.1) \ggrs.  However, there is one caveat about 
the extra transformation terms which are implicit in component, but manifest
themselves in superspace.  There is also some caution needed for 
$~d=1$~ Bianchi identities to be mentioned shortly.  

To clarify these points, we give first our results for superspace constraints at
$~0\le d \le 1/2$, which will be of relevance also for supermembrane
formulation in the next section:  
$$ \li{&T\du{\a\b}c = (\g^{c d})\low{\a\b} \nabla_d\varphi  
     + (\Pupdown)\low{\a\b} \nabla^c\varphi  ~~,  
&(6.3.1\rma) \cr  
&F_{\a\b c d} = -\frac 12 (\g_{c d}{}^e)\low{\a\b} \nabla_e\varphi 
       - \frac12\big( \Pdown \g_{\[c|} \big)_{(\a\b)} \nabla_{|d\]}\varphi~~, 
&(6.3.1\rmb) \cr 
& \big(\nabla_a\varphi\big) \big(\nabla^a\varphi\big) = 0 ~, 
     ~~~\big(\nabla_a\Tilde\varphi\big) \big(\nabla^a\Tilde\varphi\big) =0
    ~, ~~~\big(\nabla_a\varphi\big) \big(\nabla^a\Tilde\varphi\big) 
     = 1~, ~~~\nabla_\a\varphi = \nabla_\a\Tilde\varphi = 0 {~.~~~~~ ~~~~~}   
&(6.3.1\rmc) \cr } $$   
As usual all other supertorsion components at 
$~d\le 1/2$, such as $~T\du{\a\b}\g$~ are all zero.  The 
range of spinorial indices in 13D is $~{\scst \a,~\b,~\cdots ~=~
1,~2,~\cdots,~64}$.   
The first terms in (6.3.1a) and (6.3.1b) are obtained from the component
transformation rules (6.2.1a) and (6.2.1c), while the second terms can be
understood as maintaining the conditions $~T\du{\a\b} c \nabla_c\Tilde\varphi
\equiv 0$~ and $~F_{\a\b c d} \nabla^d\Tilde\varphi\equiv 0$, 
in addition to the trivial ones: $T\du{\a\b} c \nabla_c\varphi\equiv 0,~
F_{\a\b c d} \nabla^d\varphi\equiv 0$. 
This can be understood as in general terms:  Let any vector $~V_a$~ be
modified to be $~\Tilde V_a$, that satisfies the condition $~\Tilde V_a
\nabla^a\Tilde\varphi\equiv 0$~ by the simple shift:   
$$ \li{ & \Tilde V_a \equiv V_a 
     - \big(\nabla_a\varphi\big)\big(\nabla^b\Tilde\varphi\big) V_b~~.     
&(6.3.2) \cr } $$     
For example, we easily see that the second term
in (6.3.1b) is nothing else than this modification of the first term for the 
two indices $~{\scst c,~d}$~ directly obtained from the component 
transformation rule 
(6.2.1c).  The condition $~F_{\a\b c d}\nabla^d\varphi\equiv 0$~ 
is also required by
superspace Bianchi identities at $~d=1/2$.  
Technically, it is sometimes useful to use the purely 11D indices $~{\scst
i,~j,~\cdots~=~(0),~(1),~\cdots,~(9),~(10)}$~ and the extra ones $~{\scst
\pm}$~ for the bosonic indices in
$~T\du{\a\b} c,~F_{\a\b c d}$, like 
$~T\du{\a\b}i,~F_{\a\b i j}$~ in order to make superspace
computation easier. 

For the rest of this subsection, instead of going through all 
the details of Bianchi identities of $~0\le d \le 1/2$, we give important 
Fierz identities needed for these Bianchi identities.  
The first important Fierz identity is associated with the $~d=0$~ Bianchi
identity  of $~(\a\b\g d e)\-$type: 
$$ \li{& \big( \g^{a b} \big)_{(\a\b|} 
      \big( \g\du{a d} c \big)_{|\g\d)} \big(\nabla_b\varphi \big) 
      \big( \nabla_c\varphi\big)  
       - 2 \big(\g^{a b} \big)_{(\a\b|} 
      \big(\Pdown\g_a \big)_{|\g\d)} \big(\nabla_b\varphi\big)
     \big(\nabla_d\varphi\big) = 0 ~~.   
&(6.3.3) \cr } $$  
There are two principal methods to confirm this identity:  The first method 
is the direct one, namely using more basic Fierz identities in 13D in Appendix
B.  The second method uses the dimensional reduction 
from 13D into 11D.  For example, if (6.3.3) is easily seen to hold, if we 
assign explicit index ranges for all the bosonic indices
in (6.3.3).  To be more specific, we see first that  
the $~{\scst a}\-$index can take only purely 11D values due to the
factor of $~\nablasl\varphi$, while
there are three options for the index $~{\scst d~=~i,~+,~-}$.  It can be easily
seen that for each of these cases, (6.3.3) holds, in particular, the case of
$~{\scst d~=~i}$~ corresponds to the familiar 11D Fierz identity 
\ref\cf{E.~Cremmer and S.~Ferrara, \pl{91}{80}{61}; L.~Brink and P.~Howe,
\pl{91}{80}{384}.}: 
$$ \li{& \big( \g^i \big)_{(\a\b|} \big( \g_{i j}
\big)_{|\g\d)} \equiv 0 ~~,  
&(6.3.4) \cr } $$ 
where all the spinorial indices are 11D ones.  Note that even though 
we used dimensional reduction, there is no extra component overlooked in this
method, as long as we scan all the possibilities of the range of indices.    

As careful readers may have already noticed, we have mentioned neither the 
$~d=1$~ Bianchi identities, nor the constraints at $~d=1$.  This is due to
a problem with a Bianchi identity of $~(\a\b c d e)\-$type at $~d=1$~ yet to be
satisfied, associated with the extra symmetry (6.2.4).  This is caused
by the term $~\big(\g^{e f} \big)_{\a\b} F_{c d e +} \nabla_f\varphi$~ 
which does not have
any counter-terms in this Bianchi identity.  To put it differently, due to the 
extra symmetry (6.2.4), there can be additional term proportional to the
null-vector $~\nabla_a\varphi$~ 
in the anti-commutator $~\{\nabla_\a, \nabla_\b\} 
A_{a b c}$, which has no direct geometrical interpretation in
superspace.  This problem seems peculiar to this 13D system, with no 
corresponding one in 12D supergravity.  However, we  
also emphasize that this sort of problems for superspace formulation for
supergravity with extra symmetries is not a new phenomenon at
all, because we know similar supergravity/supersymmetry theories, such as the
$~N=4$~ Chern-Simons theory in 3D \ref\cs{H.~Nishino and S.J.~Gates, Jr.,
\ijmp{8}{93}{3371}.}, which is possible only in component formulation, with
some obstructions for superspace formulations caused by extra
symmetries.\footnotew{This extra symmetry arose as the result of duality 
transformation performed to get this multiplet \cs, and there seems to be a 
general close relationship between extra symmetries and duality
transformations.}  As for the present 13D supergravity, notwithstanding 
this problem to be solved at
$~d=1$~ level by some possible modifications of Bianchi identities, {\it e.g.,}
by Chern-Simons modifications, we believe that our superspace constraints at
$~0\le d \le 1/2$~ are valid for supermembrane couplings with fermionic
invariances based only on these lower-dimensional superspace constraints, 
that we will deal with next.

\bigskip\bigskip\bigskip

\centerline{\bf 7.~~Supermembrane on Background of $~D=13,\,N=1$~
Supergravity}            
\bigskip

Our next step is to put some extended objects as a probe for the consistency 
of our 13D supergravity.  The most natural extended object is the supermembrane 
\bst\ coupled to 11D supergravity \cjs, because our 13D supergravity is
a higher-dimensional generalization of the former.  The existence of the
third-rank tensor $~A_{\[3\]}$~ also suggests the natural super p-brane
\pbrane\ to be supermembrane \bst.  As usual, our next task is
to confirm the fermionic symmetries in the supermembrane \bst\ in our
13D supergravity background.  

Our total action for the supermembrane  
is similar to that for the original supermembrane \bst: 
$$\li{&S = S_\s + S_A ~~,  
&(7.1\rma) \cr 
& S_\s \equiv \int d^3 \s \, \left( - \frac12 {\sqrt{-g}} g^{i j} 
   \eta_{a b} \Pi\du i a \Pi\du j b + \frac12{\sqrt{-g}} \right)~~,  
&(7.1\rmb) \cr  
& S_A \equiv \int d^3 \s \, \left( +\frac 1 3 \e^{i j k} 
    \Pi\du i A \Pi\du j B \Pi\du k C A_{C B A} \right)~~.
&(7.1\rmc) \cr } $$ 
As in other sections of p-branes in this paper, we have switched to superspace
notation.  The indices $~{\scst i,~j,~\cdots~=~0,~1,~2}$~ are for curved
coordinates in $~D=2+1$~ world-volume, and $~\Pi\du i A\equiv \big(\partial_i
Z^M \big) E\du M A$, {\it etc.}, as in the usual supermembrane formulation
\bst.   

For the fermionic invariance of the action, we need 
superspace constraints (6.3.1) at $~0\le d \le 1/2$.   
We found the second terms in (6.3.1a) and (6.3.1b) have no  
contributions in our action invariance under the constraint (7.5) below.  

As in the cases in 12D \nishone\nishtwo, we have two sorts of fermionic 
symmetries
$$\li{ &\d E^\a = \left(I + \G \right)\ud\a\b \k^\b 
   +\big( \Pup\big)^{\a\b}
    \eta\low\b ~~,  
&(7.2\rma) \cr 
&\d E^a = 0 ~~,  
&(7.2\rmb) \cr } $$
with the 64 component fermionic parameters $~\k$~ and $~\eta$.  The  
$~\G~$ is defined similarly to \bst\ by 
$$ \li{& \G\equiv \fracm1{6\sqrt{-g}} \e^{i j k} 
       \Pi\du i a \Pi\du j b\Pi\du k c \left( \g_{a b c} \right) ~~,    
& (7.3) \cr} $$ 
satisfying the relations similar to (5.5):
$$ \li{& \G^2 = I~~,  
&(7.4\rma) \cr  
&  \e_i{}^{j k} \Pi\du j a \Pi\du k b \g_{a b} 
   \G = + 2 {\sqrt {-g}} \Pi\du i a \g_a ~~,    
&(7.4\rmb) \cr } $$
under the algebraic $~g\low{i j}\-$field equation $~g\low{i j} 
= \eta_{a b}\Pi\du i a \Pi\du j b$.  

The fermionic invariance of our action goes in the same way as
in the $~D=12,\, N=2$~ supergravity.  In particular, 
the important ingredient is the constraint condition on the
pull-back,  as in  $~D=12,\, N=2$~ supergravity \nishtwo: 
$$ \Pi\du i a \nabla_a\varphi = 0 ~~. 
\eqno(7.5) $$ 
In fact, the variation of our action under the fermionic $~\eta\-$symmetry
yields the terms 
$$ \li{ &\d_\eta S =  {\sqrt{-g}} \Pi_{i a} 
    \big(\Bar\eta \Pdown \g^{b a}\big)_\b \Pi^{i\b} \nabla_b \varphi  
     + \frac 1 4 \e^{i j k} \big( \Bar\eta \Pdown \g_{d c b} \big)_\b  
     \Pi\du i\b \Pi\du j b \Pi\du k c \nabla^d\varphi ~~.  
&(7.6) \cr } $$  
Interestingly, the second terms both in (6.3.1a) and (6.3.1b) do
not contribute under the constraint (7.5).    
Each of the terms in (7.6) further vanishes under (7.5),
due to the identity: $\Pdown \nablasl\varphi\equiv 0$.     
Similarly to the original supermembrane action \bst, the $~\k\-$symmetry 
is also easily confirmed by the use of (7.3) as well as our constraint (7.5).  
This concludes our confirmation of fermionic invariance of our total action.  

Similarly to the Green-Schwarz superstring \nishone\ on $~D=12,\,
N=1$~ supergravity or super $~(2+2)\-$brane \nishtwo\ on $~D=12,\, N=2$~
backgrounds, the $~\eta\-$invariance and $~\k\-$symmetry reduce the
total degrees of fermionic freedom into 16 which is a quarter of the original
value of 64, agreeing with the conventional supermembrane formulation \bst:  
$~64\rightarrow 32 \rightarrow 16$.

\bigskip\bigskip
\bigskip\bigskip


\centerline{\bf 8.~~Concluding Remarks} 
\bigskip

In this paper we have given rather detailed constructions of
$~N=1$~ and $~N=2$~ supergravity theories in 12D and $~N=1$~ supergravity in
13D, based on our recent technique of using scalar (super)fields that make the
system manifestly $~SO(10,2)$~ or $~SO(11,2)$~ 
Lorentz covariant, up to modified Lorentz
generators.\footnotew{We have shown even  our modified Lorentz generators
themselves can be also made `more' covariant, in terms of 
the gradients of superinvariant scalar fields, using the definition (2.2.5).}  
We have also established the fermionic invariances of superstring
on $~D=12,\,N=1$, or super $~(2+2)\-$brane 
on $~D=12,\,N=2$, and supermembrane on 
$~D=13,\,N=1$~ supergravity backgrounds, as confirmation of the consistency of
our theories.  We have seen new features of these supergravity theories in 12D
as well as parallel structure to those in 10D or 11D.  For example, we have found
how  the self-duality condition (4.2.27) in 10D is promoted to the
anti-self-duality condition (4.2.22) in 12D in an elaborate fashion.  We have
noticed how the cancellation structures among terms for supersymmetry 
in 12D are parallel to the 10D case, based on similarities  
in gamma-matrix identities in these dimensions, as in Appendix A.  

The important ingredient of our formulations is the role of the scalar 
(super)fields making our systems $~SO(10,2)$~ or $~SO(11,2)$~ Lorentz
covariant, respectively for our 12D or 13D supergravities, 
up to modified Lorentz generators.  This technique was first introduced in \lag,
where the scalar (super)fields are intact under supersymmetry.  This feature is
also consistent with the closure of local supersymmetry.  All the (super)field
equations are now made formally Lorentz covariant, but the systems realize
the spontaneous $~SO(10,2)$~ or $~SO(11,2)$~ Lorentz symmetry breakings down to
$~SO(9,1)$~ or $~SO(10,1)$, when particular solutions (2.2.2) 
are used for these scalar 
(super)fields with null-vectors.  To our knowledge, there has been no other 
example of this kind in other supersymmetric theories in 
lower-dimensions.  The advantage
of these superinvariant scalar (super)fields is more elucidated, when dealing
with globally supersymmetric Yang-Mills theories in $~D\ge 12$~ \symall\lag,
because we need no modified Lorentz generators, and these theories are entirely
Lorentz covariant.  

In addition to the gradients of scalar fields replacing the null-vectors, we 
have developed other practical techniques of formulating higher-dimensional
supergravities.  Using dimensional reduction for the confirmation of Fierz
identities for 13D supergravity is one of them.  This method
utilizes the parallel structure between 13D and 11D supergravities, saving
a lot of time for confirming complicated Fierz identities needed in practical
computations in 13D.  To our knowledge, this method 
is introduced into supergravity for the first time in this paper, and as such,
this technique has potential applications to more higher-dimensional
supergravities in $~D\ge 14$.  Armed with the lists of $~\g\-$matrix
relations also in Appendices A and B, we have now great  capability of controlling
Bianchi identities in all of these  higher-dimensions.  

In our supergravity theories in 12D, since the Lorentz symmetry
$~SO(10,2)$~ or $~SO(11,2)$~ is only formally 
recovered due to the modified Lorentz
generators, some readers may wonder if this is just a reformulation of
null-vectors violating the manifest Lorentz covariance.  Even if we admit that
the usage of scalar fields may be just a rewriting of null-vectors in
supergravity theories, in the sense that modified Lorentz generators not
totally Lorentz covariant, we stress important features of our supergravity
formulations in these dimensions.  There has been also some skepticism about the
`uniqueness' of these supersymmetry/supergravity theories, ever since the first
construction of supersymmetric theory in 12D in \ns.  According to such
a claim, the lack of Lorentz invariance makes these supergravity theories
arbitrary but not unique, {\it unlike} other conventional supergravity theories
in $~D\le 11$.  Even though this argument sounds convincing, it overlooks
important features in supergravity theories.  Such a claim is valid, only when
we are dealing with {\it non}-supersymmetric  theories without Lorentz
invariance, because we can always put any null-vector to get rid of extra 
component in any term in a field equation at our will.  Therefore the
construction of {\it non}-supersymmetric theories is always ambiguous,  when
Lorentz invariance is not manifest.  In supersymmetric theories, however,  this
is no longer the case due to the restriction by supersymmetry.  As
a matter of fact, the construction of a supergravity/supersymmetry theory
satisfying {\it all} of the following conditions is extremely difficult:  

\vfill\eject

\Item{(i)} There exist extra non-vanishing components for
physical\footnotew{`Physical' field here means a field that has
physical components in 10D.} fields.  
\Item{(ii)} These extra components have non-trivial 
dependence on coordinates.  
\Item{(iii)} Supersymmetry closes at least on-shell.    
\Item{(iv)} Supersymmetric p-Brane Formulations exist 
on such Supergravity Backgrounds. 
\vskip 0.12in 

\noindent For example, 
if we build a supergravity theory in which {\it all} of the extra
components  are deleted by constraints using null-vectors, then the system
collapses into  an ordinary supergravity theory in $~D\le 11$,
leaving nothing new.  The non-triviality of our formulation can be also seen
from the fact  that the coefficients in our supersymmetry transformation rules
and field equations are so tightly fixed that we can not shift them even by
small amounts, maintaining supersymmetry.  One can try to build an arbitrarily
new supersymmetric theories in these higher dimensions, and realize how our
systems are selected, when supersymmetry is in the game.  We emphasize that
these supersymmetric theories are strictly fixed by their proper
uniqueness, despite of the lack of total Lorentz invariance.  We also mention
that the loss of Lorentz invariance is not limited to our peculiar
system of higher-dimensional supersymmetries.  For example, the loss of 
Lorentz invariance seems 
also inevitable in the $~SL(2,\ZZ)$~ duality symmetric formulation in 
ref.~\ref\ss{J.H.~Schwarz and A.~Sen, \np{411}{94}{35}.}.   

Another important point not to be overlooked is the existence of 
super p-brane \pbrane\ actions consistently coupled to our 
$~D=12,\, N=1$~ or ~$D=12,\,N=2$~ and $~D=13,\,N=1$~ supergravity backgrounds. 
If we did not have such `probes' on our backgrounds, then one could still say
that these higher-dimensional supergravity theories were  just ordinary
supergravity in $~D\le 11$~ `in disguise'.  However, due to the  non-vanishing
components among string or membrane variables carrying the `extra' components,
we can see much more non-triviality in our total system formulated with these
extended objects.  For example, we saw that the constraints (3.12)  for the
extra string variables are required only for the components  $~\Pi\du - a
\nabla_a\varphi$~ and $~\Pi\du - a \nabla_a\Tilde\varphi$, but {\it not} for 
$~\Pi\du + a \nabla_a\varphi$~ and $~\Pi\du + a \nabla_a\Tilde\varphi$.  In
other words, the  string variables $~X^\pm(\s)$~ in the extra dimensions can
still have  non-trivial dependence on the world-sheet coordinates $~\s^+$.  In
this sense,  the Green-Schwarz superstring \gsw\ coupled to our $~N=1$~
supergravity in 12D \nishone\ has  much more content 
than just a rewriting of 10D Green-Schwarz superstring.  As a matter of fact, we
have also found an important fact that the requirement of fermionic symmetries
on the Green-Schwarz superstring world-sheet for $~D=12,\,N=1$~ supergravity, or 
on the super $~(2+2)\-$brane world-supervolume 
for $~D=12,\,N=2$~ supergravity leads to the null-vector
conditions such as  (2.3.4$\ell$) or (4.2.4e), respectively.  In other words, our
null-vector conditions are by no means artificially put by hand, but
required by the fermionic symmetries of these extended objects as probes put in
these backgrounds in 12D.  The necessity of null-vectors is also understood 
as solutions for BPS conditions for supersymmetry algebra in higher-dimensions
\stheory\barsfourteen.  We have also seen how these fermionic
symmetries reduce the degrees of freedom of these extended objects in 12D down 
to the conventional 8+8 or 16+16 physical degrees of freedom in the light-cone
coordinates for superstrings in 10D \gsw.   

Our result for 13D supergravity will motivate also other interesting
applications and directions to be explored, such as generalizing this result 
to more duality-symmetric way, like those in \ref\mfivebrane{
M.~Cederwall, B.E.W.~Nilsson, P.~Sundell, J.~High Energy Phys.~{\bf 04} 
(1998) 007, hep-th/9712059; I.~Bandos, 
N.~Berkovits and D.~Sorokin, \np{522}{98}{214}.}\ref\dual{H.~Nishino, 
{\it `Alternative Formulation for Duality-Symmetric Eleven-Dimensional 
Supergravity Coupled to Super M-5-Brane'}, UMDEPP 98--078, \hepth{9802009}.}, 
looking for lagrangian formulation like that in ref.~\lag, or considering
topological significance of the  $~F\-$field equation, or possible 
duality connections with other higher-dimensional theories, {\it etc.}   

As we have seen, it is just the beginning that higher-dimensional
supergravity revealed various unexpected
features, such as the modified Lorentz generators that had never 
been presented before in supersymmetric theories, or the power of
superinvariant scalars making the system more covariant.  
There has been some indication that the success of our supergravity
theories  based on null-vectors \nishone\nishtwo\ signals nothing but more
fundamental  theories where the null-vectors are replaced by more generalized
momenta \stheory\twotimes\ in multi-local field theories.  We are sure that the
details given in this paper  will provide us with the first step toward such
directions, leading to  more fundamental theories, with bi-local or
multi-local fields.   From this viewpoint, even though we performed rather 
detailed computation in this paper, we believe that such technicalities   
will be of practical importance, when we generalize our 
theories to more fundamental bi-local or multi-local theories, and we 
will realize that the structure of higher-dimensional supergravity is much deeper
than we had initially expected.  

\bigskip\bigskip

We are grateful to I.~Bars, S.J.~Gates, Jr., C.~Pope, J.H.~Schwarz, E.~Sezgin 
and C.~Vafa for important discussions at various stages of the works presented
in this paper.  

\bigskip\bigskip\bigskip\bigskip


\vfill\eject

\centerline{\bf Appendix A:~~Useful Relationships in $~D=10+2$}
\bigskip

In this Appendix, we list up important relationships, which are useful 
in practical manipulations.  We start with the Fierz identities 
in our $~D=10+2$~ for Weyl indices $~{\scst \a,~\b,~\g,~\d}$:
$$ \li{
&\d\du{(\a}\b \d\du{\g)}\d = \fracm1{32}
\big(\g^{a b} \big)_{\a\g} \big( \g_{a b} \big)^{\b\d} 
+ \fracm1{32(6!)} \big( \g^{\[6\]} \big)_{\a\g} \big( \g\low{\[6\]} \big)
^{\b\d} ~~, 
&(A.1) \cr 
&\d\du{\[ \a}\b \d\du{\g\]}\d = + \fracm1{16} C_{\a\g} C^{\b\d} 
     + \fracm1{16(4!)} \big( \g^{\[4\]} \big)_{\a\g} 
        \big( \g\low{\[4\]} \big)^{\b\d}~~.    
&(A.2) \cr } $$
Based on these, we can further derive the following useful relations:
$$ \li{ & \big( \g^a \big)_{(\a|\Dot\b} 
       \big(\g_a \big)\du{|\g)}{\Dot\d} = \fracm14 
      \big( \g^{b c} \big)_{\a\g}  
       \big( \g_{b c} \big)\du{\Dot\b}{\Dot\d} ~~, 
&(A.3) \cr 
&\big( \g^{a b} \big)_{(\a\b|} \big( \g_a \big) \du{|\g)}{\Dot\d}
       = + \frac1{12} \big( \g^{c d}  \big)_{(\a\b|} 
       \big( \g_{c d}\g^b \big)\du{|\g)}{\Dot\d} 
       = \frac1 8\big( \g^{c d} \big)_{(\a\b|} \big( \g^b \g_{c d}\big)
       \du{|\g)}{\Dot\d} ~~,  
&(A.4)  \cr 
& \big( \g^{a b} \big)_{(\a\b|} \big( \g_{a c}  \big)\du{|\g)}\d 
   = \frac1{10}\big( \g^{d e}\big)_{(\a\b|}\big(\g\dud{d e}b c \big)
     \du{|\g)}\d  
     + \big( \g\ud b c \big)_{(\a\b|}\d\du{|\g)}\d \cr 
& ~~~~~ ~~~~~ ~~~~~ ~~~~~ ~~~~~  + \frac1{10}\d\du c b \big( \g^{d e}
     \big)_{(\a\b|} \big( \g_{d e}\big)\du{|\g)}\d  
     -\frac1 5 \big( \g\ud d c \big)_{(\a\b|} 
      \big( \g\du d b \big)\du{|\g)}\d ~~, 
&(A.5) \cr 
& \big( \g^{a b} \big)_{\Dot\a\Dot\b} \big( \g_a \big)_{\g\Dot\d} 
     = - \big( \g_a \big)_{\g(\Dot\a|} 
      \big( \g\du a b \big)_{|\Dot\b)\Dot\d} 
      - \frac1{20} \big( \g^{c d b} \big)_{\g(\Dot\a|} 
      \big( \g_{c d}\big)_{|\Dot\b\Dot\d)} ~~, 
&(A.6) \cr 
& \big( \g_{a b} \big)_{(\a\b|} \big( \g^{a b c d e f} \big)\du{|\g)}\d 
      = - \big( \g^{a b c d e f} \big)_{(\a\b|} \big( \g_{a b} \big)
       \du{|\g)}\d
      - 2 \big( \g^{\[c d} \big)_{(\a\b} \big( \g^{e f\]} 
       \big)\du{|\g)}\d ~~,   
&(A.7) \cr 
& \big( \g_{a b} \big)_{(\a\b|} \big( \g^{a b c d e f} \big)_{|\g\d)} 
     = - \big( \g^{\[ c d } \big)_{(\a\b|}  \big(\g^{e f \]} 
        \big)_{|\g\d)} ~~, 
&(A.8) \cr 
& \big( \g^{\[ 5 \] b} \big)_{(\a\b|} \big( \g\low{\[5 \]} \big)
     \du{|\g)}{\Dot\d} 
     = - 720 \big( \g^{a b} \big)_{(\a\b|} \big( \g_a \big)
     \du{|\g)}{\Dot\d}~~, 
&(A.9) \cr
& \big( \g^{\[5\]} \big)\du{(\a|}{\Dot\b} \big( \g\low{\[ 5 \]} \big)
     \du{|\g)}{\Dot\d}
     = - 180 \big( \g^{a b} \big)_{\a\g} 
         \big( \g_{a b} \big)^{\Dot\b\Dot\d} ~~, 
&(A.10) \cr 
& \big( \g^{\[ 5\] (a|} \big)_{(\a\b|}  \big( \g\Du{\[ 5 \]}{|b)} \big)
     \du{|\g)}\d = - 120 \eta^{a b} \big( \g^{c d} \big)_{(\a\b|} 
          \big( \g_{c d} \big) \du{|\g)}\d ~~, 
&(A.11) \cr 
& \big(\g^{\[ 4 \] a b} \big)_{(\a\b|} \big(\g\low{\[ 4 \]} \big)\du{|\g)}\d 
    = + 12 \big( \g^{c d} \big)_{(\a\b|}  \big( \g\du{c d}{a b} \big)\du
     {|\g)}\d  + 360 \big( \g^{a b} \big)_{(\a\b} \d\du{\g)}\d~~, 
&(A.12) \cr  
& \big(\g^{c d} \big)_{(\a\b|} \big(\g\du{c d}{a b} \big)\du{|\g)}\d = 
     + 4 \big(\g^{\[a|c} \big)_{(\a\b|} \big(\g\ud{|b\]} c \big)\du{|\g)}\d  
     -10 \big(\g^{a b} \big)_{(\a\b} \d\du{\g)} \d~~,    
&(A.13) \cr  
& \big( \g^a \big)\du\a{\Dot\b} \big( \g_a \big)\du\g{\Dot\d} = 
    + \frac 3 8 C_{\a\g} C^{\Dot\b\Dot\d} 
     +\frac18 \big( \g^{\[2\]} \big)_{\a\g} 
     \big( \g\low{\[2\]} \big)^{\Dot\b\Dot\d} 
     + \frac1{192} \big( \g^{\[4\]}\big)_{\a\g} 
     \big( \g\low{\[4\]} \big)^{\Dot\b\Dot\d} ~~, 
&(A.14) \cr     
& \big( \g^{\[3\]} \big)\du\a{\Dot\b} \big( \g\low{\[3\]} \big)\du\g{\Dot\d} = 
    + \frac{165} 4 C_{\a\g} C^{\Dot\b\Dot\d} 
     +\frac{15}4 \big( \g^{\[2\]} \big)_{\a\g} 
     \big( \g\low{\[2\]} \big)^{\Dot\b\Dot\d} 
     - \frac3{32} \big( \g^{\[4\]}\big)_{\a\g} 
     \big( \g\low{\[4\]} \big)^{\Dot\b\Dot\d} {~~,~~~~~ ~~~~~}  
&(A.15) \cr     
& \big( \g^{\[5\]} \big)\du\a{\Dot\b} \big( \g\low{\[5\]} \big)\du\g{\Dot\d} = 
    +2970 C_{\a\g} C^{\Dot\b\Dot\d} 
     -90 \big( \g^{\[2\]} \big)_{\a\g} 
     \big( \g\low{\[2\]} \big)^{\Dot\b\Dot\d} 
     + \frac 5 4 \big( \g^{\[4\]}\big)_{\a\g} 
     \big( \g\low{\[4\]} \big)^{\Dot\b\Dot\d} {~~.~~~~~ ~~~~~}
&(A.16) \cr  } $$ 
Here since these relations are for superspace, all the (anti)symmetrizations 
are {\it not} normalized, {\it e.g.,} $~A_{\[a b\]} \equiv 
A_{a b} - A_{b a}$.\footnotew{Notwithstanding this rule in superspace, the symbol
$~{\scst \[ n \]}$~ is for {\it normalized} antisymmetrization, {\it e.g.,}
$~A_{\[3\]} B^{\[3\]} \equiv (1/6) A_{\[ a b c\]} B^{\[a b c\]}$~ as in
component notation, as has been already mentioned.}   Needless to say,
alternative identities obtained by exchanging all the dotted indices and
undotted ones also hold.   Even though this is not the exhaustive list of
relations used in our calculation, one will still find it useful for
practical computations.  

There are also other useful relations with null-vectors.  
Some typical examples are
$$ \li{&\big( \g^c \nsl \big)_{\a\b} \big( \g_c \nsl \big)\du\g\d
     = - \frac12 \big( \g^{c a} \big)_{\a\g} \big( \g\du c b \big)\du\b\d 
      n_a n_b  
     - \frac1{24} \big( \g^{\[3\] a} \big)_{\a\g}  \big( \g\du{\[3\]} b
     \big)\du\b\d n_a n_b ~~, 
&(A.17)  \cr  
&\big( \g^{\[3\]} \nsl \big)_{\a\b} \big( \g_{\[3\]} \nsl \big)\du\g\d
     = - 15 \big( \g^{c a} \big)_{\a\g} \big( \g\du c b \big)\du\b\d 
      n_a n_b  
     + \frac3 4 \big( \g^{\[3\] a} \big)_{\a\g}  \big( \g\du{\[3\]} b
     \big)\du\b\d n_a n_b {~~,~~~~~ ~~~~~} 
&(A.18) \cr  
&\big( \g^{\[5\]}\nsl \big)_{\a\b} \big( \g_{\[5\]} \nsl \big)\du\g\d
     = +360 \big( \g^{c a} \big)_{\a\g} \big( \g\du c b \big)\du\b\d 
      n_a n_b  
     - 10 \big( \g^{\[3\] a} \big)_{\a\g}  \big( \g\du{\[3\]} b
     \big)\du\b\d n_a n_b {~~. ~~~~~ ~~~~~}
&(A.19) \cr }  $$ 
Notice that these relations
hold only under the null-vector condition $~n_a n^a = 0$.        

Another crucial relation is between the $~\e\-$tensor and the $~\g\-$matrices:
$$ \g\low{\[n\]} = \fracm{(-1)^{n(n-1)/2}} {(12-n)!}
\e\Du{\[n\]} {\[ 12-n\]} 
     \g \low{\[12-n\]} \g\low{13} ~~~~(0\le n \le 12) ~~ .
\eqno(A.20) $$
In 12D we have convenient relations associated with duality of 6-th rank 
antisymmetric tensors.  For example, we can easily show that the following 
combination is identically zero:
$$ S^{\[6 \]} A_{\[6 \]} \equiv 0 ~~, 
\eqno(A.21) $$
where $~S_{\[6 \]} \equiv +(1/6!) \e\low{\[6\]}{}^{\[6\]'} S_{\[6\]'}$~ and 
$~A_{\[6\]} \equiv - (1/6!) \e\low{\[6\]}{}^{\[6\]'} A_{\[6\]'}$~ 
are respectively self-dual and anti-self-dual tensors.  
This is in a good contrast with 
the 6D case {\it e.g.,} in \ref\nssix{H.~Nishino and E.~Sezgin, 
\pl{144}{84}{187}; \np{278}{86}{353}; \np{505}{97}{497}.},  
where we had $~S^{\[3\]} S_{\[3\]}\equiv A^{\[3\]} A_{\[3\]} \equiv 0$.
Even though there is prevailing tendency nowadays regarding these $~\g\-$matrix 
manipulations in supergravity as `out of date' or `old-fashioned' methods that 
we should not be bothered, we re-emphasize here their crucial importance for
building supergravity theories {\it not} to be bypassed, with no other
alternative `quick' ways, even after 20 years
since the first discovery of supergravity \pvn.  For example, the construction 
of 12D supersymmetric Yang-Mills theory \ns, or our modified 
Lorentz generators \nishone\ would have never been achieved 
without the crucial identities in this Appendix.     

We give also basic relationships related to $~\g\-$matrices with their 
complex or hermitian conjugates.  Using the same 
notation in \kt, we list them up as
$$ \li{ &\Bar\psi = \psi^\dagger A ~~~(\hbox{for Dirac conjugate}) ~~, \cr 
& \psi = C\Bar\psi{\,}^T ~~~(\hbox{for Majorana-Weyl condition}) ~~, \cr
& \g_\m^\dagger = - A\g_\m A^{-1}~~,  
     ~~~~ A\equiv \g\low{(1)}\g\low{(12)}~~, 
          ~~~~ A^\dagger = - A~~, \cr 
& A^T = - C A C^{-1} ~~, 
					~~~~ \g^\m = - B^{-1} \g^{\m\,*} B ~~ ~~(\eta = -1
     ~~\hbox{for Majorana spinors}) ~~, \cr 
& B \equiv (A^T)^{-1} C^T = - C A~~, ~~~~C^T = - C~~~~
    (\e =+1~, ~~~\eta= - 1)~~,\cr   
& \g_\m^T = + C\g_\m C^{-1}~~, ~~~~
     C^\dagger C = + I~~, ~~~~ \psi^* = B\psi~~.  
&(A.22) \cr } $$ 
By the aid of these expressions, we can easily confirm equations in (4.1.8).  

Before concluding this Appendix, we mention other useful $~\g\-$matrix relations
which are frequently used both in superspace and component computations in 12D. 
We give below a list of such relations for readers' convenience, 
even though this list is by no means exhaustive, because other 
relations may be also easily obtained by using these identities:   
$$\li{& \g_\m \g^{\[n\]} \g^\m = (-1)^n (12 - 2n) \g^{\[n\]} ~~, 
    ~~~~ \g_\m \g^{\[6\]} \g^\m  = 0 ~~,  
&(A.23) \cr 
& \g^{\m\n}\g_{\r\s} \g_{\m\n} = - 52\g_{\r\s}~~, ~~~~ 
     \g^{\m\n} \g_\r \g_{\m\n} = - 88 \g_\r ~~, 
&(A.24) \cr 
& \g^\n\g_\m \g_{\n\r} = - 9 \g_{\m\r} - 11 g_{\m\r} ~~, ~~~~
      \g_{\n\r} \g_\m\g^\n = - 9 \g_{\m\r} + 11 g_{\m\r} ~~, 
&(A.25) \cr  
& \g_{\m\n} \g_{\s\t} \g^{\m\n\r} = - 38 \g\du{\s\t} \r 
     + 140 \d\du{\[ \s}\r \g_{\t \]} ~~, 
&(A.26) \cr    
& \g\du\m\n \g^{\s\t} \g_{\n\r} = + 6 \g\du{\m\r}{\s\t} 
     +32 \g\du{(\r} {\[ \s}\g\ud{\t \]}{\m)}   
     + 7 g_{\m\r} \g^{\s\t} + 20 \d\du \m{\[\s} \d\du \r{\t\]} ~~, 
&(A.27) \cr 
& \g^\r\g^{\m\n}\g_{\r\s} = + 7 \g\du \s{\m\n} - 18 \d\du \s{\[\m} \g^{\n\]} ~~,
     ~~~~ \g^{\r\s}\g_{\m\n} \g_\r = - 7 \g\du{\m\n}\s - 18 \d\du{\[\m}\s
     \g_{\n\]} ~~, 
&(A.28) \cr 
&\g^{\r\s\t} \g_{\m\n} \g_\r = + 6 \g\du{\m\n}{\s\t} + 32
\d\du{\[\m}{\[\s} 
     \g\du{\n\]}{\t\]} - 20 \d\du{\[\m}\s \d\du{\n\]}\t~~, 
&(A.29)  \cr
& \g^{\[2\]\s\t\o} \g_\m \g\low{\[2\]} = + 40 \g\du\m{\s\t\o} 
     - 216 \d\du\m{\[\s} \g^{\t\o\]} ~~, 
&(A.30) \cr 
& \g^{\[2\]\t\l\o} \g_{\m\n\r}\g\low{\[2\]} 
    = -144 \d\du{\[\m}{\[\t}\g\du{\n\r\]}{\l\o\]} 
     -720 \d\du{\[\m}{\[\t}\d\du\n\l \g\du{\r\]}{\o\]} 
     + 432 \d\du{\[\m}{\[\t} \d\du{\n}\l \d\du{\r\]}{\o\]}~~, 
&(A.31) \cr 
& \g^{\[2\]\m\n\r}\g_{\s\t} \g\low{\[2\]} = - 16\g\du{\s\t}{\m\n\r} 
      -240 \d\du{\[\s}{\[\m} \g\du{\t\]}{\n\r\]}  
     + 432 \d\du{\[\s}{\[\m} \d\du{\t\]}\n \g^{\r\]} ~~, 
&(A.32) \cr 
& \g^{\m\n\r\s\t} \g_{\l\o} \g_\t = + 4 \g\du{\l\o}{\m\n\r\s} 
     + 48 \d\du{\[\l}{\[\m} \g\du{\o\]}{\n\r\s\]} 
     -96\d\du{\[\l}{\[\m} \d\du{\o\]}\n \g^{\r\s\]} ~~, 
&(A.33) \cr
& \g^{\m\n\r}\g_{\l\o} \g\du\m\s = + 5\g\du{\l\o}{\n\r\s} 
     + 12 g^{\s\[\r} \g\du{\l\o}{\n\]} 
     - 14 \d\du{\[\l}\s \g\du{\o\]}{\n\r} 
     + 28 \d\du{\[\l}{\[\n} \g\du{\o\]}{\r\]\s} \cr 
&  ~~~~~ ~~~~~ ~~~~~ ~~~~~  - 18 \d\du{\[\l} {\[\n} \d\du{\o\]}{\r\]} \g^\s  
     +32\d\du{\[\l}{\[\n} g^{\r\]\s} \g_{\o\]} 
     -36 \d\du{\[\l}{\[\n|} \d\du{\o\]}\s \g^{| \r\]} ~~, 
&(A.34) \cr 
& \g^{\n\r\s\t} \g_{\l\o} \g_{\s\t} = - 26 \g\du{\l\o}{\n\r} 
     - 216 \d\du{\[\l}{\[\n} \g\du{\o\]}{\r\]} 
    + 180 \d\du{\[\l}\n\d\du{\o\]}\r ~~, 
&(A.35) \cr 
& \g^\r\g_{\s\t} \g\du\r{\l\m\n} = + 5 \g\du{\s\t}{\l\m\n} 
    -42 \d\du{\[\s} {\[\l} \g\du{\t\]}{\m\n\]} 
    - 54 \d\du\s{\[\l} \d\du\t\m \g^{\n\]} ~~, 
&(A.36) \cr
& \g^{\[\m\n} \g_{\l\o} \g^{\r\s\]} = \g\du{\l\o}{\m\n\r\s} 
     + 4 \d\du{\[\l}{\[\m} \d\du{\o\]}\n \g^{\r\s\]} ~~, 
&(A.37) \cr 
& \g_{\[\m} \g^{\r\s} \g_{\n\]} = + \g\du{\m\n}{\r\s}
       + 2 \d\du{\[\m}\r \d\du{\n\]}\s ~~, 
&(A.38) \cr 
& \g_{\[\m} \g^{\r\s\t\o} \g_{\n\]} = \g\du{\m\n}{\r\s\t\o} 
     + 12 \d\du{\[\m}{\[\r} \d\du{\n\]}\s \g^{\t\o\]} ~~, 
&(A.39) \cr 
& \g^{\[2\]} \g_{\l\o} \g\low{\[2\]}{}^{\m\n\r\s} 
      = - 8 \g\du{\l\o}{\m\n\r\s} 
     + 224 \d\du{\[\l}{\[\m} \g\du{\o\]}{\n\r\s \]} 
     + 672 \d\du{\[\l}{\[\m}\d\du{\o\]}\n \g^{\r\s\]}  ~~, 
&(A.40) \cr 
& \g^{\r\s} \g_{\m\n}\g\du{\r\s}{\t\l} = - 26 \g\du{\m\n}{\t\l} 
     + 216 \d\du{\[\m}{\[\t} \g\du{\n\]}{\l\]} 
       + 180 \d\du{\[\m}\t \d\du{\n\]} \l ~~, 
&(A.41) \cr 
& \g^{\[3\] \m\n\r} \g_{\s\t} \g\low{\[3\]} 
     = + 1008 \d\du{\[\s}{\[\m} \g\du{\t\]}{\n\r\]} 
     - 3024 \d\du{\[\s}{\[\m} \d\du{\t\]} \n \g^{\r\]}   ~~, 
&(A.42) \cr 
& \g^{\[ 2 \] \m\n\r\s} \g_{\l\o} \g\low{\[ 2 \]} 
     = - 8 \g\du{\l\o}{\m\n\r\s} 
     - 224\d\du{\[\l}{\[ \m} \g\du{\o \]}{\n\r\s \]} 
      + 672 \d\du{\[\l}{\[ \m} \d\du{\o \]}\n \g^{\r\s \]} ~~, 
&(A.43) \cr 
& \g^{ \[ 3 \] \[ \m\n\r } \g_{\l\o} \g\low{\[ 3 \]}{}^{\s \]} 
     = - 24\g\du{\l\o}{\m\n\r\s} + 336 \d\du{\[\l} {\[ \m}
     \g\du{\o\]}{\n\r\s \]} ~~, 
&(A.44) \cr  
&\g\low{\[ 3 \]} \g\low{\[ 2 \]} \g^{\[ 3 \]} = - 240 \g\low{\[ 2 \]} ~~,
     ~~~~ 
     \g\low{\[ 4 \]} \g\low{\[ 2 \]} \g^{\[ 4 \]} = + 360\g\low{\[3\]} ~~,  
&(A.45) \cr 
& \g\low{\[ 2 \]} \g\low{\[ 3\]} \g^{\[ 2 \]} = -24\g\low{\[ 3 \]} ~~, ~~~~
     \g\low{\[ 3 \]'} \g\low{\[ 3 \]} \g^{\[ 3 \]'}  
     = + 12\g\low{\[ 3 \]} ~~,\cr 
& \g\low{\[4\]} \g\low{\[ 3 \]} \g^{\[ 4 \]} = -648\g\low{\[ 3 \]} ~~, 
     ~~~~ 
   \g\low{\[ 5 \]} \g\low{\[ 3 \]} \g^{\[ 5 \]} = +4320 \g\low{\[ 3 \]} ~~, 
     ~~~~ 
   \g\low{\[ 6 \]} \g\low{\[ 3 \]} \g^{\[ 6 \]}= 0 {~~,~~~~~ ~~~~~ ~~~~~ }  
&(A.46) \cr  
& \g\low{\[ 2 \]} \g\low{\[ 4 \]} \g^{\[ 2 \]} = -4 \g\low{\[ 4 \]} ~~,
     ~~~~\g\low{\[ 3 \]} \g\low{\[ 4 \]} \g^{\[ 3 \]} = +72
     \g\low{\[ 4 \]} ~~, ~~~~
     \g\low{\[ 4 \]'} \g\low{\[ 4 \]} \g^{\[4\]'} = -408 \g\low{\[ 4 \]}
     ~~,\cr     
& \g\low{\[ 5 \]} \g\low{\[ 4 \]} \g^{\[ 5\]} = +960 \g\low{\[ 4 \]} ~~,
     ~~~~  \g\low{\[ 6 \]} \g\low{\[ 4 \]} \g^{\[ 6 \]} =
     -20160 \g\low{\[ 4 \]} ~~,  
&(A.47) \cr
& \g\low{\[ 2 \]} \g\low{\[ 5 \]} \g^{\[ 2 \]} = +8 \g\low{\[ 5 \]} ~~, ~~~~ 
     \g\low{\[ 3 \]} \g\low{\[ 5 \]} \g^{\[ 3\]} = -60
    \g\low{\[ 5 \]} ~~, ~~~~ 
    \g\low{\[ 4 \]} \g\low{\[ 5 \]} \g^{\[ 4 \]} = +120 \g\low{\[ 5 \]} ~~, \cr 
& \g\low{\[ 5 \]'} \g\low{\[ 5 \]} \g^{\[ 5 \]'} = -2400 \g\low{\[ 5 \]} ~~, 
      ~~~~ \g\low{\[ 6 \]} \g\low{\[ 5 \]} \g^{\[ 6\]} = 0 ~~, 
&(A.48) \cr 
& \g\low{\[ 2 \]} \g\low{\[ 6 \]} \g^{\[ 2 \]} = +12 \g\low{\[ 6 \]} ~~, ~~~~
     \g\low{\[ 3 \]} \g\low{\[ 6 \]} \g^{\[ 3 \]} = 0
     ~~, ~~~~
     \g\low{\[ 4 \]} \g\low{\[ 6 \]} \g^{\[ 4 \]}  = +360 \g\low{\[ 6 \]} ~~, \cr 
& \g\low{\[ 5 \]} \g\low{\[ 6 \]} \g^{\[ 5 \]} = 0 ~~, 
      ~~~~\g\low{\[ 6 \]'} \g\low{\[ 6 \]} \g^{\[ 6 \]'} 
     = +14400 \g\low{\[ 6 \]} ~~.     
&(A.49) \cr } $$
Here our (anti)symmetrizations are normalized, prepared for component
computations, as their indices show.  We also mention 
a small point that these $~\g\-$matrix algebra depends only on the total
space-time dimensions, but {\it not} on the signature.  Therefore, (A.23) -
(A.49) are useful also for theories in 12D with other signatures, such as 
$~(+,+,\cdots,+,-)$~  or the Euclidian one $~(+,+,\cdots,+)$.

\bigskip\bigskip
\bigskip\bigskip


\centerline{\bf Appendix B:~~Useful Relationships in $~D=11+2$}
\bigskip

Similarly to the previous Appendix A, we list up here some useful relationships,
in $~D=11+2$, which are by no means exhaustive collection, but will be of
great help for superspace manipulations.  Note also that these relations
for 13D should not be confused with those in Appendix A for 12D.   
$$ \li{
& \g\low{\[n\]} = \fracm{(-1)^{(n+1)(n+2)/2}}{(13-n)!}
    \e\Du{\[n\]} {\[13-n\]} \g\low{\[13-n\]} ~~~~(0\le n\le 13) ~~,
&(B.1) \cr  
\noalign{\vskip0.1in}
& \d\du{(\a}\b \d\du{\g)} \d 
    = + \frac1{64} \big(\g^{\[2\]} \big)_{\a\g} 
           \big(\g\low{\[2\]} \big)^{\b\d} 
     + \frac1{192}\big(\g^{\[3\]} \big)_{\a\g}  
          \big(\g\low{\[3\]} \big)^{\b\d} 
     + \frac1{32(6!)} \big(\g^{\[6\]} \big)_{\a\g}  
          \big(\g\low{\[6\]} \big)^{\b\d} ~~, 
&(B.2) \cr  
& \big(\g^a \big)\du{(\a|}\b \big(\g_a \big)\du{|\g)}\d 
    = \frac9{64} \big(\g^{\[2\]} \big)_{\a\g}  
          \big(\g\low{\[2\]} \big)^{\b\d} 
      - \frac7{192} \big(\g^{\[3\]} \big)_{\a\g}  
          \big(\g\low{\[3\]} \big)^{\b\d} 
       + \frac1{32(6!)} \big(\g^{\[6\]} \big)_{\a\g}  
          \big(\g\low{\[6\]} \big)^{\b\d} 
&(B.3) \cr  
& ~~~~~ ~~~~~ ~~~~~ ~~~~ = \d\du{(\a}\b \d\du{\g)}\d 
         + \frac18\big(\g^{\[2\]} \big)_{\a\g}  
          \big(\g\low{\[2\]} \big)^{\b\d} 
         - \frac1{24}\big(\g^{\[3\]} \big)_{\a\g}  
          \big(\g\low{\[3\]} \big)^{\b\d} ~~,  
&(B.4) \cr  
& \big( \g\low{\[2\]} \big)_{(\a\b|} \big( \g^{\[2\]} \big)_{|\g\d)} 
      = +\frac1 3 \big( \g\low{\[3\]} \big)_{(\a\b|} 
          \big( \g^{\[3\]} \big)_{|\g\d)} 
      = - \frac1 {6!} \big( \g\low{\[6\]} \big)_{(\a\b|} 
          \big( \g^{\[6\]} \big)_{|\g\d)} ~~, 
&(B.5) \cr  
& \g^a \g^{\[n\]} \g_a = (-1)^n (13-2n) \g^{\[n\]} ~~, 
&(B.6) \cr  
& \g^{a b} \g^{c d} \g_{a b} = - 68
\g^{c d} ~~, ~~~~
     \g_{a b c} \g^{d e} \g^{a b c} = - 396\g^{d e}~~, 
&(B.7) \cr 
& \g_{a d} \g^{b c} \g^d = + 8 \g\du a {b c} + 10 \d\du a{\[b} \g^{c\]} ~~, 
    ~~~~ \g^d \g^{b c} \g_{a d} 
     = - 8 \g\du a{b c} + 10 \d\du a {\[b} \g^{c\]} ~~, 
&(B.8) \cr    
& \g_{a d} \g^{b c d} = +10\g\du a{b c} + 11\d\du a{\[ b}\g^{c\]} ~~,  
& (B.9) \cr    
& \g_{a e} \g^{c d} \g\du b e = - 7 \g\du{a b}{c d} 
    - 8 \eta_{a b} \g^{c d} + 9 \d\du{(a}{\[c} \g\du{b)}{d\]} 
     - 11\d\du{\[a}c \d\du {b\]}{d} ~~, 
&(B.10) \cr 
& \g_e \g_{a b} \g^{c d e} = + 7 \g\du{a b}{c d} 
     - 9 \d\du{\[a}{\[c} \g\du{b\]}{d\]} - 11 \d\du{\[a} c \d\du{b\]} d~~, 
&(B.11) \cr 
& \g^{c d e} \g_{a b} \g_e = + 7 \g\du{a b}{c d} 
     + 9 \d\du{\[a}{\[c}\g\du{b\]}{d\]} - 11 \d\du{\[a}c\d\du{b\]} d ~~, 
&(B.12) \cr   
& \g^d\g^{a b c} \g_{d e} = + 6 \g\du e{a b c} - 4\d\du e{\[a} \g^{b c\]} ~~,
&(B.13) \cr 
& \g^{a b} \g^{d e} \g_{a b c} = - 52 \g\ud{d e} c 
     + 88 \d\du c{\[d} \g^{e\]} ~~, 
    ~~~~\g_{d e a} \g^{b c} \g^{d e} 
       = - 52\g\du a{b c} - 88 \du a{\[b} \g^{c\]} {~~,~~~~~ ~~~~~}   
&(B.14) \cr 
& \g\low{\[3\]} \g^{b c} \g\ud{\[3\]} a 
     = - 240 \g\du a{b c} + 660\d\du a{\[b} \g^{c\]}~~,   
&(B.15) \cr 
& \g\low{\[2\] a} \g^{c d} \g\ud{\[2\]} b = - 38 \g\du{a b}{c d} 
    + 70 \d\du{(a}{\[c} \g\Du{b)}{d\]} 
     - 110 \d\du a{\[c} \d\du b{d\]} - 52 \eta_{a b} \g^{c d} ~~,  
&(B.16) \cr 
& \g\low{\[2\]} \g^{c d} \g\du{a b}{\[2\]} = - 38 \g\du{a b}{c d} 
     - 70 \g\d\du{\[a}{\[c} \g\du{b\]} {d\]} 
     + 110 \d\du a{\[c} \d\du b{d\]} ~~, 
&(B.17) \cr    
& \g\low{\[3\]} \g^{a b} \g\ud{\[3\]}{c d} = - 126 \g\ud{a b}{c d} 
     - 450 \d\du{\[c} {\[a} \g\du{d\]}{b\]} 
     + 990 \d\du c{\[a}\d\du d{b\]} ~~
&(B.18) \cr    
& \g^b\g^{a_1\cdots a_6} \g_{b c} 
    = - \frac1{60} \d\du c{\[a_1} \g^{a_2\cdots a_6\]} ~~. 
&(B.19) \cr 
} $$
As their indices show, these formulae are prepared in the superspace notation,
and as such, the antisymmetrization is like $~B_{\[a b) } \equiv B_{a b} 
\mp B_{b a}$.

\bigskip\bigskip
\bigskip\bigskip

\centerline{\bf Appendix C:~~Lorentz Algebra with Modified $~\TildeM_{a b}$}  
\bigskip

In this Appendix, we examine the significance of our algebra with $~\TildeM$~ 
defined by (2.2.1), which is highly non-trivial.  We know that the Lorentz 
covariance in the extra dimensions out of 12D is lost, and therefore we need to 
confirm at least the ordinary 10D.  To this end, we use the local Lorentz 
indices $~{\scst i,~j,~\cdots~=~(0),~(1),~\cdots,~(9)}$~ for the purely 10D
part.  

We first note the relationships
$$R\du{A B \g}\d = - \frac12 R\du{A B}{i j} 
         \big( \TildeM_{i j}\big)\du\g\d~~, ~~~~
R\du{A B\Dot\g}{\Dot\d} = - \frac12 R\du{A B}{i j} 
         \big( \TildeM_{i j}\big)\du{\Dot\g}{\Dot\d} ~~,
\eqno(C.1) $$
which is confirmed also by the use of extra condition (2.3.6).  This implies 
that $~R\du{A B \g}\d$~ behaves as if 
it were within the 10D sub-manifold, realizing only $~SO(9,1)$~ subgroup of 
$~SO(10,2)$.  This feature is also valid for the combination $~\phi\du M{a b}
\TildeM_{a b}$, because only the combination
$~\phi\du M{i j}\big(\TildeM_{i j}\big)
\du{\ul\a}{\ul\b}$~ survives, while other components  
$~\phi\du M{\pm i}\big(\TildeM_{\pm i}\big)\du{\ul\a}{\ul\b}$~ and 
$~\phi\du M{+ -}\big(\TildeM_{+ -}\big)\du{\ul\a}{\ul\b}$~
vanish,\footnotew{Here 
the indices $~{\scst\ul\a,~\ul\b,~\cdots}$~ denote general spinorial indices 
both dotted and undotted.} due to either 
the definition (2.1.6), or the extra constraint (2.3.6).  We now see how our 
system realizes only the $~SO(9,1)$~ sub-algebra in the total 12D.    

We next compute the commutators among $~\TildeM'$s.  Out of $~\[ \TildeM_{a b}, 
\TildeM_{c d} \]\du\g\d$, there are six different combinations, when 10D 
indices are distinguished from the extra coordinates $~\pm$:  
(i) $\[ \TildeM_{i j}, \TildeM_{k l} \]\du{\un\g}{\un\d}$, 
(ii) $\[ \TildeM_{i j}, \TildeM_{+k} \]\du{\un\g}{\un\d}$, 
(iii) $\[ \TildeM_{i j}, \TildeM_{+-} \]\du{\un\g}{\un\d}$, 
(iv) $\[ \TildeM_{+i}, \TildeM_{+j} \]\du{\un\g}{\un\d}$, 
(v) $\[ \TildeM_{+i}, \TildeM_{+-} \]\du{\un\g}{\un\d}$, 
(vi) $\[ \TildeM_{+-}, \TildeM_{+-} \]\du{\un\g}{\un\d}$.  
The first combination is easy to satisfy the 10D algebra, when we use (2.2.6b).
The sector (ii) is also straightforward, which agrees with 
the fully covariant 12D algebra
$$ \[ \calM_{a b}, \calM^{c d} \] 
     = - \d\du{\[a|}{\[ c|} \calM\du{|b\]}{|d \]} ~~. 
\eqno(C.2) $$
All of the sectors (iii), (iv) and (vi) vanish, satisfying (C.2), while (v) 
yields the result $~(1/2) \eta\low{+-} \big( \TildeM_{i +} \big) 
\du{\un\g}{\un\d}$, 
with the factor 2 discrepancy compared with (C.2), which, however,  
causes no problem, when we inspect the Jacobi identities next.  

We can confirm the Jacobi identities for our non-trivially 
modified $~\TildeM_{a b}$, which form the most important foundation for  
Bianchi identities in superspace:
$$ \[ \TildeM_{a b},\[\TildeM_{c d}, \TildeM_{e f} \] \] 
+ \[ \TildeM_{c d},\[\TildeM_{e f}, \TildeM_{a b} \] \] 
+ \[ \TildeM_{e f},\[\TildeM_{a b}, \TildeM_{c d} \] \] \equiv 0~~. 
\eqno(C.3) $$
There are ten different combinations for the indices 
$~{\scst \[ a b \] \[c d\] \[e f\]}$~ when 10D indices are
distinguished from the extra ones, symbolically categorized as
(i) ${\scst \[ i j\] \[k l\] \[ m n \]}$, 
(ii) ${\scst \[ i j\] \[k l\] \[ + m\]}$, 
(iii) ${\scst \[ i j\] \[+ k\] \[ + l \]}$, 
(iv) ${\scst \[ + i \] \[+ j\] \[ + k\]}$, 
(v) ${\scst \[ i j\] \[k l\] \[ + - \]}$, 
(vi) ${\scst \[ i j\] \[+ k\] \[ + -\]}$, 
(vii) ${\scst \[ i j\] \[+ -\] \[ + - \]}$, 
(viii) ${\scst \[ + i\] \[+ - \] \[ + - \]}$, 
(ix) ${\scst \[ + i \] \[+ j \] \[ + - \]}$, 
(x) ${\scst \[ + - \] \[+ - \] \[ + - \]}$.  
Among these (i) is easy to see, because $~\TildeM_{a b}$~ 
satisfies the $~SO(9,1)$~ 
sub-algebra, when all the indices are 10D.  The sectors (ii) - (x) are
all easily shown to vanish, when the results in the basic commutators are used.
In particular, despite of the factor 2 discrepancy mentioned in the previous
paragraph, we can confirm the satisfaction of the sectors (vi) and (ix). 

There is, however, some caveat needed about the 
basic algebra structure in our formulation, associated with the Jacobi
identities among our generators $~\TildeM,~P,~Q$.  As careful readers may
have already noticed, the success of our superspace formulation does not
necessarily corresponds to the satisfactions of all of these Jacobi identities
among generators, because our superspace Bianchi identities hold only modulo our
constraint, {\it e.g.,} (2.3.9) for $~D=12,\, N=1$.  As a matter of
fact, among the ten possible sectors (I) $\TildeM \TildeM
\TildeM$, ~(II) $ P P P $, ~(III) $ Q Q Q  $, ~(IV) $ \TildeM 
\TildeM P$, ~(V) 
$\TildeM \TildeM Q $, ~(VI) $ P P \TildeM $, ~(VII) $P P Q $, ~(VIII)
$ Q Q \TildeM $, ~(IX) $Q Q P$, ~(X) $\TildeM P Q$~ of Jacobi identities, we 
can easily confirm that all of these identities are satisfied, 
as long as $~\TildeM$'s carry only purely 10D indices $~{\scst i,~j,~\cdots}$,
while there are some  non-vanishing components, {\it e.g.,} in the cases of 
(V) $\[ \TildeM_{+i}, \[ \TildeM_{j k}, Q_\a\]\] + (\hbox{2 perms.})\neq
0$~ and ~(VIII) $ \[ \TildeM_{+i}, \{ Q_\a,Q_\b \} \] + (\hbox{2 perms.})\neq
0$.  This poses no problem, as we have stressed also in subsection 2.1 as well 
as in this Appendix, because these non-vanishing Jacobi identities become
irrelevant under our constraints such as $~\phi\du A{+ i} = 0$, at the
superspace Bianchi identity level $~\[ \nabla_A, \[ \nabla_B, \nabla_C\} \} +
(\hbox{2 perms.}) \equiv 0$~ in terms of $~\nabla_A$~ instead of the
generators.   This feature is one of the most peculiar and important aspects in
our formulation with no other analogs in other theories, which should be always
kept in mind in future applications.

\bigskip\bigskip
\bigskip\bigskip

\centerline{\bf Appendix D:~~Variation of Extra Constraints 
under Supersymmetry}  
\bigskip

In this Appendix, we analyze the consistency between our extra constraints imposed 
on our fields and supersymmetry.  Here we concentrate on the $~D=12,\,N=2$~ 
supergravity in section 4.2, giving some typical examples.  As
such an example,  we consider the variation of the following equation 
in (4.2.4a) under supersymmetry at the lowest order:    
$$ \li{ \d_Q \big( R\du\m\n{}_{r s} \phinu \big) 
     & ~ = + (D^\n\varphi) D_\m\big(\d\omega_{\n r s} \big)  -
       (D^\n\varphi) D_\n\big(\d\omega_{\m r s} \big) \cr 
& ~ = - 2 (D^\n\varphi) \big( \Bar\e \g\du{\[ r|} \t D_{|s\]}{\cal R}_{\m\n} 
     \big) \phitau +\hbox{c.c.}  = 0 ~~.  
&(D.1) \cr } $$
Here we have used another extra constraint (4.2.4c).  Another interesting example
involving the coset $~SU(1,1)/U(1)$~ is the variation of the 
first equation in (4.2.4b):  
$$ \li{\d_Q \big( P^\m \phimu\big) = & \, 
      - \e_{\a\b} \big(D^\m\varphi\big) (\d_Q V\du+\a) \partial_\m V\du+\b
      - \e_{\a\b} V\du+\a \big(D^\m\varphi\big)\partial_\m(\d_Q V\du+\a ) \cr   
= & \,- \e_{\a\b} V\du+\a V\du-\b \big(D^\m\varphi\big) \big( \Bar\e^* \g^\n
     D_\m \l\big)\phinu = 0 ~~.  
&(D.2) \cr } $$
Here use is made of the second equation in (4.2.4d).  In a very similar fashion, 
we can see that the variations of all of our extra conditions (4.2.4) under
supersymmetry vanish, upon using other extra constraints, 
up to higher-order terms
which we skip in this paper.  Even though we skipped similar analysis for 
the $~N=1$~ supergravity, it can be easily performed in a more
direct manner in superspace.

\vfill\eject

\immediate\closeout\rfile\writestoppt
\baselineskip=11pt\centerline{{\bf References}}
\font\smallreffonts=cmr9 \font\it=cmti9 \font\bf=cmbx9%
\bigskip{ {\smallreffonts%
\parindent=15pt\escapechar=` \input refs.tmp\vfill\eject}}

\end{document}